%% file: main.tex
\documentclass[longauth]{aa}

\usepackage{euclid}
\usepackage{graphicx}
\setcitestyle{semicolon}
\usepackage{scalerel}
\usepackage{ulem}

\usepackage[table]{xcolor}
\usepackage[bottom]{footmisc}

\makeatletter
\renewcommand{\eprint}[2][]{\unskip}
\AtBeginDocument{}
\makeatother

\usepackage{txfonts}
\usepackage{glossaries}
\usepackage[pdfencoding=auto,psdextra]{hyperref}
\hypersetup{
    colorlinks=true,
    linkcolor=blue,
    filecolor=magenta,      
    urlcolor=blue,
    citecolor=blue
}
\makeatletter
\renewcommand*\aa@pageof{, page \thepage{} of \pageref*{LastPage}}
\makeatother

\usepackage[utf8]{inputenc}

\usepackage[switch, modulo]{lineno}
\renewcommand{\linenumbers}[0]{}
\usepackage{multirow}

\newacronym{2pcf}{2PCF}{2-point correlation function}
\newacronym{3pcf}{3PCF}{3-point correlation function}
\newacronym{hod}{HOD}{halo occupation distribution}
\newacronym{fob}{FoB}{figure of bias}
\newacronym{dic}{DIC}{deviance information criterion}
\newacronym{pt}{PT}{perturbation theory}
\newacronym{eft}{EFTofLSS}{Effective Field Theory of Large Scale Structure}
\newacronym{bao}{BAO}{baryon-acoustic-oscillation}
\newacronym{mcmc}{MCMC}{Markov Chain Monte Carlo}

\renewcommand{\vec}{\mathbf}
\newcommand{\nn}{\nonumber}

\newcommand{\xv}{\mathbf{x}}

\newcommand{\kv}{\mathbf{k}}
\newcommand{\nv}{\mathbf{n}}

\newcommand{\qv}{\mathbf{q}}

\newcommand{\Pl}{P_{\rm L}}

\newcommand{\kmax}{k_{\rm max}}
\newcommand{\kmaxP}{k_{\rm max}^{P}}
\newcommand{\kmaxB}{k_{\rm max}^{B}}

\newcommand{\omegac}{\omega_{\rm c}}
\newcommand{\As}{A_{\rm s}}
\newcommand{\G}{{\mathcal G}}
\newcommand{\dd}{{\mathrm d}}

\newcommand{\Ms}{\, h^{-1} \, M_\odot}
\newcommand{\Mpc}{\, h^{-1} \, {\rm Mpc}}

\newcommand{\cGpc}{\, h^{-3} \, {\rm Gpc}^3}
\newcommand{\kMpc}{\, h \, {\rm Mpc}^{-1}}
\newcommand{\kcMpc}{\, h^3 \, {\rm Mpc}^{-3}}
\newcommand{\eq}[1]{Eq.~(\ref{#1})}

\newcommand{\Ps}{P_{s}}

\newcommand{\Bs}{B_{s}}
\newcommand{\bGtwo}{b_{\mathcal{G}_2}}

\usepackage{fontawesome5}
\usepackage{hyperref}
\makeatletter
\newcommand{\github}[1]{%
   \href{#1}{\faGithubSquare}%
}
\makeatother

\begin{document}
%
%
\title{\Euclid preparation}
\subtitle{Galaxy power spectrum and bispectrum modelling}

\include{authors} 
%
%
\abstract{Higher-order correlation functions of the large-scale galaxy distribution offer access to information beyond that contained in standard 2-point statistics such as the power spectrum. In this work we assess this potential for the \Euclid mission using synthetic catalogues of H$\alpha$ galaxies based on the 54 $\cGpc$ Flagship I simulation, designed to reproduce the \Euclid spectroscopic sample. We comprehensively validate the one-loop galaxy power spectrum and tree-level bispectrum predictions from perturbation theory in both real and redshift space. Assuming scale cuts consistent with our previous power spectrum study on the same catalogues, this modelling yields unbiased cosmological constraints for the bispectrum up to $k_{\rm max} = 0.15\,\kMpc$ in real space and $0.08 \, (0.1)\,\kMpc$ at the lowest (highest) redshift, corresponding to $z=0.9$ ($z=1.8$), for the monopole and quadrupole in redshift space using statistical uncertainties corresponding to the full simulation volume. With these scale cuts, adding bispectrum information to the power spectrum improves constraints on the amplitude of scalar perturbations and the matter density by up to 30\%, increasing the overall figure of merit for key cosmological parameters by a factor of about 2.5. Similar conclusions hold when statistical uncertainties are rescaled to a \Euclid-like volume, highlighting the importance of the bispectrum for fully exploiting the forthcoming \Euclid data. Our analysis also provides the first detailed characterisation of the nonlinear bias model of H$\alpha$ emitters, showing that bias relations calibrated on low-resolution \textit{N}-body simulations do not adequately describe the clustering of H$\alpha$ galaxies at low redshift, whereas excursion-set and co-evolution relations for tidal biases remain accurate. Finally, we benchmark six independent modelling codes, finding excellent agreement under matched assumptions, which testifies to the robustness of our results and provides a benchmark for the official analysis pipeline.}
%
%
%
%

\titlerunning{Galaxy power spectrum and bispectrum modelling}
\authorrunning{Euclid Collaboration: K. Pardede et al.} 
   
\maketitle
%
%
%
%
\section{\label{sec:Intro} Introduction}

Stage IV missions such as \Euclid  \citep{LaureijsEtal2011, MellierEtal2025}, the Dark Energy Spectroscopic Survey \citep[DESI;][]{AghamousaEtal2016}, the Vera Rubin Observatory \citep{IvezicEtal2009}, and the \textit{Nancy Grace Roman} Telescope \citep{WangEtal2022} are expected to address several fundamental questions in cosmology, including the physics of dark energy, the properties of dark matter, and the mass of neutrinos.

Whether spectroscopic surveys aiming at the analysis of the 3-dimensional galaxy distribution or photometric catalogues providing weak lensing maps, the primary observables for all such surveys consist of 2-point correlations of cosmological perturbations.  The \gls{2pcf}, or the power spectrum in Fourier space, captures indeed all the information available on the statistical properties of a given Gaussian random field. However, the large-scale structure, and specifically the distribution of galaxies at large scales, is a highly non-Gaussian field, as a consequence of gravitational instability, nonlinear galaxy bias, and redshift-space distortions.  

The galaxy bispectrum, that is, the counterpart of the \gls{3pcf} in Fourier space, is the lowest-order correlator encoding the non-Gaussian properties of the galaxy distribution. The original interest in this probe lies in its ability to disentangle nonlinear effects of different origin, providing, for instance, independent constraints on galaxy bias parameters and on the amplitude of linear matter perturbations  \citep{FryGaztanaga1993, MatarreseVerdeHeavens1997, ScoccimarroEtal1998, ScoccimarroCouchmanFrieman1999}. In addition, it represents a direct test of the initial conditions, potentially discriminating inflationary models \citep{Scoccimarro2000A, VerdeEtal2000, ScoccimarroSefusattiZaldarriaga2004, SefusattiKomatsu2007}     

The galaxy bispectrum was measured from the first galaxy surveys more than twenty years ago \citep{BaumgartFry1991, ScoccimarroEtal2001B, FeldmanEtal2001, VerdeEtal2002} all the way to the most recent Baryonic Oscillation Spectroscopic Survey \citep[BOSS;][]{GilMarinEtal2017} and DESI \citep{NovellMasotEtal2025} observations. Yet, despite this relatively long history, the analysis of the bispectrum is only recently becoming part of a standard, complete analysis of spectroscopic data sets. This is due to the higher complexity of its estimation, its modelling in \gls{pt}, and the accurate description of finite-volume effects. In addition, since the signal is distributed across a larger number of measurable configurations, obtaining a purely numerical estimate of its covariance properties is significantly more demanding than for the power spectrum, while an analytical prediction is even more challenging. 

Due to different methodological assumptions in tackling these problems, joint power spectrum-bispectrum analysis in the recent literature has led to somehow different results on the improvement on the power spectrum-only analysis, even on the same data set. The broad picture that is emerging, however, is that the bispectrum indeed provides, on one hand, tight constraints on linear and nonlinear galaxy bias, providing an important consistency check for the perturbative model. On the other hand, it allows for improvements, with respect to the power spectrum-only case, on the determination of cosmological parameters of the order of tens of percent, depending on the overall signal corresponding to a given data vector (including redshift-space anisotropies), on the modelling choice, and on the covariance estimate \citep{GagraniSamushia2017, YankelevichPorciani2019, HahnEtal2020, OddoEtal2020, GualdiVerde2020, HahnVillaescusaNavarro2021, MoradinezhadDizgahEtal2021, GualdiGilMarinVerde2021,  AgarwalEtal2021, OddoEtal2021, HahnEtal2024}. In addition, a similar improvement can be obtained on constraints of local primordial non-Gaussianity, while non-local models can only be constrained by bispectrum measurements \citep{DAmicoEtal2024, CabassEtal2022, CabassEtal2022B, CagliariEtal2025}.    

In this work, we present a test of the joint modelling of the galaxy power spectrum and bispectrum against a synthetic H$\alpha$ galaxy catalogue, representative of the \Euclid spectroscopic sample, performed in snapshots of the Flagship I simulation \citep{PotterStadelTeyssier2017}. This is part of the series of \Euclid preparation papers to validate the modelling used to analyse the full shape of 2- and 3-point clustering measurements from the final data sample. In a previous study, \cite{PezzottaEtal2024} examined the performance of the galaxy power spectrum in real space, while ongoing work is focused on the galaxy power spectrum in redshift space (Euclid Collaboration: Camacho et al., in prep.) and on the joint analysis of power spectrum and bispectrum in the specific case of local non-Gaussian initial conditions (Euclid Collaboration: Linde et al., in prep.). A similar analysis of the galaxy \gls{2pcf} in redshift space is performed in Euclid Collaboration: Kärcher et al. (in prep.), while the combined analysis of the galaxy \gls{2pcf} and \gls{3pcf} is investigated in real and redshift space, respectively in \citet{GuidiEtal2025} and Euclid Collaboration: Pugno et al. (in prep.). Additionally, a parallel investigation is ongoing for the BAO signal extracted from the redshift-space \gls{3pcf} (Euclid Collaboration: Moresco et al., in prep.).

In our analysis, we consider the standard estimator of \citet{ScoccimarroCouchmanFrieman1999} for the bispectrum multipoles in redshift space. Alternative choices have been explored in the literature \citep{HashimotoRaseraTaruya2017, SugiyamaEtal2019} along with several compression methods to reduce the size of the data vector \citep[see, e.g.,][and references therein]{SchmittfullBaldaufSeljak2015, ByunEtal2017, GualdiEtal2018, SchmittfullMoradinezhad2021, PhilcoxEtal2021}. We adopt the theoretical description of the galaxy bispectrum based on perturbation theory at tree-level \citep{Fry1984, HivonEtal1995, ScoccimarroCouchmanFrieman1999, ChanScoccimarroSheth2012, BaldaufEtal2012}. This is the leading order contribution, expected to provide an accurate description at large scales.\footnote{Predictions at one-loop level have been also explored over the years \citep{ScoccimarroFrieman1996, Sefusatti2009, AnguloEtal2015B, BaldaufEtal2015A, HashimotoRaseraTaruya2017, EggemeierScoccimarroSmith2019, EggemeierEtal2021, AlkhanishviliEtal2022, PhilcoxEtal2022, DAmicoEtal2024B, DAmicoEtal2024}, although only very recently robust and complete methods for an efficient evaluation are being developed \citep{AnastasiouEtal2024, BaksEtal2025A}.} A recent assessment of the range of validity of this approximation for the matter bispectrum and varying simulation volume can be found in \citet{AlkhanishviliEtal2022},
indicating a reach of $\kmax\simeq 0.12\kMpc$ for a volume of $8\cGpc$, corresponding to the typical size of a redshift bin in the \Euclid full mission \citep{MellierEtal2025}. Similar simulation-based tests of the tree-level model for the bispectrum of tracers in real space indicate a limit of about $\kmax\simeq 0.15\kMpc$ for a BOSS-like volume of $6\cGpc$ \citep{OddoEtal2021, EggemeierEtal2021}. In contrast, in redshift space, the scale cut for the tree-level bispectrum monopole and quadrupole has been set to approximately $0.08\kMpc$ in several BOSS analyses, with a similar range of validity holding as well for the analysis of simulated samples with much larger volumes \citep{DAmicoEtal2020, IvanovEtal2022B, PhilcoxIvanov2022,  PhilcoxEtal2022, RizzoEtal2023, IvanovEtal2023, DAmicoEtal2024}.

In this work, we reassess these results in view of the upcoming analysis of the data collected by the spectroscopic survey of \Euclid, making use of mock H$\alpha$ catalogues over the relevant redshift range, $0.9<z<1.8$. Additionally, we evaluate the validity of different relations among galaxy bias parameters, potentially able to reduce the dimensionality of the model parameter space, in the context of a \Euclid-specific galaxy population.

This article is organised as follows. In Sect~\ref{sec:theo_model} we describe the theoretical model for both the redshift-space galaxy power spectrum and bispectrum. In Sect.~\ref{sec:data} we present the simulated H$\alpha$ galaxy samples used as testing ground, along with the corresponding measurements of the data vectors and covariance matrices of both the power spectrum and bispectrum used in this work. Section~\ref{sec:method} outlines the methodology employed in the fitting procedure, including the performance metrics used to define the range of validity of the model. In Sect.~\ref{sec:real} and Sect.~\ref{sec:redshift} we present the final results for the real- and redshift-space analyses, respectively. Finally, we draw our conclusions in Sect.~\ref{sec:conclusion}.

\section{\label{sec:theo_model} Theoretical model}

We model the galaxy power spectrum and bispectrum in redshift space using perturbation theory \citep[see][for a classical review]{BernardeauEtal2002} within the framework of the Effective Field Theory of Large-Scale Structure \citep[EFTofLSS,][]{BaumannEtal2012, CarrascoHertzbergSenatore2012}. This formulation captures the nonlinear gravitational evolution on mildly nonlinear scales, incorporates all relevant contributions from galaxy bias and stochasticity \citep[see][for a review]{DesjacquesJeongSchmidt2018}, and includes a systematic treatment of the impact of unresolved small-scale physics.

Below, we present the expressions for the redshift-space predictions of the one-loop galaxy power spectrum and the tree-level bispectrum. The real-space counterparts correspond to the same expressions, except that all terms explicitly depending on the logarithmic growth rate, $f$, are set to zero.

\subsection{Definitions and conventions}
\label{sec:definitions}

We adopt the following convention for the Fourier Transform and its inverse,
\begin{align}
        \delta(\kv) & \equiv \int\!\! \dd^3x\, \rm e^{-\rm i\kv\cdot\xv}\,\delta(\xv)\;, \quad{\rm and}\quad 
        \delta(\xv) \equiv \int\!\! \frac{\dd^3k}{(2\pi)^3}\, \rm e^{\,\rm i\kv\cdot\xv}\,\delta(\kv)\;.
        \label{eq:fourier_to_conf_transform}
\end{align}
This leads to the definition of the power spectrum $P(k)$ of a generic density contrast $\delta(\kv)$ given by
\be
    \ave{\delta(\kv_1)\,\delta(\kv_2)} \equiv (2\pi)^3\, \delta_{\rm D}\left(\kv_{12}\right)\, P(k_1)\;,
    \label{eq:def_power_spectrum}
\ee
where $\delta_{\rm D}$ is the Dirac delta function and where we introduced the notation $\kv_{ij}\equiv\kv_i+\kv_j$. The bispectrum is then defined as 
\be
    \ave{\delta(\kv_1)\,\delta(\kv_2)\,\delta(\kv_3)} \equiv (2\pi)^3\, \delta_{\rm D}\left(\kv_{123}\right)\, B(k_1,k_2,k_3)\;.
    \label{eq:def_bispectrum}
\ee

\subsection{Power spectrum}

The one-loop prediction for the power spectrum in redshift space consists of four contributions,
\begin{align}
\label{eq:Ps_not_IRresummed}
\Ps(\kv) &= \Ps^{\rm tree}(\kv) + \Ps^{\rm 1-loop}(\kv) + \Ps^{\rm ctr}(\kv) + \Ps^{\rm stoch}(\kv)\;. 
\end{align}
Here, the first two terms correspond to the linear (tree-level) and one-loop contributions, given by
\begin{equation}
    \label{eq:Ps_tree}
    P^{\rm tree}_{s}(\kv) =  Z_1^2(\kv) \, \Pl(k) \;,
\end{equation}
and 
\begin{align}
P^{\rm 1-loop}_{s}(\kv)  =& \,2 \!\int\! \frac{\dd^3 q}{(2\pi)^3} \, Z_2^2(\qv, \kv-\qv) \, \Pl(q) \, \Pl(|\kv-\qv|) 
\nonumber \\
& + 6 ~ Z_1(\kv) \Pl(k) \!\int\! \frac{\dd^3 q}{(2\pi)^3} \,\, \!Z_3(\qv, -\qv, \kv) \,\!\Pl(q)\;,
\end{align}
with $\Pl(k)$ representing the linear matter power spectrum while the kernels $Z_n$ account for nonlinearities from gravitational instability, bias, and redshift-space distortions \citep[see, e.g.,][]{BernardeauEtal2002}.\footnote{We refer the reader to \cite{PerkoEtal2016A} for the explicit expressions for the redshift-space kernels.}  The nonlinear bias contributions, in particular, have been first derived in \citet{McDonaldRoy2009}, \citet{ChanScoccimarroSheth2012}, and \citet{BaldaufEtal2012}, and are captured by the following bias expansion
\begin{equation}\delta_{\rm g}(\xv)=b_1\,\delta(\xv)+\frac12\,b_2\,\delta^2(\xv)+b_{\G_2}\,\G_2(\xv)+b_{\Gamma_3}\,\Gamma_3(\xv)\;,
\end{equation}
where $b_1$ and $b_2$ are the local bias parameters while $b_{\G_2}$ and $b_{\Gamma_3}$ weight the contributions from the non-local-in-matter-density operators \citep[see, e.g.,][for the explicit expressions]{DesjacquesJeongSchmidt2018}. The third term in 
\eq{eq:Ps_not_IRresummed} refers to the contribution from the EFTofLSS counterterms that capture the impact of small-scale physical effects on the large-scale power spectrum,
\begin{align}
\label{P_ct}
\Ps^{\rm ctr}(\kv) =& - 2 \left(c_0 + c_2 f \mu^2 + c_4 f^2 \mu^4\right) k^2 \Pl(k) 
\nonumber \\& +
\tilde{c}_{\nabla^2 \delta} f^4 \mu^4 \left(b_1 + f\mu^2\right)^2 k^4 \Pl(k)\;,
\end{align}
where $\mu=\hat{\kv}\cdot\hat{\nv}$, with $\hat{\kv}$ denoting the wavevector direction, $\hat{\nv}$ the line-of-sight (LOS) unit vector. The standard counterterms in redshift-space are denoted by $c_0$, $c_2$, and $c_4$, while $\tilde{c}_{\nabla^2 \delta}$ is the next-to-leading order counterterm accounting for higher-derivative terms from the nonlinear redshift-space distortion mapping. Finally, we model the stochastic contribution to the power spectrum as
\begin{align}
\Ps^{\rm stoch}(\kv) &= \Big(1+\alpha_{\rm P} + \epsilon_{0, k^2} k^2 \Big) \frac{1}{\bar{n}}\;,
\end{align}
where $\bar{n}$ is the galaxy density and $\alpha_{\rm P}$ and $\epsilon_{0,k^2}$  represent a constant and a scale-dependent deviations from Poisson shot-noise.  We neglect an additional anisotropic shot-noise term $\bar{n}^{-1} \epsilon_{2, k^2} \mu^2 k^2$ due to its strong degeneracy with the next-to-leading order counterterm $\tilde{c}_{\nabla^2 \delta}$.

\subsection{Bispectrum}

The redshift-space galaxy bispectrum is described at tree-level in perturbation theory as
\be
\label{eq:Bmodel}
\Bs(\kv_1,\kv_2,\kv_3)=\Bs^{\rm tree}(\kv_1,\kv_2,\kv_3)+\Bs^{\rm stoch}(\kv_1,\kv_2,\kv_3)\;,
\ee
where the deterministic part is given by
\begin{align}
\label{eq:B_det}
    \Bs^{\rm tree}(\kv_1, \kv_2, \kv_3) = & \,2\, Z_1(\kv_1)\,Z_1(\kv_2)\,Z_2(\kv_1, \kv_2)\,\Pl(k_1)\,\Pl(k_2)
    \nonumber\\ & 
    + {\rm 2~ perms.}\;,
\end{align}
where `perms.' denotes the two cyclic permutations over the wavevectors $\kv_1, \kv_2$, and $\kv_3$, while for the stochastic contribution we adopt the approximation, motivated by comparison with simulations in \citet{RizzoEtal2023}, given by
\begin{align}
\label{eq:B_stochastic}
\Bs^{\rm stoch}(\kv_1, \kv_2, \kv_3) = & \frac{b_1}{\bar{n}}(1+\alpha_1)\,Z_1(\kv_1)\,\Pl(k_1)+2~{\rm perms.} 
\nonumber \\ 
& + \frac{1+\alpha_2}{\bar{n}^2}\;,
\end{align}
where the parameters $\alpha_1$ and $\alpha_2$ vanish in the Poisson limit.

While our bispectrum model is tree-level, we allow an additional, formally one-loop, term which serves as a phenomenological modelling of Finger-of-God (FoG) effects, by modifying the $Z_1$ kernel as \citep{IvanovEtal2022B}
\begin{equation}
\label{Z1_FoG}
Z_1(\kv) \rightarrow Z_1^\mathrm{FoG}(\kv) = \left[ b_1 + f\mu^2 - c_1^\mathrm{FoG} \mu^2 \left(\frac{k}{k^r_\mathrm{NL}}\right)^2 \right]\;,
\end{equation}
with an additional counterterm $c_1^\mathrm{FoG}$ and nonlinear scale $k^r_\mathrm{NL} = 0.3 \kMpc$.

\subsection{Infrared resummation}
\label{sec:IR-resummation}

The model above fails to accurately capture the smoothing of acoustic features caused by large-scale bulk flows. A proper description requires the resummation of all infrared (IR) modes larger than the observed scales of interest \citep{EisensteinSeoWhite2007, CrocceScoccimarro2008, BaldaufEtal2015B, SenatoreZaldarriaga2015, LewandowskiSenatore2020}. 

We account for this effect in the power spectrum using the method outlined in \citet{BlasEtal2016} and \citet{IvanovSibiryakov2018}, based on the decomposition of the linear power spectrum into a smooth no-wiggles part $P_{\rm nw}(k)$ and a wiggly part $P_{\rm w}(k)$ that captures the baryon acoustic oscillations,
\begin{equation}
    \Pl(k)=P_{\rm nw}(k)+P_{\rm w}(k)\;.
\end{equation}
We implement specifically the recipe for the wiggle-no-wiggle split proposed in \citet{VlahEtal2016}, but see \citet{MoradinezhadDizgahEtal2021} and \citet{PezzottaEtal2024} for a comparison with alternative methods. 

For the bispectrum we consider the simple replacement $\Pl \rightarrow P^{\rm IR,\, LO}$, where $P^{\rm IR,\, LO}$ denotes the leading-order contribution to the infrared-resummed power spectrum, both in the tree-level expression as well as in the stochastic contribution \citep{IvanovEtal2022B}. However, we expect that the effect of IR resummation on the bispectrum is negligible over the relevant range of scales \citep{AlkhanishviliEtal2022}.

\subsection{Redshift-space multipoles}

Our observables are the redshift-space multipoles for the power spectrum and bispectrum. These are obtained from the predictions above as  
\begin{equation}
    P_\ell(k) = \frac{2\ell+1}{2} \int_{-1}^{1} \dd\mu ~ P_{s}(\kv) \mathcal{L}_\ell(\mu)\;,
\end{equation}
where $\mathcal{L}_\ell$ are the Legendre polynomials, and 
\begin{align}
\label{e:Bl}
B_\ell(k_1,k_2,k_3) & = \frac{2\ell+1}{4\pi}\! \int_{-1}^{1}\!\!\dd\mu_1 \!\int_{0}^{2 \pi} \dd\phi \,B_{s}(\kv_1, \kv_2, \mu_1,\phi)\,{\mathcal L}_{\ell}(\mu_1)\;,
\end{align}
where now, following the definition of \citet{ScoccimarroCouchmanFrieman1999}, $\mu_1 \equiv \hat{\kv}_1 \cdot \hat{\nv}$ is the cosine of the angle between $\kv_1$ and the LOS, while $\phi$ is the azimuthal angle describing a rotation of $\kv_2$ around $\kv_1$, here averaged over.

\subsection{Alcock--Paczyński effect}

Another important element that we incorporate into our theoretical model is the Alcock--Paczyński (AP) effect \citep{AlcockPaczynski1979}. This comes from the assumption of a fiducial cosmology, different from the true one, which distorts the underlying wavenumbers and angles ($q, \nu$) into the observed coordinates ($k, \mu$) as follows
\begin{align}
    q^2 &= k^2 \left[ \alpha_{\parallel}^{-2} \mu^2 + \alpha_{\perp}^{-2}\left(1-\mu^2\right) \right]\;,
    \nonumber \\
    \nu^2 &= \alpha_\parallel^{-2} \mu^2 \left[ \alpha_\parallel^{-2} \mu^2 + \alpha_\perp^{-2}\left(1-\mu^2\right)\right]^{-1}\;,
\end{align}
where 
\begin{equation}
    \alpha_\parallel = \frac{H_\mathrm{fid}(z) \, H_{0, \mathrm{true}}}{H_\mathrm{true}(z) \, H_{0, \mathrm{fid}}}\,, \quad{\rm and}\quad 
    \alpha_\perp = \frac{D_{A,\mathrm{true}}(z) \, H_{0, \mathrm{true}}}{D_{A, \mathrm{fid}}(z) \, H_{0, \mathrm{fid}}}\;,
\end{equation}
with $D_A$ being the angular diameter distance.\footnote{These $\alpha$ parameters denote the AP scaling factors and should not be confused with the shot-noise parameters introduced earlier.} In this case, the power spectrum multipoles and bispectrum multipoles read
\begin{equation}
\label{P_with_AP}
    P_\ell(k) = \frac{2 \ell+1}{2 \alpha_\parallel \alpha_\perp^2} \int_{-1}^{1} \dd\mu ~ \mathcal{L}_\ell(\mu) P_s(q[k, \mu], \nu[\mu])\;,
\end{equation}
and 
\begin{align}
\label{B_with_AP}
    B_\ell(k_1, k_2, k_3) 
    =& \frac{2 \ell+1}{4 \pi \alpha_\parallel^2 \alpha_\perp^4} \int_0^{2\pi} \dd\phi \int_{-1}^{1} \dd\mu_1 ~ \mathcal{L}_\ell(\mu_1) 
    \nonumber \\ & 
    \times B_s(q_1, q_2, q_3, \nu_1, \nu_2, \nu_3)\;,
\end{align}
where we omit the implicit dependence on $k_i$ and $\mu_i \equiv \hat{\kv}_i \cdot \hat{\nv}$ (for $i = 1, 2, 3$) for readability such that $q_i\equiv q[k_i, \mu_i]$ and $ \nu_i\equiv\nu[\mu_i]$. In practice, we use the Gauss--Legendre quadrature method to perform the AP integrals.

\section{\label{sec:data} Data}

\subsection{Flagship I simulation}

As in the companion paper \citep{PezzottaEtal2024} we test our models against power spectrum and bispectrum measurements from galaxy catalogues built upon the \Euclid Flagship I simulation. This simulation was generated using the \texttt{PKDGRAV3}~\citep{PotterStadelTeyssier2017} \textit{N}-body code to follow the evolution of 2 trillion dark matter particles in a periodic cubic box with a side length of $L=3780\,\Mpc$. It assumes a flat $\Lambda$ cold dark matter ($\Lambda$CDM) cosmological model characterised by the scaled Hubble parameter $h=0.67$, the physical cold dark matter and baryon density parameters $\omega_{\rm c}=0.12$ and $\omega_{\rm b}=0.022$, a scalar spectral index $n_{\rm s}=0.97$ and the amplitude of scalar perturbations $\As=2.09427\times 10^{-9}$, corresponding to $\sigma_8=0.83$, the root mean square of density fluctuations inside a sphere of radius $8\Mpc$. The total neutrino mass is set to zero.

The mass resolution of the simulation, characterised by the single particle mass $m_{\rm p} \simeq 2.4 \times 10^9 \Ms$, allows one to resolve the haloes with a mass of few $\times \  10^{10} \Ms$, which host the majority of ${\rm H}\alpha$ emission line galaxies,  the main targets of the \Euclid spectroscopic sample. Dark matter haloes are identified with a friends-of-friends algorithm in the snapshots corresponding to redshifts $z=\{0.9,1.2,1.5,1.8\}$, covering the relevant range expected for the sample. The haloes are then populated using a \gls{hod} model described in detail in \citet{PezzottaEtal2024}. We notice that while in that reference two distinct models are considered \citep[Model 1 and 3 from][obtained from direct fits to ${\rm H}\alpha$ surveys]{PozzettiEtal2016}, here we limit ourselves to Model 3, characterised by a lower number density. Some specifications of the ${\rm H}\alpha$ galaxy catalogues are summarised in Table \ref{tab:catalogue_params}.

\begin{table}[t!]
\centering
\caption{Details of the Flagship I galaxy catalogues. For each snapshot redshift $z$ the table lists the number of galaxies $N_{\rm g}$, the mean number density $\bar{n}$ and
 the scales at which the Poisson shot-noise contribution equals the signal of the real-space power spectrum, $k_{\rm sn}^{(r)}$, and the redshift-space power spectrum monopole, $k_{\rm sn}^{(s)}$.
\label{tab:catalogue_params}}
\begin{tabular}{c|rcc}
\hline \hline
   & \\[-9pt]
 $z$ & $N_{\rm g}$  & $\bar{n} \left[\kcMpc\right]$ & $k_{\rm sn}^{(r)}$  $\left(k_{\rm sn}^{(s)}\right)~[\kMpc]$  \\ [0.3em]
\hline
   & \\[-9pt]
$0.9$ & $110\,321\,755$ & 0.0020 & 0.51~(0.50) \\[3pt]
$1.2$ & $55\,563\,490$& 0.0010 & 0.39~(0.42)\\[3pt]
$1.5$ & $31\,613\,213$ & 0.0006 & 0.26~(0.30)\\[3pt]
$1.8$ & $16\,926\,864$ & 0.0003 & 0.22~(0.26)\\[3pt]
\hline
\end{tabular}
\end{table}

\subsection{Power spectrum and bispectrum measurements}

All measurements have been obtained employing the codes developed by the Euclid Collaboration for the analysis pipeline (Euclid Collaboration: Salvalaggio et al., in prep. and Euclid Collaboration: Rizzo et al., in prep.). These adopt a fourth-order scheme for the density estimation on the grid and the interlacing technique to reduce aliasing \citep{SefusattiEtal2016}. Preliminary tests of an earlier independent version of the bispectrum estimator can be found in \citet{RizzoEtal2023}.  

The estimator of the redshift-space power spectrum multipoles is given by
\begin{align}
\label{eq:P_ell_estimator}
\hat{P}_\ell(k) = \frac{2\ell+1}{V\, N_k} \sum_{\qv \in k} |\delta_{\rm g}(\qv)|^2
\,\mathcal{L}_\ell(\hat{\qv} \cdot \hat{\nv})\;,
\end{align}
where $V$ is the volume of the simulation box. The sum over $\qv \in k$ includes all wavenumbers $\qv$ falling in the $k$-bin of size $\Delta k$, with $k-\Delta k/2\le |\qv|<k+\Delta k/2$. The quantity $N_k=\sum_{\qv\in k}$ denotes the number of modes within that bin. We chose for all measurements $\Delta k = 2 \, k_{\rm f}$, where $k_{\rm f}$ is the fundamental frequency of the simulation box. The LOS $\hat{\nv}$ is assumed constant, equivalent to the distant observer approximation. Following the choice of \citet{ScoccimarroCouchmanFrieman1999}, for the bispectrum multipoles we have 
\begin{align}
\label{B_ell_estimator}
\hat{B}_\ell(k_1,k_2,k_3) = & \frac{2
\ell+1}{V\,N_B} \sum_{\qv_1\in k_1}\sum_{\qv_2\in k_2}\sum_{\qv_3\in k_3} \delta_{\rm K}(\qv_{123})
\\ \nonumber  
& \times \delta_{\rm g}(\qv_1) \, \delta_{\rm g}(\qv_2) \, \delta_{\rm g}(\qv_3) \, \mathcal{L}_\ell(\hat{\qv}_1 \cdot \hat{\nv})\;,
\end{align}
where $\delta_{\rm K}(\qv)$ is the Kronecker delta defined to be equal to 1 for $\qv=0$ and zero otherwise, while $k_1\ge k_2\ge k_3$ is assumed without loss of generality and
\begin{equation}
\label{eq:NB_def}
N_B(k_1,k_2,k_3) \equiv \sum_{\qv_1 \in k_1} ~ \sum_{\qv_2 \in k_2} ~ \sum_{\qv_3 \in k_3} ~ \delta_{\rm K}(\vec{q}_{123})\;,
\end{equation}
is the number of fundamental triangles $\left\{\qv_1,\qv_2,\qv_3\right\}$ in each triangular bin centred on $\left\{k_1, k_2, k_3\right\}$ and the same bin size $\Delta k$. Real-space measurements correspond to the case of the monopole estimator, $\ell=0$.

\begin{figure*}[t!]
  \centering
  \includegraphics{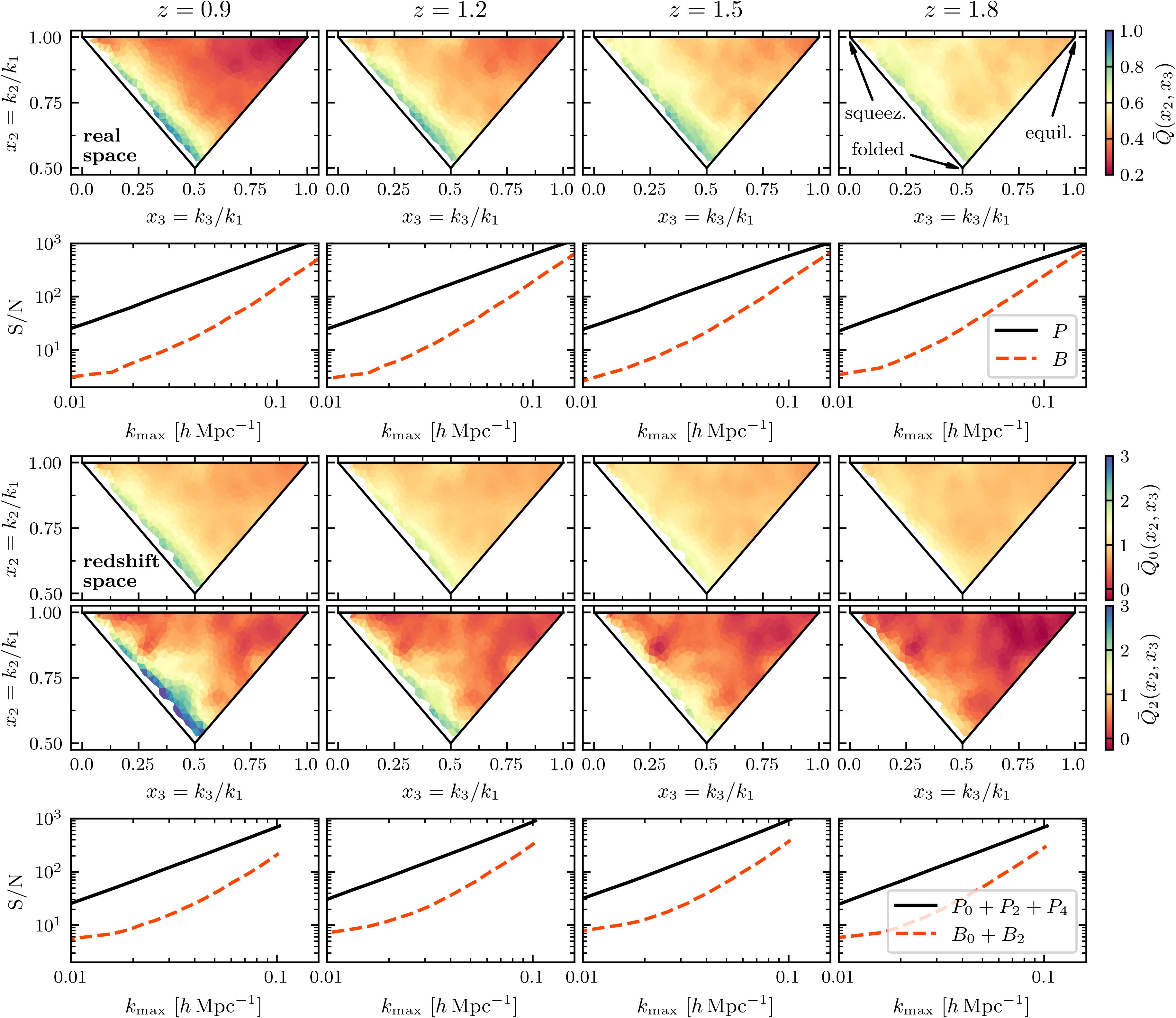}
  \caption{The first row shows, for the four redshifts, the dependence on the triangle shape of the bispectrum measurements in real space in terms of the reduced bispectrum $Q_{\rm g}$ of \eq{eq:Q_real}, averaged over the values of $k_1$ from  $0.02$ to $0.16\kMpc$, as a function of the ratios $x_2\equiv k_2/k_1$ and $x_3\equiv k_3/k_1$. The same quantity in redshift space for the reduced bispectra defined in \eq{eq:Q_redshift} is displayed in the third and fourth rows for monopole and quadrupole respectively, with $k_1$ averaged here from $0.02$ to $0.11\kMpc$. The second and fifth row present the $\mathrm{S}/\mathrm{N}$ as function of $\kmax$ in the power spectrum and bispectrum respectively in real space -- \eq{eq:StoN_P} and \eq{eq:StoN_B} -- and redshift space -- \eq{eq:StoNPell} and \eq{eq:StoNBell}.}
  \label{fig:Bdata_HOD3}
\end{figure*}

In the first row of Fig.~\ref{fig:Bdata_HOD3} we show, for the four redshifts, the dependence on the shape of the triangular configuration of the bispectrum measurements in real space in terms of the reduced bispectrum 
\begin{equation}
\label{eq:Q_real}
    Q_{\rm g}(k_1,k_2,k_3)\equiv\frac{B_{\rm g}(k_1,k_2,k_3)}{P_{\rm g}(k_1)P_{\rm g}(k_2)+ 2~{\rm perms.}}\;,
\end{equation}
averaged over the values of $k_1$ from  $0.02$ to $0.16\kMpc$, as a function of the ratios $x_2\equiv k_2/k_1$ and $x_3\equiv k_3/k_1$. The third and fourth rows of Fig.~\ref{fig:Bdata_HOD3} show the same quantity in redshift space in terms of the reduced bispectrum multipoles
\begin{equation}
\label{eq:Q_redshift}Q_\ell(k_1,k_2,k_3)\equiv\frac{B_\ell(k_1,k_2,k_3)}{P_0(k_1)P_0(k_2)+ 2~{\rm perms.}}\;,
\end{equation}
now averaged over $k_1$ from $0.02$ to $0.11\kMpc$. The plots show a strong dependence on shape, particularly approaching squeezed and folded configurations at low redshift, for both real and redshift space.

\subsection{Binning effect}
\label{sec:bining}

To account for the binning effects on our measurements in the theoretical predictions, we evaluate the power spectrum model for the bin $k$ at the `effective' wavenumbers $k_\text{eff}$,
\begin{equation}
    P^\text{binned}_\ell(k) \simeq P_\ell(k_\text{eff})\;,
\end{equation}
where 
\begin{equation}
    k_\text{eff} \equiv \frac{1}{N_k} \sum_{\qv \in k} q\;.
\end{equation}
Similarly, we evaluate the bispectrum model at the `ordered' effective triangles $\left\{k_{1, \rm eff}, k_{2, \rm eff}, k_{3, \rm eff}\right\}$ as described in \cite{OddoEtal2020} such that
\be
B_{\ell}^{\text{binned}}(k_1, k_2, k_3) \simeq B_{\ell}(k_{1, \rm eff}, k_{2, \rm eff}, k_{3, \rm eff})\;,
\ee
where the wavenumbers $\{q_1, q_2, q_3\}$ are ordered before being averaged, such that, for the triplet  $\left\{k_1, k_2, k_3\right\}$
\begin{align}
k_{1,\rm eff} &=\frac{1}{N_{B}}\sum_{\textbf{q}_1\in k_1}\sum_{\textbf{q}_2\in k_2}\sum_{\textbf{q}_3\in k_3} \delta_{\rm K}\left(\textbf{q}_{123}\right)\max(q_1,q_2,q_3)\;,
 \\
k_{2,\rm eff} &=\frac{1}{N_{B}}\sum_{\textbf{q}_1\in k_1}\sum_{\textbf{q}_2\in k_2}\sum_{\textbf{q}_3\in k_3} \delta_{\rm K}\left(\textbf{q}_{123}\right) \,{\rm med}(q_1,q_2,q_3)\;,
\\ \nonumber
\text{and} \quad
\\
k_{3,\rm eff} &=\frac{1}{N_{B}}\sum_{\textbf{q}_1\in k_1}\sum_{\textbf{q}_2\in k_2}\sum_{\textbf{q}_3\in k_3} \delta_{\rm K}\left(\textbf{q}_{123}\right)\min(q_1,q_2,q_3)\;.
\end{align}

\subsection{Power spectrum and bispectrum covariance}

We rely on analytical estimates of the power spectrum and bispectrum covariance matrices, obtained in the Gaussian approximation. The power spectrum multipoles covariance, defined as 
\be
\label{eq:cov_pk}
C^{P}_{\ell_1\ell_2}(k_i,k_j)\equiv \langle \hat{P}_{\ell_1}(k_i)\,\hat{P}_{\ell_2}(k_j)\rangle-\langle \hat{P}_{\ell_1}(k_i)\rangle\langle\hat{P}_{\ell_2}(k_j)\rangle\;,
\ee
is given by  \citep{GriebEtal2016}
\begin{align}
\label{eq:C^P_l1l2_def}
C^{P}_{\ell_1\ell_2}(k_i,k_j)  \simeq & 
\frac{(2\ell_1+1)(2\ell_2+1)}{N_{k_i}} \delta^{\rm K}_{ij} 
\nn \\  
& \times \sum_{\ell_3}\sum_{\ell_4=0}^{\ell_3}\,P_{\ell_3,\,{\rm tot}}(k_i)\,P_{\ell_4,\,{\rm tot}}(k_i)
\nn \\  
& \times
\int_{-1}^{1} \dd\mu \, \mathcal{L}_{\ell_1}(\mu) \mathcal{L}_{\ell_2}(\mu)  \mathcal{L}_{\ell_2}(\mu) \mathcal{L}_{\ell_3-\ell_4}(\mu)\;,
\end{align}
where $\delta^{\rm K}_{ij}$ is the standard Kronecker delta. In Eq.~\eqref{eq:C^P_l1l2_def}, we adopted the thin-shell approximation and the monopole $P_\mathrm{0,tot}$ includes the shot-noise contribution while $P_\mathrm{\ell,\,tot}=P_\mathrm{\ell}$ for $\ell\ne 0$ \citep[see the Appendix of][for an explicit expression of the last integral]{GriebEtal2016}. In real space this simplifies to
\begin{align}
C^{P}(k_i,k_j) &\simeq \frac{2}{N_{k_i}^2} \delta^{\rm K}_{ij} \,
 P_\mathrm{g,\,tot}(k_i) \, P_\mathrm{g,\,tot}(k_i) \equiv \delta^{\rm K}_{ij} \,
\Delta^2_P(k)\;.
\end{align}
In these equations, for consistency, we use the approximate expression
\be
N_k\simeq 4\pi\,k_i^2\,\Delta k / k_{\rm f}^3\;.
\ee

The bispectrum multipoles covariance is defined as
\be
C^{B}_{\ell_1\ell_2}(t_i,t_j)\equiv \langle \hat{B}_{\ell_1}(t_i)\,\hat{B}_{\ell_2}(t_j)\rangle-\langle \hat{B}_{\ell_1}(t_i)\rangle\langle\hat{B}_{\ell_2}(t_j)\rangle\;,
\ee
where $t_i=\{k_{1,i},k_{2,i},k_{3,i}\}$ and $t_j=\{k_{1,j},k_{2,j},k_{3,j}\}$ represent two triangle configurations. In this case, the Gaussian and thin-shell approximations lead to \citep{RizzoEtal2023, IvanovEtal2023}
\begin{align}
\label{cov_bisp_gaussian_thinshell}
C^{B}_{\ell_1\ell_2}(t_i,t_j) \simeq &  
  \frac{(2\ell_1+1)(2\ell_2+1)}{N_B}V\,\delta_{ij}^{\rm K}
\nn \\&  \times
\sum_{\ell_3,\ell_4,\ell_5} 
P_{\ell_3,\,\mathrm{tot}}(k_1)\,P_{\ell_4,\,\mathrm{tot}}(k_2)\,P_{\ell_5,\,\mathrm{tot}}(k_3)
\nn \\ & \times 
R_{\ell_1,\ell_2;\ell_3\ell_4,\ell_5}(k_1,k_2,k_3)
\;,
\end{align}
where
\begin{align}
R_{\ell_1,\ell_2;\ell_3,\ell_4,\ell_5}(k_1,k_2,k_3)
 \equiv &\frac{1}{4\pi} \int_{-1}^{1} \dd \mu_1 \int_{0}^{2 \pi} \dd \xi\, 
 \nn \\ &  
\Big[\left(1+\delta^{\rm K}_{k_2 k_3}\right)\,\mathcal{L}_{\ell_1}(\mu_1)\mathcal{L}_{\ell_2}(-\mu_1) 
\nn \\ &  +
\delta^{\rm K}_{k_1 k_2}\left(1+\delta^{\rm K}_{k_2 k_3}\right)\,\mathcal{L}_{\ell_1}(\mu_1)\mathcal{L}_{\ell_2}(-\mu_2) 
\nn \\ & +
2\,\delta^{\rm K}_{k_1 k_3}\,\mathcal{L}_{\ell_1}(\mu_1)\mathcal{L}_{\ell_2}(-\mu_3)\Big]
\nn \\ & \times
\mathcal{L}_{\ell_3}(\mu_1)\mathcal{L}_{\ell_4}(\mu_2)\mathcal{L}_{\ell_5}(\mu_3)
\;,
\end{align}
with 
\begin{equation}
\mu_2(\mu_1, \xi) = \mu_1 \mu_{12}  + \sqrt{1- \mu_1^2}\sqrt{1-\mu_{12}^2} \cos ~\xi \;,
\end{equation}
and
\begin{equation}
\mu_3(\mu_1, \xi) = - \frac{k_1 \, \mu_1 + k_2 \, \mu_2(\mu_1, \xi)}{k_3} \;,
\end{equation}
having defined $\mu_{12} \equiv \hat{\kv}_1 \cdot \hat{\kv}_2$. We note that for consistency, we also have computed \eq{eq:NB_def} in the thin-shell approximation
\begin{equation}
    N_B(k_1, k_2, k_3) \simeq 8 \pi^2 k_1 k_2 k_3 ~\Delta k^3 /k_{\rm f}^6\;. 
\end{equation}
In real space, the Gaussian covariance reduces to
\begin{align}
\label{eq:var_B}
C^{B}(t_i,t_j)\simeq &  
  \frac{s_B(t_i)}{N_B}V\,\delta_{ij}^{\rm K}
P_{\rm g,\,\mathrm{tot}}(k_1)\,P_{\rm g,\,\mathrm{tot}}(k_2)\,P_{\rm g,\,\mathrm{tot}}(k_3)
\nn \\ \equiv & \, 
\delta_{ij}^{\rm K} \, \Delta^2_B(t_i)
\;,
\end{align}
where $s_B = 6, 2, 1$ for equilateral, isosceles, and general triangles, respectively.

Figure~\ref{fig:Bdata_HOD3} shows as well a comparison, for each redshift, between the cumulative signal-to-noise-ratio ($\mathrm{S}/\mathrm{N}$) in the power spectrum measurements and the bispectrum one. Specifically, the second row shows the real-space quantities
\begin{equation}
\label{eq:StoN_P}
\left(\mathrm{S}/\mathrm{N}\right)^2_P=\sum_k^{\kmax}\frac{P_{\rm g}^2(k)}{\Delta^2_{P}(k)}\;,
\end{equation}
and
\begin{equation}  
\label{eq:StoN_B}
\left(\mathrm{S}/\mathrm{N}\right)^2_B=\sum_{\kmax \, \geq \, k_1 \, \geq \, k_2 \, \geq \, k_3}\frac{B_{\rm g}^2(k_1, k_2, k_3)}{\Delta^2_{B}(k_1, k_2, k_3)}\;,
\end{equation}
where the second sum is assumed to extend to all triangles $t$ with sides smaller than or equal to $\kmax$, while the fifth row shows instead the cumulative $\mathrm{S}/\mathrm{N}$ for the combination of power spectrum and bispectrum multipoles, where now 
\begin{equation}
\label{eq:StoNPell}
     \left(\mathrm{S}/\mathrm{N}\right)^2_{P_\ell} =
    \sum_{\ell_1,\ell_2=0,2,4}
    \sum_k^{\kmax}
    \,P_{\ell_1}(k)\,\left[C^{P}_{\ell_1,\ell_2}(k,k)\right]^{-1}\,P_{\ell_2}(k)\;, 
\end{equation}
and 
\begin{equation}
\label{eq:StoNBell}
     \left(\mathrm{S}/\mathrm{N}\right)^2_{B_\ell}=\sum_{\ell_1,\ell_2=0,2}\sum_t^{\kmax}\,B_{\ell}(t)\,\left[C^{B}_{\ell_1,\ell_2}(t,t)\right]^{-1}\,B_{\ell_2}(t) 
    \;.
\end{equation}
In both cases the overall bispectrum signal grows relatively faster with $\kmax$ with respect to the power spectrum \citep{SefusattiScoccimarro2005}, but it never reaches equality over the range of scale relevant for our analysis, keeping a relative difference of the order of a few percent across all redshifts. 

Non-Gaussian terms constitute the main contributions to off-diagonal elements of the power spectrum and bispectrum covariance matrices \citep{ScoccimarroZaldarriagaHui1999, SefusattiEtal2006}. In the power spectrum case they are, however, subdominant with respect to the Gaussian contribution to the variance considered here, considering the relevant range of scales and the absence of super-sample effects for measurements in a box with periodic boundary conditions \citep[see, e.g.,][]{WadekarScoccimarro2020}. The situation is different, to some extent, for the bispectrum where non-Gaussian contributions to the variance can be larger than Gaussian contributions for squeezed triangular configurations \citep{Barreira2020, BiagettiEtal2022, SalvalaggioEtal2024}. In addition, they constitute the leading contribution to the cross-covariance between power spectrum and bispectrum. We will explore with an approximate recipe how the prominent non-Gaussian contributions can affect our results in Sect.~\ref{sec:ng_cov}, keeping in mind that a consistent analytical description of the full covariance for power spectrum and bispectrum is beyond the scope of this work. Our main results assume the common  Gaussian approximation for power spectrum and bispectrum multipoles, ignoring cross-covariance.

\section{\label{sec:method} Methodology and performance metrics}

In this section, we describe the Bayesian likelihood analysis we use to assess the validity of our model, including the parameter space explored and the associated priors. We will also outline various performance metrics used to evaluate the accuracy, constraining power, and goodness of fit of the joint power spectrum and bispectrum analysis as a function of scale cuts and redshifts.

\begin{table}[t!]
\caption{List of model free parameters. The notation $\mathcal{U}[a,b]$ stands for a uniform distribution over the interval $[a, b]$. When the priors differ between real space and redshift space, we adopt their intersection.}
  \renewcommand{\arraystretch}{1.3}
  \centering
  \begin{tabular}{lcc}
    \hline \hline
     & Parameter & Prior\\
     \hline
    \multirow{3}{*}{Cosmology} & $h$ & \small{$\mathcal{U}\,[0.57,0.85]$}\\
    & $\omega_\mathrm{c}$ & \small{$\mathcal{U}\,[0.085,0.15]$} \\
    & $10^9\times \As$ & \small{$\mathcal{U}\,[1.05, 2.95]$} \\
    \hline
    \multirow{4}{*}{Galaxy bias} & $b_1$ & \small{$\mathcal{U}\,[0.9,3.5]$}\\
    & $b_2$ & \small{$\mathcal{U}\,[-4,4]$}\\
    & $b_{\mathcal{G}_2}$ & \small{$\mathcal{U}\,[-4,4]$}\\
    & $b_{\Gamma_3}$ & \small{$\mathcal{U}\,[-8,8]$}\\
    \hline
    \multirow{5}{*}{Counterterms} & $c_0\,\big[({\rm Mpc}/h)^2\big]$ & \small{$\mathcal{U}\,[-500,500]$}\\    
    & $c_2\,\big[({\rm Mpc}/h)^2\big]$ & \small{$\mathcal{U}\,[-500,500]$}\\    
    & $c_4\,\big[({\rm Mpc}/h)^2\big]$ & \small{$\mathcal{U}\,[-500,500]$}\\    
    & $c_{\nabla^4 \delta}\,\big[({\rm Mpc}/h)^4\big]$ & \small{$\mathcal{U}\,[-500,500]$}\\    
    & $c_1^\mathrm{FoG}$ & \small{$\mathcal{U}\,[-10,10]$}\\    
    \hline
    \multirow{3}{*}{Shot noise} & $\alpha_{\rm P}$ & \small{$\mathcal{U}\,[-1,2]$} \\
    & $\epsilon_{0, k^2}\,\big[({\rm Mpc}/h)^2\big]$ & \small{$\mathcal{U}\,[-500,500]$}\\
    & $\alpha_1$ & \small{$\mathcal{U}\,[-1,2]$}\\
    \hline
  \end{tabular}
  \label{tab:priors}
\end{table}

\subsection{Fitting procedure}

We denote as $\mathcal{P}_\alpha \equiv \left\{P_\ell(k_i), B_\ell(t_j)\right\}$ the full data vector consisting of the multipoles of the galaxy power spectrum $P_\ell(k)$, with $\ell = 0, 2, 4$, and the multipoles of the galaxy bispectrum $B_\ell(k_1, k_2, k_3)$, with $\ell = 0, 2$. The index $\alpha$ spans all multipoles, all wavenumbers $k_i$ with $i=1, \dotsc, N_\mathrm{bins}$, and all triangles $t_j=\left\{k_{1,j}, k_{2,j}, k_{3, j}\right\}$ with $j=1, \dotsc, N_\mathrm{tri}$.\footnote{We impose $k_1 < k_2 + k_3$, since the analytical expression we adopt for the covariance of the bispectrum is evaluated in thin-shell approximation, and this fails to properly account for collinear triangles where $k_1 = k_2 + k_3$ \citep{BiagettiEtal2022, RizzoEtal2023}.}

We assume a Gaussian likelihood $\mathcal{L}(\boldsymbol{\theta})$ given by
\begin{align}
\label{eq:chisquare}
-2 \ln \mathcal{L}(\boldsymbol{\theta})  + \mathrm{const.} &=  \chi^2(\boldsymbol{\theta})
\nonumber \\&=  
\sum_{a, \,b} \Big [\mathcal{P}_a(\boldsymbol{\theta}) - \hat{\mathcal{P}}_a \Big] ~C_{a b}^{-1} ~\Big [\mathcal{P}_b(\boldsymbol{\theta}) - \hat{\mathcal{P}}_b \Big]\,,
\end{align}
where $\chi^2$ denotes the chi-square statistic, $\hat{\mathcal{P}}_a$ is the measured data vector, $\mathcal{P}_a(\boldsymbol{\theta})$ is the corresponding theoretical model, function of the parameter set $\boldsymbol{\theta}$, and $C_{a b}$ is the covariance matrix. 

We sample the likelihood with Monte Carlo Markov Chains (MCMC) using the nested sampling method implemented in \texttt{MultiNest} \citep{FerozHobson2008, FerozHobsonBridges2009, FerozEtal2019} and its Python wrapper \texttt{PyMultinest} \citep{BuchnerEtal2014}. For our main results, we obtain the theoretical model predictions for the one-loop power spectrum and tree-level bispectrum either from the \texttt{PBJ} code \citep{MorettiEtal2023}, or \texttt{COMET} \citep{EggemeierEtal2023, EggemeierEtal2025, PezzottaEtal2025}. While the former exploits the \texttt{FAST-PT} algorithm \citep{McEwenEtal2016} to speed up loop evaluation, the latter makes use of Gaussian processes to emulate the linear and nonlinear components of the power spectrum. A comparison of the two codes, together with other publicly available codes from the literature, is presented in Sect.~\ref{sec:code_comparison}.

Table~\ref{tab:priors} presents the priors for both cosmological and nuisance parameters of the theoretical model. We use flat priors with wide intervals to ensure they remain as uninformative as possible.

\begin{figure*}[t!]
  \centering
  \includegraphics[width=\textwidth]{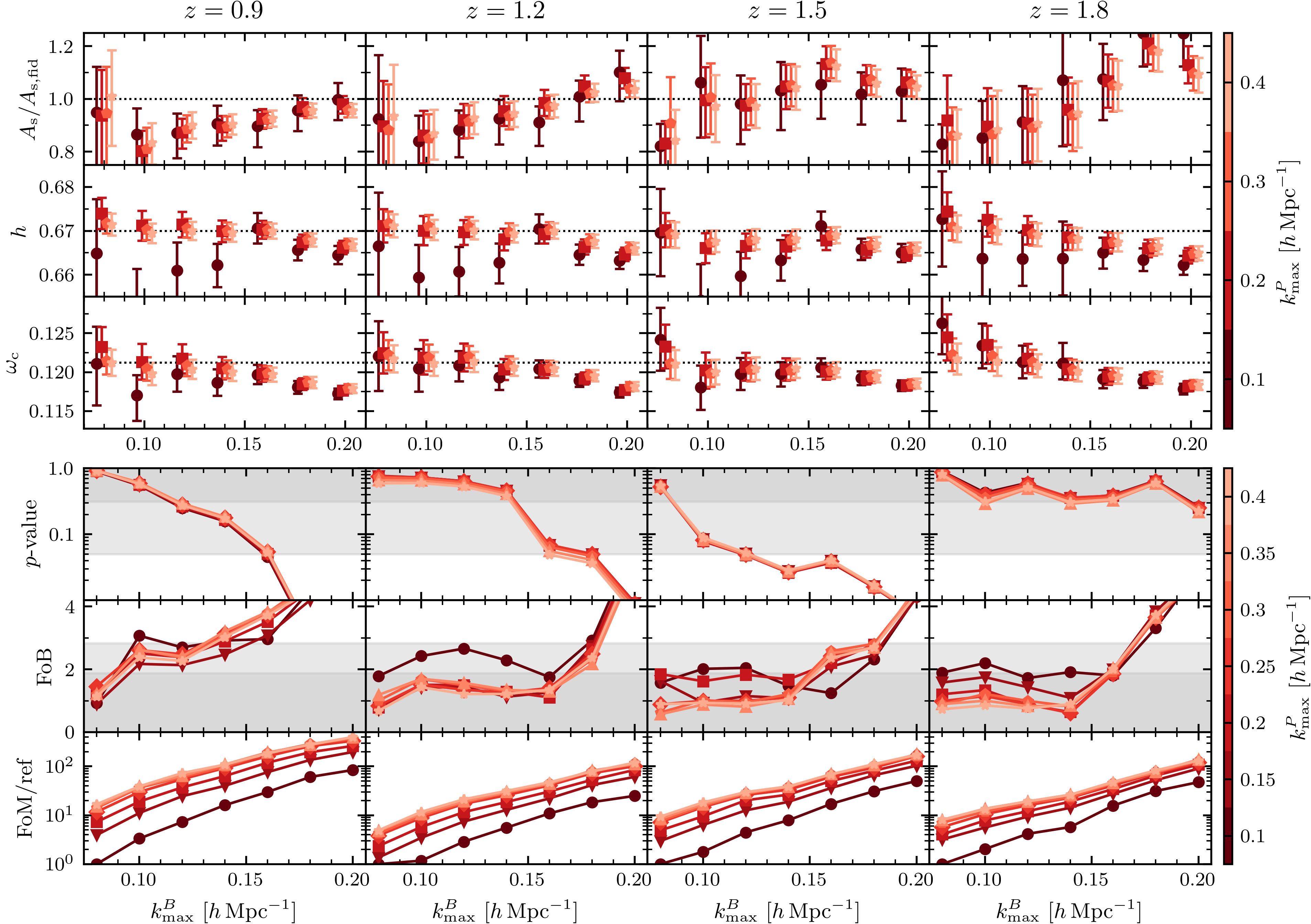}
  \caption{Results of the power spectrum and bispectrum analysis in real space assuming the maximal model as a function of the bispectrum scale cut $\kmaxB$ and for different values of the power spectrum scale cut $\kmaxP$, denoted by the colour gradient. Each column corresponds to a different redshift. The top three rows show posterior means and 68\% credible intervals for the three cosmological parameters (slightly displaced along the $x$-axis for readability), while the bottom three show respectively the goodness of fit in terms of the $p$-value, the FoB, and FoM. While not explicitly plotted, we have verified that the maximum a posteriori (MAP) values for the cosmological parameters fall within the 68\% credible intervals. The dashed lines in the top three rows denote the fiducial values of the three cosmological parameters. The grey bands in the $p$-value and FoB panels represent the 68 and 95 percentiles of the associated distributions. The FoM values are normalised to the value computed at $\kmaxP=0.1\kMpc$ and $\kmax^B=0.08\kMpc$.}
  \label{fig:cosmo_no_relation}
\end{figure*}

\subsection{Performance metrics}

We quantify systematic errors on recovered parameters induced by an inaccurate theoretical model in terms of the figure of bias (FoB) defined as
\begin{equation}
    \mathrm{FoB}(\boldsymbol{\theta}) \equiv \left[\Big(\langle \boldsymbol{\theta} \rangle - \boldsymbol{\theta}_\mathrm{fid} \Big)^\top ~ S^{-1} (\boldsymbol{\theta}) ~ \Big(\langle \boldsymbol{\theta} \rangle - \boldsymbol{\theta}_\mathrm{fid} \Big) \right]^{1/2}\,,
\end{equation}
where $\langle \boldsymbol{\theta} \rangle$ indicates the posterior mean of the parameters, $\boldsymbol{\theta}_\mathrm{fid}$ their fiducial values, and $S(\boldsymbol{\theta})$ is their covariance matrix. Based on the parameter covariance matrix we further define the figure of merit (FoM) to quantify the constraining power of a given observable (or combination of observables) as \citep{Wang2008}
\begin{equation}
  \mathrm{FoM}(\boldsymbol{\theta}) \equiv  \mathrm{det}\,\Big(S(\boldsymbol{\theta})\Big)^{-1/2}\,.
\end{equation}
We evaluate both the FoB and FoM for the three cosmological parameters included in our analysis, setting $\boldsymbol{\theta} = \{h,\,\omegac,\,\As\}$ in the two expressions above. Finally, to assess the goodness of fit, we compute the posterior averaged $\langle \chi^2 \rangle$ from \eq{eq:chisquare}, which is converted into a $p$-value using the corresponding degrees of freedom of the fit.

\section{\label{sec:real} Real-space analysis}

This section investigates the validity of the perturbative model in real space. We begin with the maximal model, where all parameters are allowed to vary. Subsequently, we explore whether the bias and shot-noise parameters of the H$\alpha$ galaxies satisfy any relations that would enable us to reduce the size of the parameter space without incurring biases in the cosmological parameters.

\subsection{Maximal model}

We perform joint fits of the power spectrum and bispectrum using the maximal model, varying the scale cuts $\kmaxP$ and $\kmaxB$ applied to the two observables. Specifically, we test seven values of $\kmaxP$ between $0.1$ and $0.4\,\kMpc$ and seven values of $\kmaxB$ between $0.08$ and $0.2\,\kMpc$. The results for each redshift bin are shown in Fig. \ref{fig:cosmo_no_relation}, where the posterior means and 68\% credible intervals of the cosmological parameters $\As$, $h$, and $\omegac$ are plotted in the first three rows as a function of $\kmaxB$, with the colour gradient indicating different choices of $\kmaxP$. The fiducial values of the parameters (dotted horizontal lines) are generally recovered within $2\sigma$ up to $\kmaxB \simeq 0.15\,\kMpc$, beyond which the posterior means show a stronger dependence on $\kmaxB$.  In contrast, the dependence on the power spectrum scale cut is weak, except for the lowest value of $\kmaxP$, where there are notable differences in the recovered parameters. We attribute this to the limited constraining power at this scale cut, which, combined with the strong degeneracy between $\As$ and $b_1$, leads to significant projection effects in the marginalised constraints.

The $p$-values and FoB metric, shown alongside the FoM in the bottom three rows of Fig.~\ref{fig:cosmo_no_relation}, fully support this picture. While the $p$-values decrease with $\kmaxB$ and indicate inadequate fits for the first three redshift bins beyond $\kmaxB \simeq 0.15\,\kMpc$, the FoB remains within the 95th percentile over the same range but rises sharply at smaller scales. This is also the case for the highest redshift bin, even though the goodness of fit remains acceptable. The FoM panels demonstrate that including the bispectrum yields a substantial gain in constraining power by roughly an order of magnitude within the model's validity range. This is driven primarily by tighter constraints on $\As$ due to the strong degeneracy between $\As$ and $b_1$ for the power spectrum in real space. Redshift-space distortions (as we will discuss in Sect.~\ref{sec:redshift}) allow this degeneracy to be significantly reduced already at the power spectrum level, leading to a smaller impact by the bispectrum.

These findings are consistent with previous analyses of the power spectrum and bispectrum in real space \citep[see, e.g.,][]{EggemeierEtal2021, OddoEtal2021}. Importantly, however, they are applied here to a galaxy population that resembles more closely the one observed by \Euclid. 

\subsection{Testing galaxy bias relations}
\label{ssec:bias_relations}

Previous studies have shown that the bias parameters of dark matter haloes and luminous red galaxies are not independent, but instead follow certain relations to good approximation \citep[e.g.,][]{ChanScoccimarroSheth2012,BaldaufEtal2012,ShethChanScoccimarro2013,LazeyrasEtal2016, AbidiBaldauf2018, EggemeierEtal2021}. Our aim is to test 1) whether such relations also hold for our H$\alpha$ galaxy populations, and 2) how many model parameters can be robustly constrained through the joint analysis of the power spectrum and bispectrum. 

Specifically, we test the following bias relations:
\begin{itemize}
    \item a relation between the combination of our local quadratic and tidal bias parameters as a function of the linear bias,
    \begin{equation}
    \label{eq:b2_hod}
         b_2 - \frac{4}{3}\,\bGtwo = -0.015 - 1.58\,b_1 + 0.809\,b_1^2 + 0.025\,b_1^3\;,
    \end{equation}
    which was obtained by fitting corresponding quantities measured from haloes in simulations \citep{LazeyrasEtal2016} and subsequently adapted to the specific \gls{hod} of our simulated data (for more details, see Appendix~\ref{app:b2tilde});
    \item the local-Lagrangian relation for the tidal bias parameter \citep{ChanScoccimarroSheth2012,BaldaufEtal2012},
    \begin{equation}
    \label{eq:bGtwo_LL}
       b_{\mathcal{G}_2} =-\frac{2}{7}\left(b_1-1\right)\;,
     \end{equation}
     derived under the assumption that the initial (Lagrangian) patches from which the galaxies formed have vanishing $b_{{\cal G}_2}$ and that their comoving number density is conserved in their subsequent evolution \citep[`co-evolution', ][]{Fry1996};
    \item a quadratic fit to the excursion set prediction of the tidal bias~\citep{ShethChanScoccimarro2013, EggemeierEtal2020, PezzottaEtal2021},
    \begin{equation}
    \label{eq:bGtwo_fit}
        b_{\mathcal{G}_2} = 0.524 -0.547\,b_1 + 0.046\,b_1^2\; ;
    \end{equation}
    \item a relation for the third order tidal bias parameters \citep{ChanScoccimarroSheth2012},
    \begin{equation}
      \label{eq:bG3_relation}
      b_{\Gamma_3} = -\frac{1}{6}(b_1 -1 ) - \frac{5}{2}b_{\mathcal{G}_2}\;,
    \end{equation}
    based also on co-evolution and vanishing Lagrangian $b_{\Gamma_3}$ (but not necessarily $b_{{\cal G}_2}$)\;.
\end{itemize}

\begin{figure}
  \centering
  \includegraphics[width=0.98\columnwidth]{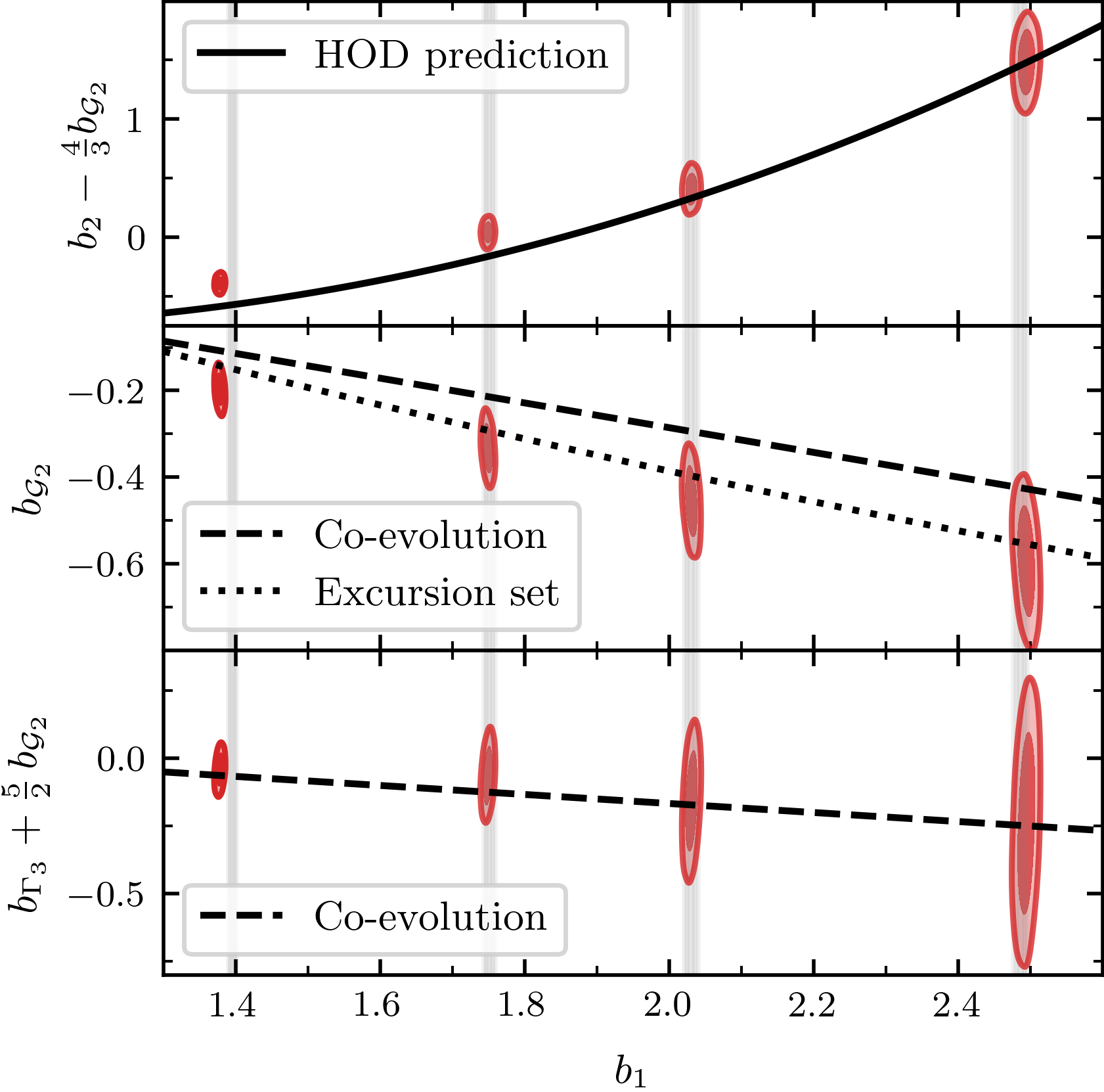}
  \caption{Constraints on nonlinear galaxy bias parameters for four different redshift bins as a function of the corresponding recovered value of the linear bias. These are compared to the predictions of the bias relations: Eq.~\ref{eq:b2_hod} in the first panel; Eqs.~(\ref{eq:bGtwo_LL}, \ref{eq:bGtwo_fit}) (based on co-evolution and excursion set, respectively) in the second panel; and Eq.~\eqref{eq:bG3_relation} (based on co-evolution) in the third panel. The grey vertical bands correspond to the $b_1$ estimates (and their 1$\sigma$ uncertainty) from the galaxy and matter power spectra described in \citet{PezzottaEtal2024}. While not shown explicitly, we have verified that the MAP values for the bias parameters fall within the 68\% credible intervals.}
  \label{fig:bias_relations_fixed_cosmology}
\end{figure}

\begin{figure*}
  \centering
  \includegraphics[width=\textwidth]{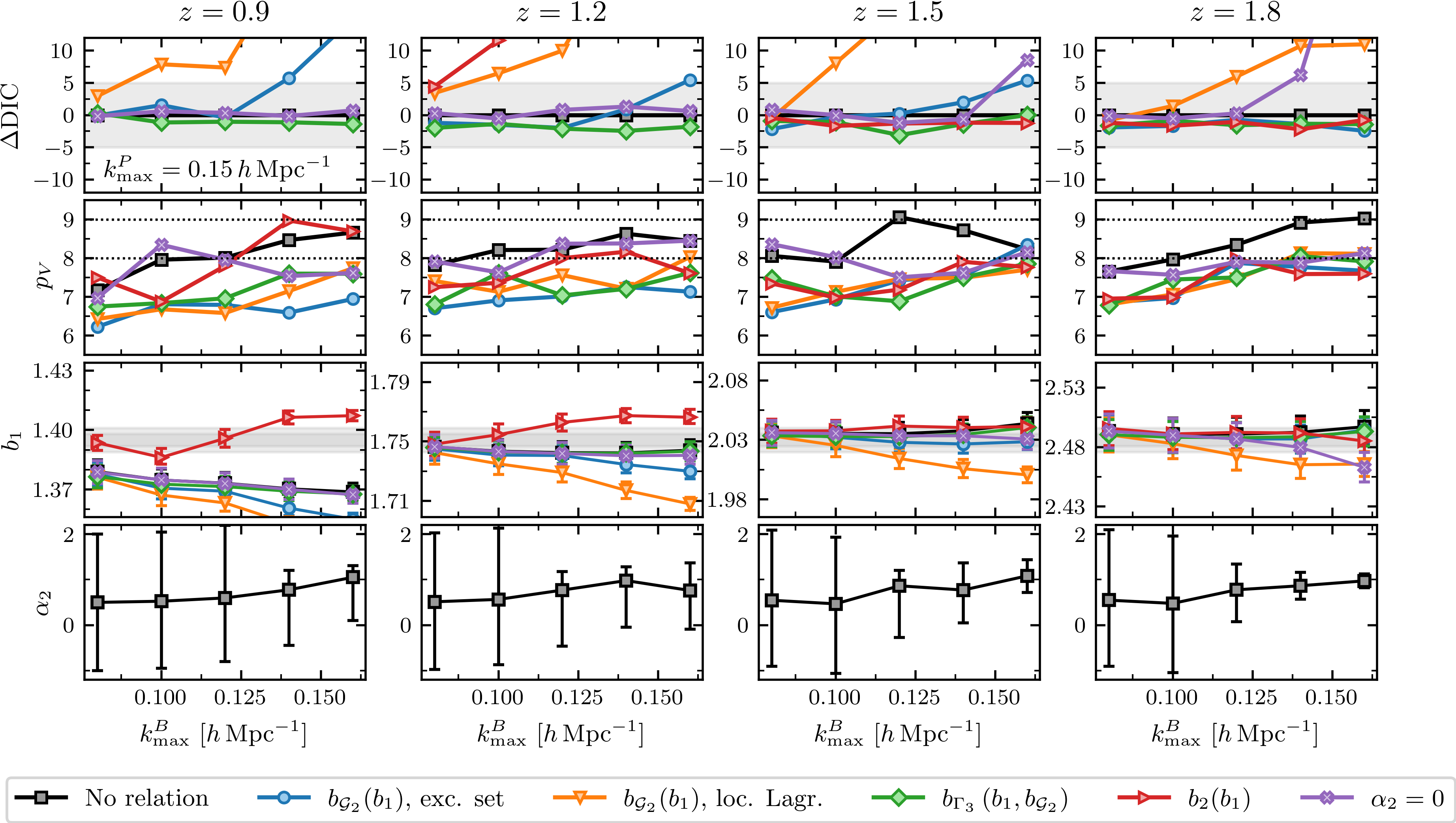}
  \caption{Performance of various 8-parameter models obtained either by imposing a relation among the bias parameters or by setting $\alpha_2=0$ in Eq.~\eqref{eq:B_stochastic}, compared to the 9-parameter maximal model. First row: difference in the DIC with statistically significant deviations lying outside the grey band. Second row: effective number of parameters $p_V$ constrained by the data. Third and fourth rows: marginalised constraints on the parameters $b_1$ and $\alpha_2$ (the grey band in the $b_1$ panels is identical to that in Fig.~\ref{fig:bias_relations_fixed_cosmology}). All results are shown as a function of $\kmaxB$ at fixed $\kmaxP=0.15\kMpc$, with columns corresponding to the four redshift bins.}
\label{fig:DIC_bias_relations}
\end{figure*}

As a first assessment of the validity of the bias relations, Fig.~\ref{fig:bias_relations_fixed_cosmology} compares their predictions with the posterior constraints on the relevant parameters obtained from the maximal model, while keeping the cosmological parameters fixed to their fiducial values. We find that the fit of Eq.~\eqref{eq:b2_hod} reproduces the recovered quadratic bias parameters for galaxy samples with large $b_1$ (i.e., the two highest redshift bins), but fails in the two lowest redshift bins. This discrepancy could be related to the fact that the relation of \citet{LazeyrasEtal2016} was calibrated on haloes with masses above $\logten[M/(\Ms)]=12.55$, whereas the Flagship I catalogues are dominated by haloes of typical mass on the order of $ 10^{10}\Ms$, which host most of the H$\alpha$ galaxies. For the $\bGtwo$ parameter, the excursion-set prediction of Eq.~\eqref{eq:bGtwo_fit} agrees well with the measured values, while the local-Lagrangian prediction, Eq.~\eqref{eq:bGtwo_LL}, systematically overestimates them. Finally, the combination $b_{\Gamma_3}+(5/2)\,\bGtwo$ is accurately described by the co-evolution relation of Eq.~\eqref{eq:bG3_relation}.

We now turn to the use of bias relations in the context of likelihood evaluations. Fixing cosmology, we compare the maximal model with alternative prescriptions in which each relation is imposed individually, thereby reducing the number of free parameters from nine to eight. To assess their performance, we compute the difference in \gls{dic} relative to the maximal model, taken as reference. The \gls{dic} is defined as
\begin{equation}
    \mathrm{DIC} = \ave{D}_\mathrm{post} + p_{\rm V}\;,
\end{equation}
where $D=-2 \ln \mathcal{L}$ is the deviance averaged over the posterior, and $p_{\rm V} = 0.5 \,\mathrm{Var}(D)$, where $\mathrm{Var}(D)$ denotes the variance of $D$, is the effective number of parameters that is constrained by the data vector. A difference $\Delta \mathrm{DIC} > 5$ indicates a strong preference for the reference model, disfavouring the model with the imposed relation. We plot this quantity together with $p_V$ for various redshifts and as a function of $\kmaxB$ in the first two rows of Fig.~\ref{fig:DIC_bias_relations}. In addition to the four bias relations discussed above, we also test a model in which the parameter $\alpha_2$, describing a correction to the Poisson shot-noise contribution to the bispectrum (Eq.~\ref{eq:B_stochastic}) is set to zero. 

Among these prescriptions, only the co-evolution relation $b_{\Gamma_3}(b_1)$ yields fits comparable to the maximal model across all redshifts and values of $\kmaxB$, although it is not significantly favoured by the DIC. In contrast, the fit for $b_2 - 4/3b_{{\cal G}_2}$, performs poorly for the two lowest redshift bins, consistent with the findings of Fig.~\ref{fig:bias_relations_fixed_cosmology}. The local Lagrangian relation $b_{\G_2}(b_1)$ is also strongly disfavoured at nearly all redshifts and scale cuts \citep[see also][]{OddoEtal2020}. On the other hand, the assumption $\alpha_2=0$ appears to be viable except at the largest $\kmaxB$ values at $z=1.5$ and 1.8. Indeed, the $p_V$ values show that in the maximal model one parameter remains unconstrained on large scales. This turns out to be $\alpha_2$ as illustrated in the bottom panels of Fig.~\ref{fig:DIC_bias_relations}, where its marginalised posterior distribution fully spans the prior range $[-1,2]$ for $\kmax^B<0.12\,\kMpc$. Moreover, fixing $\alpha_2=0$ (violet line) brings the value of $p_V$ closer to the actual number of model parameters, while all other 8-parameter models consistently give $p_V < 8$. Finally, the third row of Fig.~\ref{fig:DIC_bias_relations} shows the marginalised constraints on the linear bias parameter $b_1$ as a function of $\kmaxB$, compared to the estimates of \citet{PezzottaEtal2024} using the galaxy and matter power spectra (shown as grey horizontal bands). We note that every model underestimates $b_1$ at $z=0.9$ (see also Fig.~\ref{fig:bias_relations_fixed_cosmology}), whereas at higher redshifts a good match is found for the maximal model (black line), the $b_{\Gamma_3}(b_1,\bGtwo)$ relation (green line), and the $\alpha_2=0$ prescription (violet line).

\section{\label{sec:redshift} Redshift-space analysis}

We now proceed to the analysis of the power spectrum and bispectrum multipoles in redshift space. We consider here a reference model corresponding to the maximal model but fixing $\alpha_2=0$, a sensible choice given the results of the previous section. We compare the redshift-space results to those obtained from the real-space measurements, we estimate the performance expected for a volume corresponding to the final Data Release 3 (DR3) of \Euclid, and explore several aspects of the analysis related to the relevance of the bispectrum quadrupole and the non-Gaussianity of the full covariance matrix. We finally provide a comparison of several different codes available within the collaboration to demonstrate the robustness of our results.

\subsection{Reference model}
\label{ssec:maxmod} 

\begin{figure*}[t!]
  \centering
  \includegraphics[width=\textwidth]{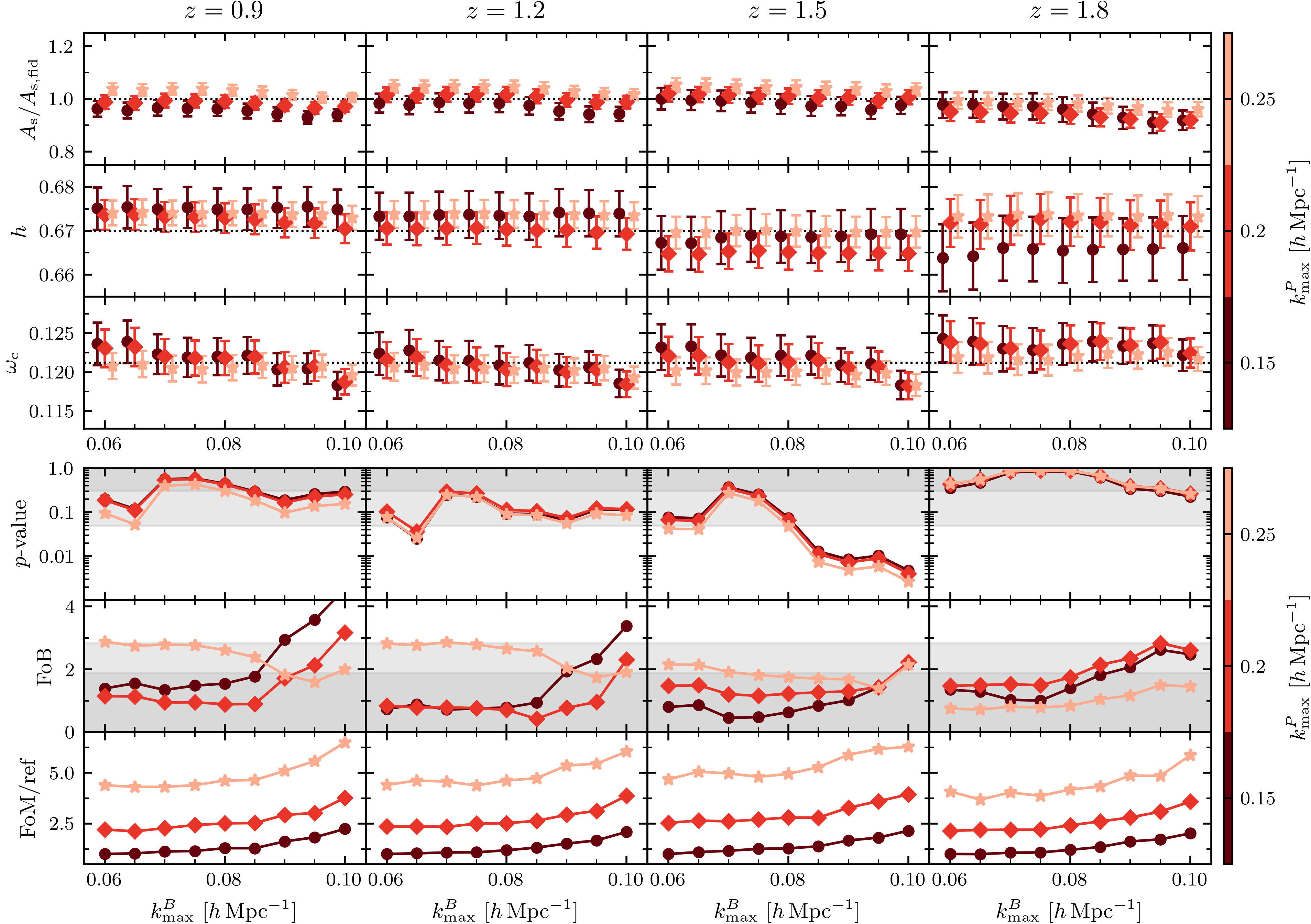}
  \caption{Same as Fig.~\ref{fig:cosmo_no_relation}, but for the results of the joint power spectrum and bispectrum analysis in redshift space. While identical scale cuts are applied to the bispectrum multipoles, i.e., $\kmaxB \equiv \kmax^{B_0} = \kmax^{B_2}$, the power spectrum multipoles satisfy $\kmaxP \equiv \kmax^{P_0} = \kmax^{P_2} = \kmax^{P_4} + 0.05 ~ \kMpc$. In each redshift bin the FoM reference value is computed for the case $\kmaxP=0.1\kMpc$ and $\kmax^B=0.06\kMpc$. Similarly to Fig.~\ref{fig:cosmo_no_relation}, we have checked that the MAP values for the cosmological parameters fall within the 68\% credible intervals.}
  \label{fig:FoM_FoB_pval_MaxMod}
\end{figure*}

Figure~\ref{fig:FoM_FoB_pval_MaxMod} presents the results of the analysis of the power spectrum and bispectrum multipoles in redshift space assuming the reference model, shown as functions of $\kmaxB \equiv k_{\rm max}^{B_0} = k_{\rm max}^{B_2}$ for different values of $\kmaxP \equiv k_{\rm max}^{P_0} = k_{\rm max}^{P_2} = k_{\rm max}^{P_4} + 0.05 ~ \kMpc$, since the power spectrum hexadecapole typically has a lower reach. 

The first three rows present, respectively, the marginalised constraints on the three cosmological parameters $\As$, $h$, and $\omegac$. In this case, we notice some significant deviation only at the largest value of $\kmaxB$ considered, that is around $0.1\kMpc$, specifically for $\omegac$. The last three rows show the $p$-value, FoB, and FoM results.

As shown in the real-space results, the variation in the $p$-value is primarily influenced by the bispectrum scale cuts, with the power spectrum range having a minimal impact. Overall, the $p$-values fall within the 95th percentile, with some exceptions at the intermediate redshifts, $z=1.2$ and $z=1.5$ that can be interpreted as stochastic fluctuations (see also the comparatively low $p$-values for the $z=1.5$ bin in real space).

The FoB determined in terms of the fiducial values of the parameters $h$, $\omegac$, and $\As$ -- using the 95th percentile as the threshold -- indicates that at the lowest redshift, $z = 0.9$, the tree-level bispectrum model remains valid up to $\kmaxB \simeq 0.08~\kMpc$, while it reaches $\kmaxB \simeq 0.1~\kMpc$ at $z = 1.8$. The intermediate redshift bins exhibit patterns similar to those observed at the highest redshift. In particular, when considering the power spectrum cutoff $\kmaxP = 0.2~\kMpc$, almost all $\kmaxB$ values satisfy the 95th percentile FoB criterion across all redshifts. For a more conservative choice, $\kmaxP = 0.15~\kMpc$, the FoB criterion implies $\kmaxB \simeq 0.08$, $0.095$, $0.1$, and $0.1~\kMpc$ for $z=0.9$, $1.2$, $1.5$, and $1.8$, respectively. Since we expect the power spectrum modelling to remain reliable for $\kmaxP \le 0.2~\kMpc$, we adopt as our fiducial bispectrum cutoff the minimum $\kmaxB$ that satisfies the FoB criterion within this range, leading to the final choices quoted above. At smaller scales, the FoB varies quite significantly with $\kmaxB$, and more so at lower redshifts, indicating a breakdown of the model. The FoB diagnostic also validates the scale cuts for the power spectrum. For instance, it shows that $\kmaxP$ cannot exceed $0.2~\kMpc$ for $z=0.9$ and $1.2$, while $\kmaxP=0.25~\kMpc$ is safe for $z=1.8$, with more noisy results at $z=1.5$, consistent with the results obtained from the power spectrum-only analysis of Euclid Collaboration: Camacho et al. (in prep.).

The FoM clearly indicates that the constraining power on the cosmological parameters is primarily driven by the power spectrum. However, for large values of $\kmaxB$ we find a relevant increase in the FoM due to the bispectrum. For instance, at the lowest redshift, $z=0.9$, the analysis with $k_\mathrm{max}^P = 0.15 \kMpc$ and $k_\mathrm{max}^{B} = 0.1 \kMpc$ shows results comparable to the case with  $\kmaxP = 0.2 \kMpc$ but a lower $\kmaxB = 0.06 \kMpc$. This trend is consistent across all redshifts, with a slightly less pronounced gain at higher redshift. Overall, the bispectrum increases the FoM by a factor of about 2.5.

The marginalised relative uncertainties on the cosmological parameters for the joint analysis with $\kmaxP = 0.2 \kMpc$ and $\kmaxB = 0.1 \kMpc$ are listed in Table~\ref{tab:1sigma_Euclidlike} and compared with those from the power spectrum-only analysis. While the inclusion of the bispectrum does not yield any noticeable improvement in the constraint on $h$, it reduces the uncertainties on $\As$ and $\omegac$ by approximately $15$ and up to $30\,\%$, respectively. These improvements are slightly larger than those found in \citet{IvanovEtal2022B}, but qualitatively agree with the findings of \citet{EggemeierEtal2025}.

\begin{table}[t!]
\centering
\caption{Relative uncertainties on cosmological parameters from the joint analysis for either the Flagship I volume, or the \Euclid-like volume  (relative differences with respect to the power spectrum-only runs are given in parenthesis). Results are derived using $\kmaxP = 0.2 \kMpc$ and the $\kmaxB$ values in the table (in units of $\kMpc$). \label{tab:1sigma_Euclidlike}}
\begin{tabular}{ccccc}
\hline \hline
  $z$ & $\kmaxB$ & $\sigma_{\As}/\As$ [\%] & $\sigma_h/h$ [\%] & $\sigma_{\omega_{\rm c}}/\omega_{\rm c}$ [\%] \\
  \hline\\[-0.8em]
  \multicolumn{5}{l}{Flagship I volume} \\[0.2em]
 
  $0.9$ & 0.08 & 2.5~(\textbf{$-$15}) & 0.5~(\textbf{$-$11}) & 1.8~(\textbf{$-$6}) \\ 
  $1.8$ & 0.1 & 3.1~(\textbf{$-$17}) & 0.8~(\textbf{$-$3}) & 1.5~(\textbf{$-$27}) \\
  \hline\\[-0.8em]
  \multicolumn{5}{l}{\Euclid-like volume} \\[0.2em]
  $0.9$ & 0.1 & 5.0~(\textbf{$-$18})& 1.2~(\textbf{2}) & 3.4~(\textbf{$-$23}) \\
  $1.8$ & 0.11 & 5.1~(\textbf{$-$17}) & 1.4~(\textbf{$-$1}) & 2.5~(\textbf{$-$35}) \\ 
\hline
\end{tabular}
\end{table}

It is possible to further boost the overall FoM by imposing relations among the bias parameters. In line with the real-space findings, we observe that a significant increase in the FoM is obtained when the co-evolution relation $b_{\Gamma_3}(b_1, b_{\mathcal{G}_2})$ is assumed in the likelihood evaluation. In contrast, applying the quadratic fit to the excursion set prediction of the tidal bias $b_{\mathcal{G}_2}(b_1)$ results in only a minor improvement in the constraining power of the cosmological parameters. Additionally, when examining the FoB metric, particularly at the highest redshift $z=1.8$, we find that imposing these bias relations do not introduce any systematic error on the recovered  cosmological parameters relative to their fiducial values.

\subsection{\label{ssec:PBvsP} Comparison with real-space results}

\begin{figure}[b]
  \centering
  \includegraphics[width=\columnwidth]{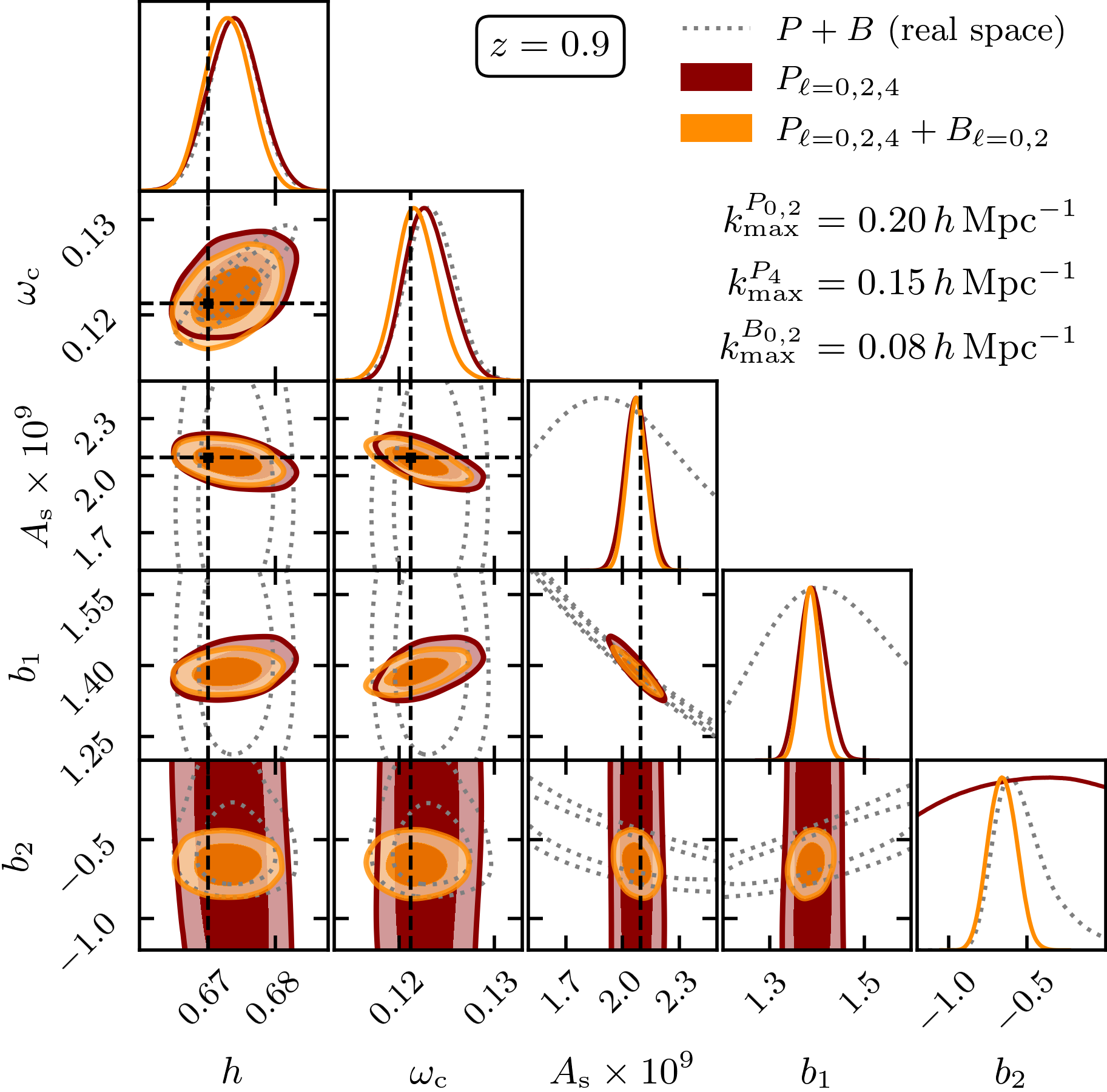}
  \caption{Comparison of the joint power spectrum and bispectrum analysis in redshift space of the $z=0.9$ snapshot (filled, orange contours), with the results from the redshift-space power spectrum alone (filled, red contours) and from the joint correlators in real space (dotted curves). Dashed lines denote the fiducial values for the cosmological parameters.}
  \label{fig:vs_PB_and_Pell}
\end{figure}

An interesting comparison can be made between the constraints from the power spectrum multipoles alone, the joint real-space power spectrum and bispectrum analysis, and the combined power spectrum and bispectrum multipoles analysis. This is shown in Fig.~\ref{fig:vs_PB_and_Pell} for the $z=0.9$ snapshot. For a fair comparison, we choose the same scale cuts for the real-space correlators and their corresponding redshift-space multipoles.  

As expected, the degeneracy between the linear bias $b_1$ and the scalar amplitude $\As$ is lifted in redshift space, already with the power spectrum multipoles alone, leading to significantly tighter constraints on these parameters compared to the real-space analysis. However, the information from the bispectrum aids in constraining nonlinear bias parameters. Figure~\ref{fig:vs_PB_and_Pell} shows the improvement in $b_2$, but similar results are obtained for other nonlinear bias parameters ($b_{\mathcal{G}_2}$ and $b_{\Gamma_3}$). In this respect, the bispectrum provides an additional consistency test on the perturbative model, since we do have qualitative predictions for these parameters that can be compared with these results.

\subsection{Performance for \Euclid-DR3 (full mission)}
\label{ssec:euclid_dr3}

\begin{figure}[t!]
  \centering
  \includegraphics{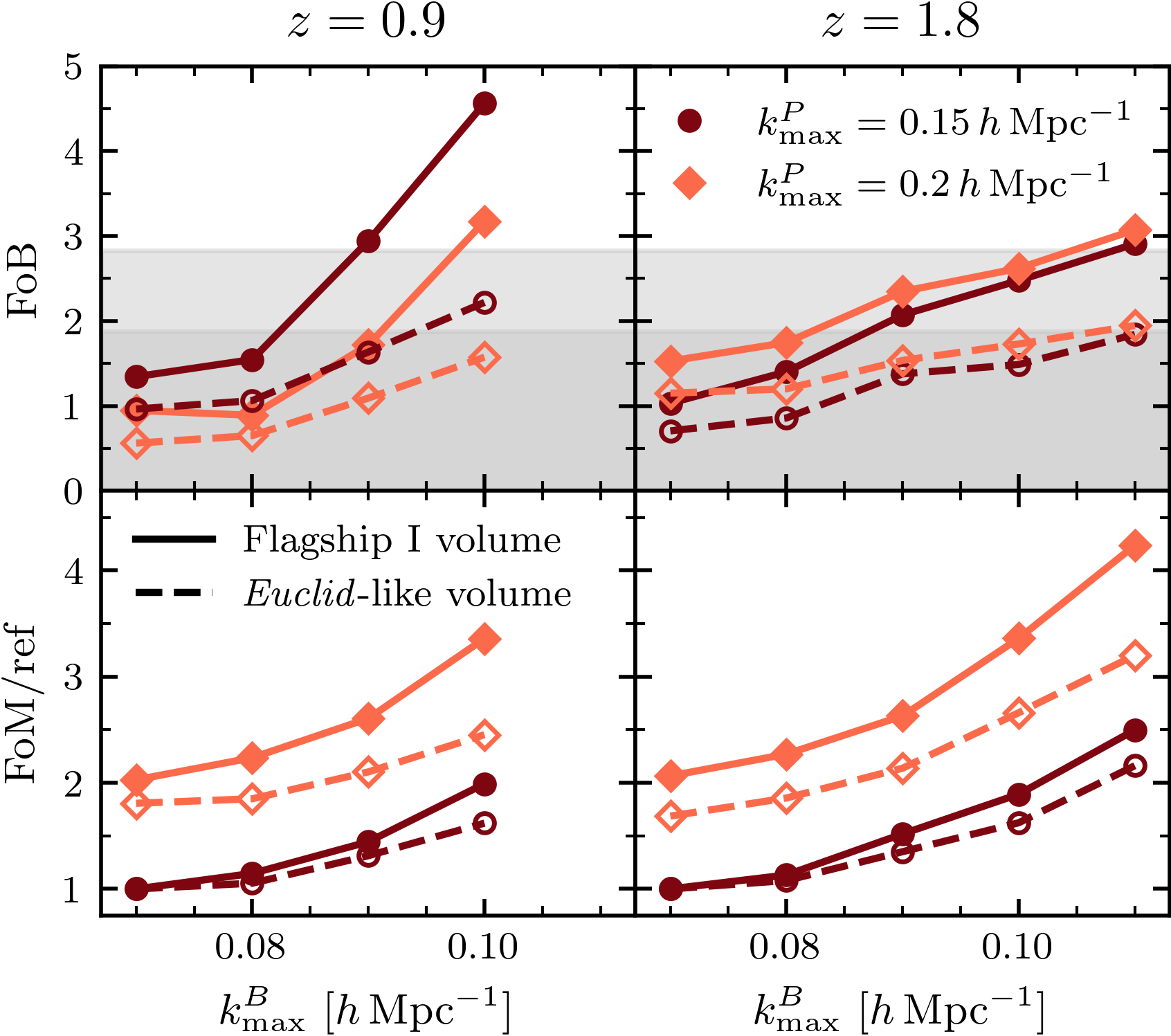}
  \caption{Comparison of the FoB and FoM from analyses using either the full Flagship I volume (solid lines, filled symbols) or a \Euclid-like survey volume at mission end (dashed lines, open symbols), shown for two values of $\kmaxP$ and for the lowest and highest redshift bins. FoM reference values are defined from the analysis performed at the lowest $\kmaxP$ and $\kmaxB$ separately in the two cases.}
  \label{fig:euclid_volume_runs}
\end{figure}

We now estimate the performance of the joint power spectrum and bispectrum multipole analysis on the projected full-mission \Euclid survey volume by rescaling the covariance for the Flagship I simulation.\footnote{We note that while volume rescaling provides an order-of-magnitude estimate of the constraining power, it fails to account for the inherent scatter of the new realisation. These limitations apply to bispectrum as well, as for example it neglects the distinct scaling of the Gaussian and non-Gaussian contributions to the covariance. Consequently, this approach renders the $\chi^2$ statistic unusable as a goodness-of-fit metric.} Specifically, we adopt a sky area of $14\,000$ deg$^2$ and two redshift bins defined as $0.8<z<1$ and $1.5<z<1.8$, corresponding to the first and last bins assumed in the forecasts of \citet{BlanchardEtal2020}. These bins are associated with the measurements from the Flagship I simulation snapshots at $z=0.9$ and $z=1.8$, respectively, and correspond to volumes that are smaller than the simulation volume by factors of 6.67 and 3.33.

The resulting FoB and FoM are shown in Fig.~\ref{fig:euclid_volume_runs}. As expected, for a \Euclid-like volume, the FoB is smaller due to the larger statistical uncertainty, and can be considered acceptable across all the range considered, up to $\kmaxB=0.11$ at large redshift.

In both cases, the FoM is shown in terms of the relative difference with respect to the reference value obtained for the lowest values of $\kmaxP$ and $\kmaxB$. In the \Euclid-like case, we find that the FoM improves by a factor of up to 2.45 (3.20) relative to the reference value at $z=0.9$ ($z=1.8$). The marginalised relative 1$\sigma$ errors for each cosmological parameter are shown in Table~\ref{tab:1sigma_Euclidlike}, demonstrating percent-level precision even in the \Euclid-like case. Moreover, the improvements from including the bispectrum remain roughly consistent with those found for the Flagship I volume, since the reduced volume is compensated by the extended model reach.

\subsection{Constraining power of the bispectrum quadrupole}
\label{ssec:bisp_quad}

Our analysis so far included the bispectrum quadrupole $B_2(k_1,k_2,k_3)$. Since this observable is defined for the same number of triangular configurations as the bispectrum monopole, it is interesting to investigate whether the effort required by the larger data vector is justified, particularly considering the limited number of mocks usually available to estimate numerically the full covariance matrix.

\begin{figure}[t!]
  \centering
  \includegraphics{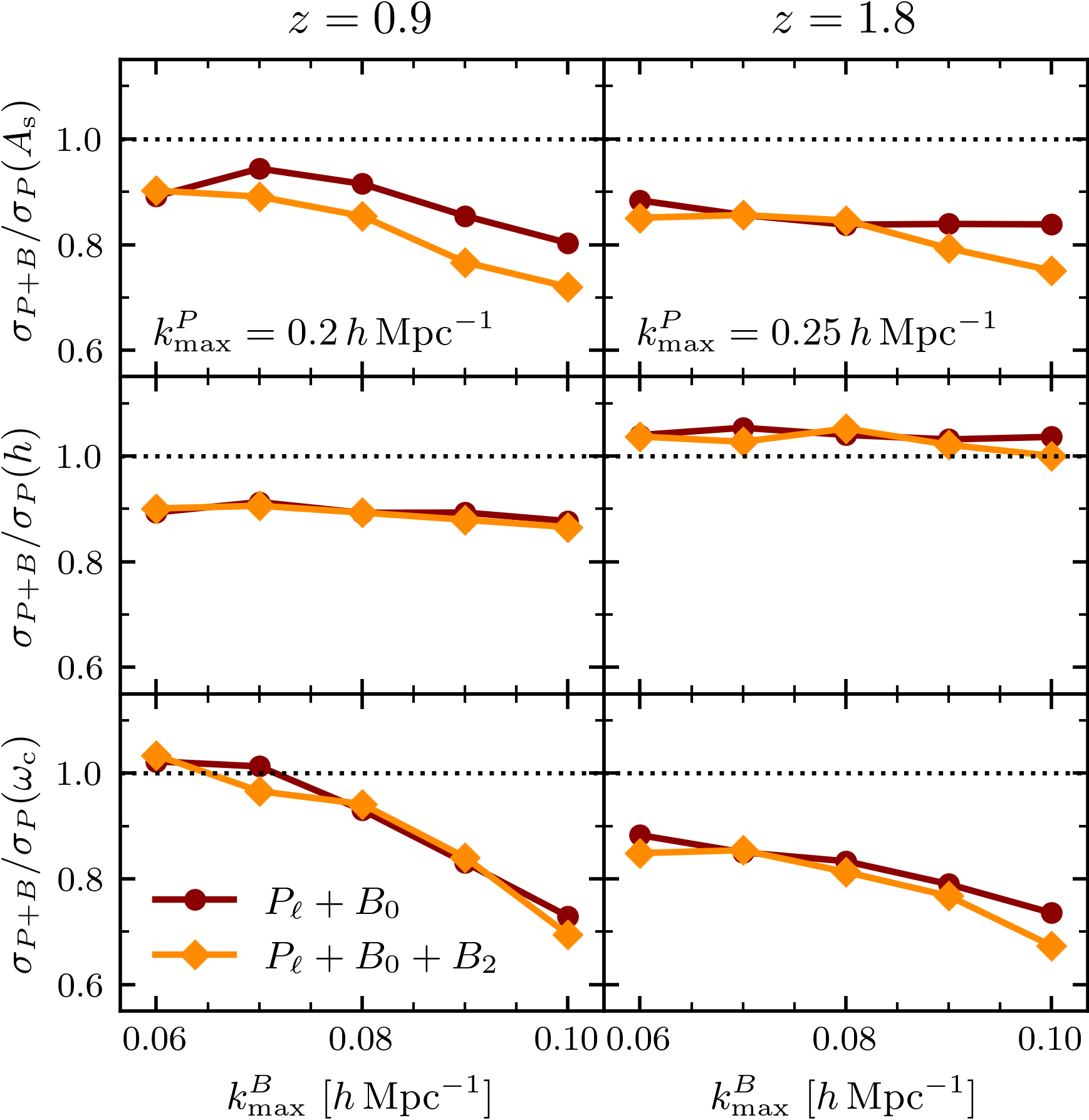}
  \caption{Reduction in the 1$\sigma$ constraints on the three cosmological parameters from including the bispectrum, using either the monopole alone (red), or monopole and quadrupole together (orange). Results at $z=0.9$ (left) are for $\kmaxP = 0.2\,\kMpc$, and at $z=1.8$ (right) for $\kmaxP = 0.25\,\kMpc$.}
  \label{fig:B2_constraining_power}
\end{figure}

Figure~\ref{fig:B2_constraining_power} shows the ratio of the marginalised constraints on the three cosmological parameters from the joint analysis of the power spectrum and bispectrum with and without quadrupole, to the power spectrum-only results. We find that the quadrupole provides additional constraining power, up to $12\%$, only on the amplitude parameter $\As$, while it does not yield significant improvements on the determination of the other two cosmological parameters, $h$ and $\omega_{\rm c}$. We notice that the joint analysis yields a slightly wider constraint on $h$ compared to the analysis using only the power spectrum at $z=1.8$. This can be understood considering that the joint analysis covers a larger parameter space including the additional shot noise parameter $\alpha_1$ and the counterterm $c_1^\mathrm{FoG}$ in the bispectrum model.

Our results are consistent with those of \citet{IvanovEtal2023} and \citet{EggemeierEtal2025}, who analysed synthetic catalogues designed to match the clustering of BOSS-like galaxies. The two studies used a cumulative volume of 566 $\cGpc$ and a \Euclid-like volume, respectively, and reported approximately \,10\% improvements in constraints on $\As$ from the bispectrum quadrupole, with no noticeable gains on other cosmological parameters such as $h$, $\omega_{\rm c}$, and $n_{\rm s}$. Application to BOSS data in \citet{IvanovEtal2023} confirmed this picture, whereas the analysis of \citet{DAmicoEtal2024} found no appreciable impact of the quadrupole, possibly due to using a one-loop model for the monopole and a correspondingly larger reach in $k$ compared to the quadrupole. More significant improvements, of the order 30 to 70\%, are instead reported by \citet{GualdiVerde2020} and \citet{GualdiGilMarinVerde2021}, who analysed synthetic data with a different bispectrum estimator and a different modelling.

\subsection{Non-Gaussian covariance}
\label{sec:ng_cov}

\begin{figure}
  \centering
   \includegraphics{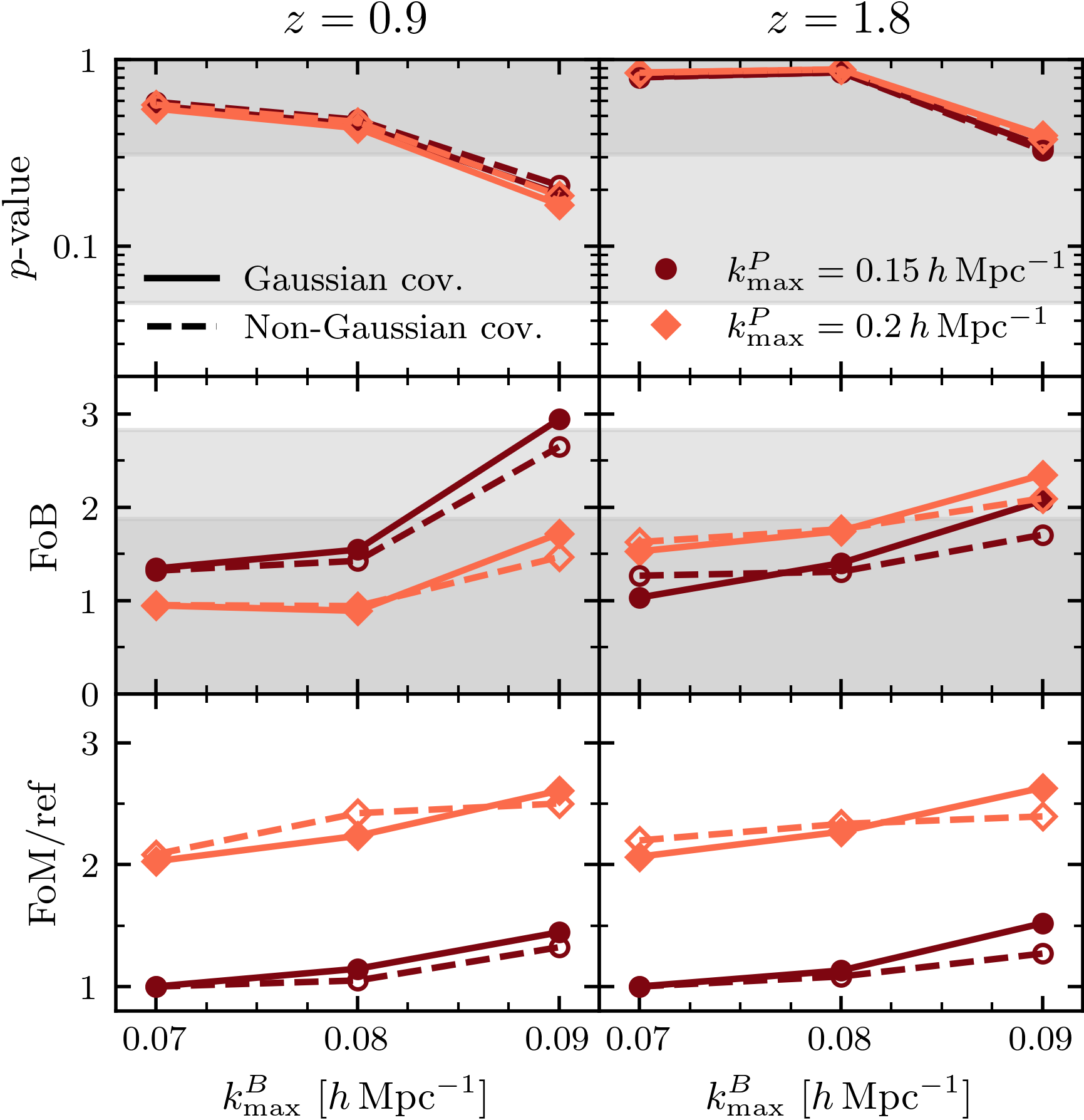}
  \caption{Same as the lower panels of Fig.~\ref{fig:FoM_FoB_pval_MaxMod}, but comparing results obtained with Gaussian (solid lines, filled symbols), or non-Gaussian (dashed lines, open symbols) covariance matrices. Results are shown for two values of $\kmaxP$ and for the lowest and highest redshift bins. FoM reference values correspond to the Gaussian covariance case and the lowest $\kmaxP$ and $\kmaxB$ values.}
  \label{fig:FoM_FoB_pval_NGCov}
\end{figure}

\begin{figure}
  \centering
  \includegraphics{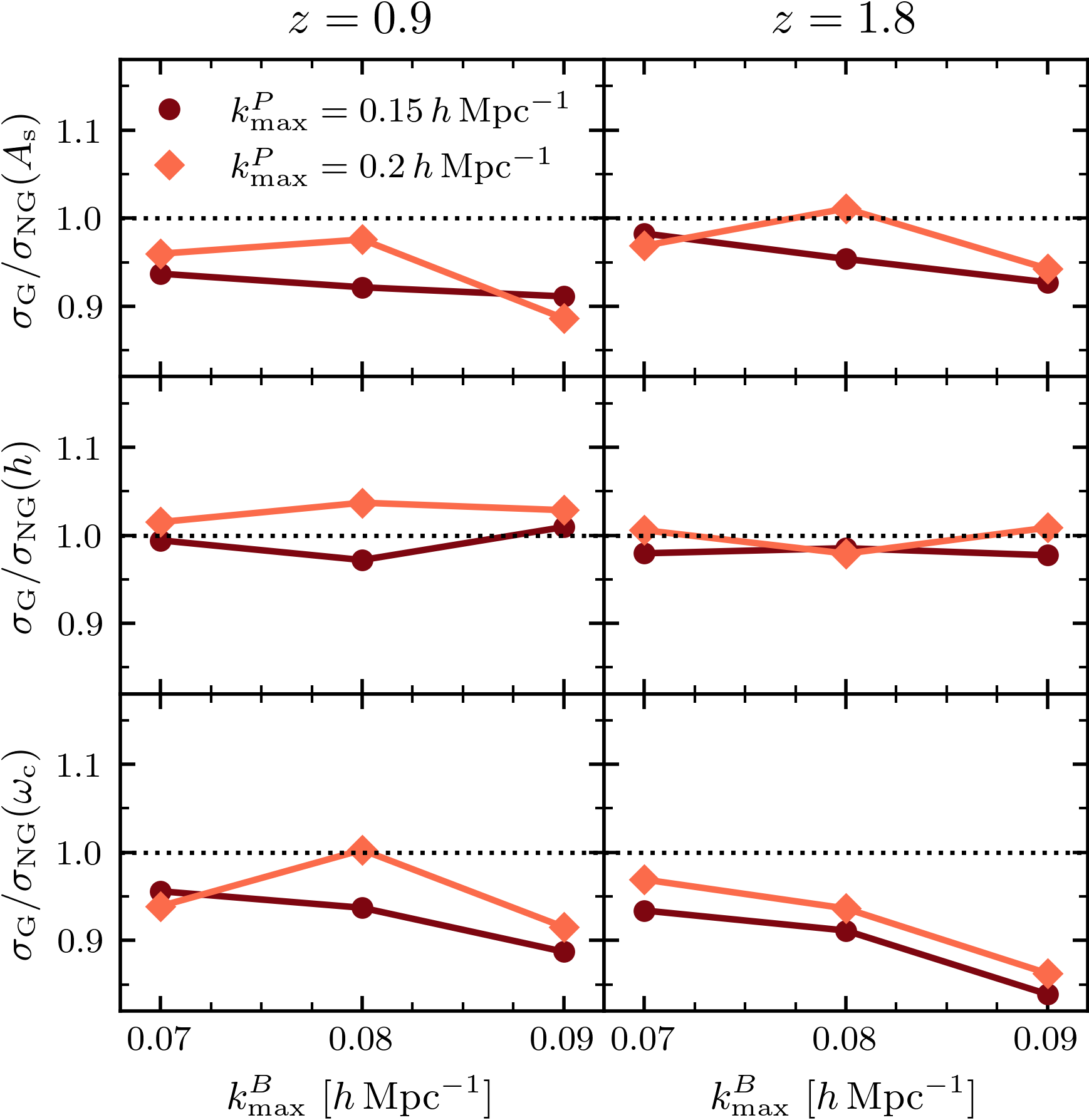}
  \caption{Ratios of the 1$\sigma$ constraints obtained from the joint power spectrum and bispectrum analysis using a Gaussian (G) covariance to those obtained assuming the non-Gaussian (NG) one, shown for two values of $\kmaxP$ and for the lowest (left) and highest (right) redshifts considered.}\label{fig:1sigma_NGCov}
\end{figure}

We now repeat the main analysis performed so far, taking into account, at least approximately, the non-Gaussian contribution to the bispectrum covariance and the cross-covariance between power spectrum and bispectrum multipoles. 

Such terms are usually neglected in the analysis of measurements from simulations in boxes with periodic boundary conditions. Yet, the non-Gaussian contribution is, in fact, the leading term for the covariance of the squeezed, triangular configurations that share the smallest momenta $\{k_{3,i},k_{3,j}\}$, as discussed in \citet{Barreira2019} and \citet{ BiagettiEtal2022}. In general, the full non-Gaussian covariance corresponds to the sum of a term from the product of two bispectra ($B$ -- $B$), a term from the product of a power spectrum and trispectrum ($P$ -- $T$), and a term due to the galaxy six-point function \citep[see, e.g.,][]{SefusattiEtal2006}. Here, we follow \citet{SalvalaggioEtal2024} and adopt the approximation where the $P$ -- $T$ term is replaced by an additional $B$ -- $B$ contribution as they have the same value in the squeezed-triangle limit (where they are the most relevant) and the 6-point function term is neglected altogether. In addition, we approximate the power spectrum-bispectrum cross-covariance with its leading contribution, given by the product of the power spectrum and bispectrum, neglecting the 5-point function contribution.\footnote{All such quantities are numerically evaluated with the public \texttt{Python} code \texttt{bisque}, \href{https://gitlab.com/jacopo.salvalaggio/bisque}{https://gitlab.com/jacopo.salvalaggio/bisque}.} 

\begin{table*}
\centering
\caption{List of the codes participating in the comparison test, including the main references and public repository (if available).}
\label{tab:code_list}
\begin{tabular}{l|l}
\hline \hline
   & \\[-9pt]
Codes name and description & Reference(s)  \\ [3pt]
\hline
& \\[-9pt]
\texttt{PBJ} (\textbf{P}ower spectrum \& \textbf{B}ispectrum \textbf{J}oint analysis) & \cite{MorettiEtal2023} \\[3pt]
\texttt{COMET} (\textbf{C}osmological \textbf{O}bservables \textbf{M}odelled by \textbf{E}mulated perturbation \textbf{T}heory) & \cite{EggemeierEtal2023} \github{https://gitlab.com/aegge/comet-emu} \\[3pt]
\texttt{CLASS-PT}, non-linear perturbation theory extension of the Boltzmann code \texttt{CLASS} & \cite{ChudaykinEtal2020} \github{https://github.com/Michalychforever/CLASS-PT}\\[3pt]
\texttt{CosmoSIS-gClust}, new library for \texttt{CosmoSIS} to perform galaxy clustering analysis & Linde et al. (in prep.) \\[3pt]
\texttt{CLASS-OneLoop}, non-linear pertubation theory extension of the Boltzmann code \texttt{CLASS} & \citet{LindeEtal2024} \\[3pt]
\texttt{PyBird} (\texttt{\textbf{P}ython} code for \textbf{B}iased tracers in redshift space) & \cite{DAmicoSenatoreZhang2021} \github{https://github.com/pierrexyz/pybird/} \\[3pt]
\hline
\end{tabular}
\end{table*}
 
Figure \ref{fig:FoM_FoB_pval_NGCov} presents the comparison of the FoM, FoB, and $p$-value for the analysis assuming the non-Gaussian contribution to the results already shown in  Fig.~\ref{fig:FoM_FoB_pval_MaxMod}.  
We observe that, overall, accounting for the non-Gaussian contribution in the covariance does not lead to significantly different results. The FoB is slightly reduced at the largest $\kmaxB$ values (with a maximum difference of approximately $21\%$), while the goodness of fit is essentially unaffected. The FoM values change mildly, with the relative difference at most approximately 33\%, again for $\kmaxB=0.09\kMpc$. The effect on the individual cosmological parameters is explored in more detail in Fig.~\ref{fig:1sigma_NGCov}. The Gaussian approximation underestimates specifically the uncertainties of the $\omega_{\rm c}$ and $\As$ parameters, the parameters more directly related to the signal amplitude, with respect to the analysis with the non-Gaussian covariance. The difference increases as we move to smaller scales, reaching a maximum of approximately 16\% in the case of the highest bispectrum scale cut considered ($k_\mathrm{max}^{B} = 0.09~\kMpc$). 

These simple results should, of course, be taken with a grain of salt, because of the approximations assumed but also, and more importantly, because in the realistic case of a survey footprint, additional super-sample contributions should be taken into account \citep[see, e.g.,][for the power spectrum case]{WadekarScoccimarro2020}.

\subsection{Code comparison}
\label{sec:code_comparison}

\begin{figure*}
\centering
   \includegraphics[width=17cm]{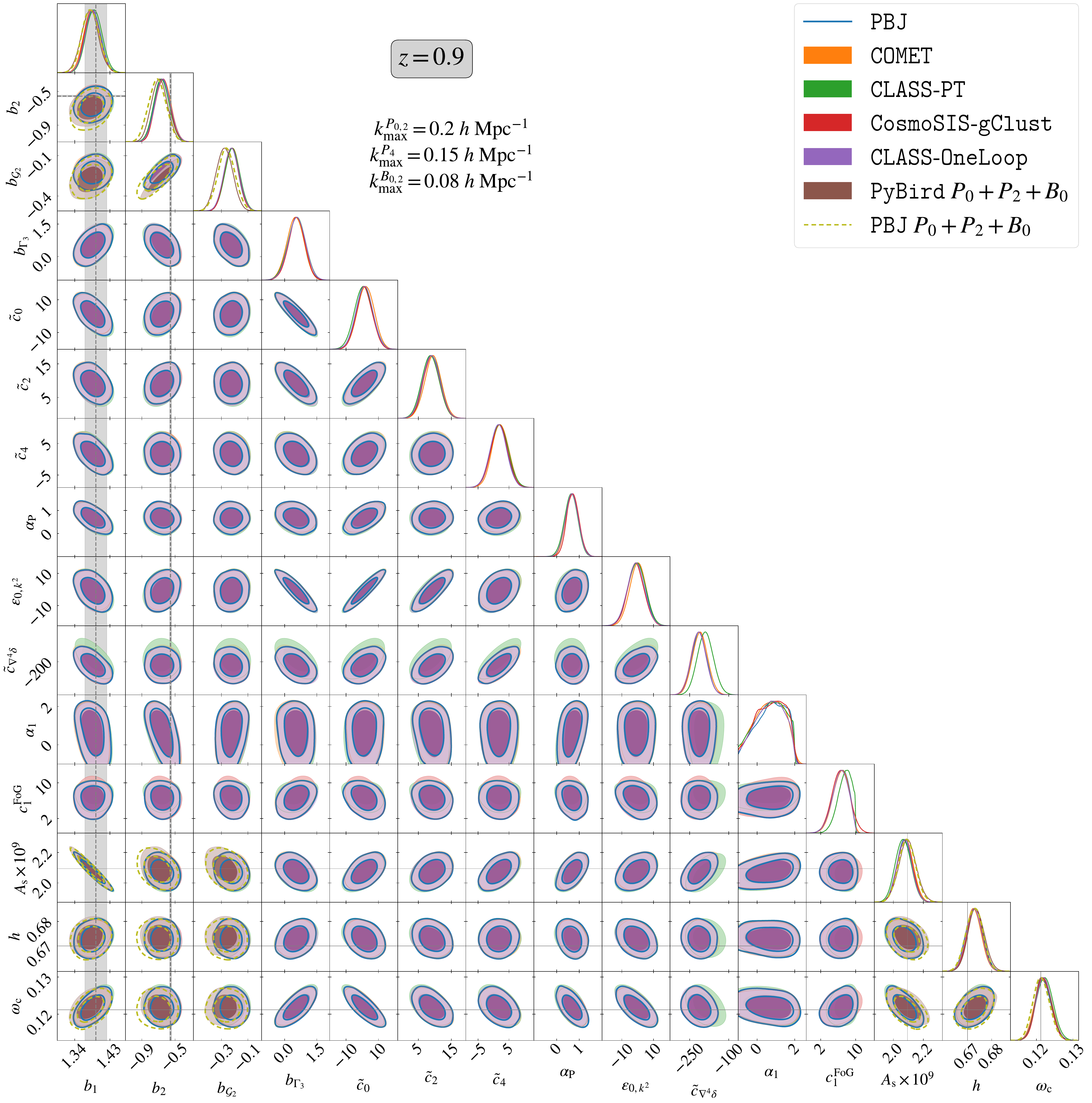}
   \caption{Comparison between different codes at the level of 2-dimensional posteriors of the nuisance and cosmological parameters from the analysis done at $z=0.9$. Solid lines denote the fiducial values for the cosmological parameters and \gls{hod} prediction, based on the bias relations from \cite{TinkerEtal2010} and \cite{LazeyrasEtal2016} for linear bias and quadratic bias, respectively. The grey bands represent the propagation of an assumed 2\% uncertainty in the bias relations. Additionally, we present a comparison of the marginalised 2-dimensional parameters from the joint analysis of $P_0 + P_2 + B_0$ between $\mathtt{PyBird}$ and $\mathtt{PBJ}$. We note that parameters entering the model linearly are analytically marginalised in $\mathtt{PyBird}$ and are therefore not shown in the posterior contours.}
\label{fig:codes_comparison}
\end{figure*}

To ensure the reliability of the results presented in this paper and, at the same time, provide a robust benchmark for the validation of the analysis pipeline that will be applied to \Euclid data, we conduct a comparison of different codes available within the collaboration. In addition to the primary codes used in the main study, \texttt{PBJ} and \texttt{COMET}, the other codes are listed in Table \ref{tab:code_list} and described briefly in Appendix~\ref{app:code_desc}. 

To ensure a fair comparison, all codes employ a consistent analysis setup, including the priors specified in Table\,\ref{tab:priors}, as well as identical scale cuts, given by $\kmax^{P_0} = \kmax^{P_2} = \kmax^{P_4} + 0.05 ~ \kMpc=0.2~(0.25)~\kMpc$, respectively for the two snapshots at $z=0.9~(1.8)$, while $\kmax^{B_0} = \kmax^{B_2} = 0.08 \kMpc$ for both. 
As shown in Fig.\,\ref{fig:codes_comparison}, the resulting 2-dimensional posteriors for the bias and cosmological parameters from the analysis at $z=0.9$ show excellent agreement among the participating codes. Quantitatively, the root-mean-square (RMS) shift of the posterior means relative to the \texttt{PBJ} 1D marginalised uncertainty is $0.16\sigma$. Importantly, this agreement extends not only to cosmological parameters but also to nuisance parameters. A similar level of consistency is observed for the analysis at $z=1.8$.

We perform a slightly different comparison between $\mathtt{PBJ}$ and $\mathtt{PyBird}$, limited to the power spectrum multipoles and the bispectrum monopole. This is motivated by the fact that although \texttt{PyBird} can output all Legendre quadrupoles \citep[see, e.g.,][]{DAmicoEtal2024}, for the comparison we use \texttt{PyBird} and \texttt{PBJ} defined in terms of the angle with the shortest side $k_3$, following the choice in Eq.~\eqref{B_ell_estimator}. In addition, it adopts a different basis for the bias and shot-noise parameters, detailed in Appendix~\ref{app:PyBirdBias} along with the corresponding priors assumed in our runs. The marginalised 2-dimensional parameter constraints, also shown in Fig. \ref{fig:codes_comparison}, demonstrate good agreement between $\mathtt{PyBird}$ and $\mathtt{PBJ}$.\footnote{In Fig. \ref{fig:codes_comparison}, we use the transformations $b_2 = 2 (\tilde{b}_2 + \tilde{b}_5 - \tilde{b}_1)$ and $b_{\mathcal{G}_2} = 2/7 (\tilde{b}_2 - \tilde{b}_5)$, where $\tilde{b}_1, \tilde{b}_2$, and  $\tilde{b}_5$ denote the set of bias parameters implemented in $\texttt{PyBird}$. See Appendix~\ref{app:PyBirdBias} for details.}

\section{Conclusions}
\label{sec:conclusion}

In this work, we presented a comprehensive validation of the joint modelling of the galaxy power spectrum and bispectrum using a synthetic H$\alpha$ galaxy population based on a high-resolution \textit{N}-body simulation designed to reproduce the \Euclid spectroscopic sample. We assessed the performance of the one-loop galaxy power spectrum and tree-level bispectrum predictions from perturbation theory, in both real and redshift space, employing a FoM, FoB, and the $p$-value to quantify the constraining power, accuracy, and goodness of fit of the model. Our results demonstrate that combining power spectrum and bispectrum measurements yields significant gains of up to 30\% in parameter constraints, underlining the importance of their joint analysis for fully exploiting the constraining power of the forthcoming \Euclid data.

Our real-space analysis provides the first detailed characterisation of the nonlinear bias model for a synthetic population of H$\alpha$ galaxies across different redshifts. We showed that bias relations calibrated on high-mass dark matter haloes \citep[e.g.][]{LazeyrasEtal2016} fail to capture the behaviour of our low-redshift samples (at $z=0.9$ and $z=1.2$), since the \gls{hod} reaches haloes of significantly smaller masses (below the calibration threshold $\log_{10}[M/(h^{-1}M_\odot)]=12.55$). On the other hand,  the second-order tidal bias, $b_{{\cal G}_2}$, is well described by the excursion set prediction of \citet{ShethChanScoccimarro2013} and the third-order tidal bias by the co-evolution expression for $b_{\Gamma_3}$ \citep{ChanScoccimarroSheth2012} as a function of the linear bias and $b_{{\cal G}_2}$.

Assuming statistical uncertainties corresponding to the full Flagship I simulation volume of 54~$\cGpc$, we found that the tree-level EFTofLSS approximation accurately models the large-scale redshift-space bispectrum up to $\kmaxB\simeq 0.08\kMpc$ and $\kmaxB\simeq 0.1\kMpc$, respectively, at the lowest ($z=0.9$) and highest ($z=1.8$) redshifts considered here. In contrast, the validity ranges in real space are significantly higher ($0.14$ to $0.16\kMpc$), indicating that redshift-space distortions are the main source of theoretical systematics.  Within these ranges,  addition of the bispectrum improves constraints on $\As$ and $\omega_{\rm c}$ by up to 30\% (see Table~\ref{tab:1sigma_Euclidlike}), with the bispectrum quadrupole contributing as much as 10\%, while constraints on $h$ are mostly saturated by the power spectrum. The resulting FoM, defined in terms of these three parameters, increases by about a factor of 2.5 across all redshifts. 

Rescaling the covariance matrix to the expected \Euclid redshift-bin volumes (smaller than the Flagship I volume by factors of approximately 3--6) leads to comparable relative gains due to a slightly extended modelling reach that compensates for the reduced volume. Under these conditions the FoM improves by up to a factor of 2.45 (3.20) at $z=0.9$ ($z=1.8$) compared with the power spectrum alone. Further improvements in the modelling of redshift-space distortions may extend the validity range of the bispectrum model and produce even larger gains, as recently shown by \citet{EggemeierEtal2025}.

We examined the effect of non-Gaussian terms in the bispectrum covariance and their leading contribution to the cross covariance between the two probes by means of the simple approximation proposed in \citet{SalvalaggioEtal2024}, where the $P$--$T$ term is replaced by an additional $B$--$B$ contribution. We found that the bounds on cosmological parameters from analyses using Gaussian covariances are generally tighter than those using non-Gaussian covariances, with differences increasing at smaller scales and reaching a maximum of approximately 16\% at the highest bispectrum scale cut considered ($k_\mathrm{max}^{B} = 0.09~\kMpc$). Of course, the full covariance of lightcone measurements will include additional finite-volume contributions that we are neglecting here and that will be considered elsewhere. Still, it is important to keep in mind that the usual Gaussian approximation for the covariance matrix might be particularly limited in the bispectrum case.  

Finally, we provided a detailed comparison among six different codes for the joint analysis of the power spectrum and bispectrum available within the \Euclid collaboration. We found remarkable agreement in all cases where the comparison is performed under the exact same assumptions, with an RMS shift of the posterior means of $0.16\sigma$ relative to the reference uncertainty. This represents an important benchmark for the validation of the official collaboration code. 

This work represents the first of a series of papers in preparation for the bispectrum analysis of \Euclid data. It presents a study of synthetic data from the Flagship I simulation along with similar work, of imminent publication, focused on 2-point statistics 
(\citealt{PezzottaEtal2024};
Euclid Collaboration: Camacho et al., in prep.;
Euclid Collaboration: Kärcher et al., in prep.) and on the 3-point correlation function (\citealt{GuidiEtal2025}; Euclid Collaboration: Pugno et al., in prep.; Euclid Collaboration: Moresco et al., in prep.). A direct extension of our results to the case of cosmologies with non-Gaussian initial conditions will be given elsewhere (Euclid Collaboration: Linde et al., in prep.). 

Our next steps will be directed toward tackling the systematic errors characterising the \Euclid galaxy sample based on slitless spectroscopy. These efforts will be carried out in the context of Data Release I.

\begin{acknowledgements}

\AckEC
This research benefits from the High Performance Computing facility of the University of Parma, Italy (HPC.unipr.it).

\end{acknowledgements}


\bibliography{cosmologia.bib}


\begin{appendix}

\section{Local, quadratic bias fit}
\label{app:b2tilde}

\citet{LazeyrasEtal2016} provides the following relation between the linear and quadratic, local bias parameters, obtained from `separate Universe' simulations 
\begin{equation}
\tilde{b}_2^{\rm h} =0.412-2.143\, b_1^{\rm h} + 0.929\, \left(b^{\rm h}_1\right)^2+0.008\, \left(b_1^{\rm h}\right)^3\;,
\end{equation}
where $\tilde{b}_2$ represents the quadratic bias definition when the tidal operator in the expansion is assumed to be $K^2=\G_2+(2/3)\,\delta^2$ leading to the relation
   \begin{equation}
         \tilde{b}_2  \equiv b_2-\frac{4}{3}\,\bGtwo\;,
    \end{equation}
$b_2^{\rm h}$ being the quadratic parameter associated to the tidal operator $\G_2$ assumed in this work.

Given predictions for the halo bias functions, galaxy bias parameters can be calculated using the halo model and the \gls{hod} \citep[see, e.g.,][for a review]{CooraySheth2002}. In particular, we have 
\begin{align}
\label{eq:bp}
b_1 & = \frac{\int_0^\infty {\rm d}M\, n(M)\,N_{\rm g}(M)\,b_1^{\rm h}(M)}{\int_0^\infty {\rm d}M \, n(M)N_{\rm g}(M)}\;,\\
\label{eq:b2hm}
b_2 &=  \frac{\int_0^\infty {\rm d}M\, n(M)\,N_{\rm g}(M)\,b_2^{\rm h}(M)}{\int_0^\infty {\rm d}M \, n(M)N_{\rm g}(M)}\;,\\
\nonumber \text{and} \quad \\
\label{eq:g2hm}b_{K^2} & = \frac{\int_0^\infty {\rm d}M \,n(M)N_{\rm g}(M)\,b_{K^2}^{\rm h}(M) }{\int_0^\infty {\rm d}M \,n(M)N_{\rm g}(M)}\;,
\end{align}
where $n(M)$ is the halo mass function, $N_{\rm g}(M)$ the \gls{hod} model, $b_1^{\rm h}$, $b_2^{\rm h}$, and $b_{K^2}^{\rm h}$ are the corresponding halo bias functions. Here we neglect the response of the \gls{hod} to large-scale perturbations \citep{VoivodicBarreira2021, BarreiraLazeyrasSchmidt2021} since these effects were not accounted for in the construction of the catalogues. We use these equations to obtain the relation for the galaxy bias parameters specific to our H$\alpha$ galaxies given by
   \begin{align}
    \label{eq:b2tilde_hod}
         \tilde{b}_2 &  = -0.015 - 1.58\,b_1 + 0.809\,b_1^2 + 0.025\,b_1^3\;,
    \end{align}
leading to Eq.~\eqref{eq:b2_hod}.    

\section{Description of participating codes}
\label{app:code_desc}

\texttt{CosmoSIS-gClust} extends \texttt{CosmoSIS} \citep{ZuntzEtal2015} with two modular C libraries for EFTofLSS predictions (Eulerian moment expansion, \citealt{ChenEtal2021}) and likelihood inference of the real- and redshift-space power spectrum and bispectrum of matter and biased tracers. It computes one-loop power spectrum and tree-level 
bispectrum with \texttt{FFTLog} \citep{Hamilton2000}, supports Gaussian and non-Gaussian initial conditions, and includes analytic marginalisation and posterior reconstruction for linear nuisance parameters, and can be used with all \texttt{CosmoSIS} samplers. Full details and an application to \Euclid-like mocks including local primordial non-Gaussianity (PNG) will appear in Linde et al. (in prep.) and Euclid Collaboration: Linde et al. (in prep.), and the code will be publicly released thereafter.

\texttt{CLASS-OneLoop} \citep{LindeEtal2024} is an extension of the Boltzmann solver \texttt{CLASS},\footnote{Independent of \texttt{CLASS-PT} and fully integrated into \texttt{CLASS}.} providing EFTofLSS predictions of the one-loop power spectrum (using the Eulerian moment expansion of \citealt{ChenEtal2021}) and the tree-level bispectrum of matter and biased tracers in real and redshift space. These functionalities will be released as part of the main public branch of \texttt{CLASS}. As elsewhere in \texttt{CLASS}, the extension can be used directly with \texttt{MontePython} \citep{AudrenEtal2012, BrinckmannLesgourgues2018} for likelihood inference. Analytic marginalisation and posterior reconstruction of linear nuisance parameters are implemented, along with a fast–slow split that considerably accelerates MCMC performance.

\texttt{PyBird} \citep{DAmicoSenatoreZhang2021} is a code written in \texttt{Python} 3, designed for evaluating multipoles of the power spectrum of biased tracers in redshift space based on the EFTofLSS formulations found in \citet{PerkoEtal2016A} and \citet{DAmicoEtal2020}. The main technology used for the fast loop and IR resummation evaluation is the \texttt{FFTLog} code \citep{Hamilton2000, SimonovicEtal2018, DAmicoSenatoreZhang2021}. While the power spectrum module is publicly available at the address in Table~\ref{tab:code_list}, here we used the \texttt{PyBird} extension that also includes the computation of the bispectrum multipoles \citep{DAmicoEtal2024, DAmicoEtal2025}.\footnote{Additionally, \texttt{PyBird} can also compute the configuration-space one-loop correlation function of biased tracers \citep{ZhangEtal2022}.}

\section{PyBird bias parameters}
\label{app:PyBirdBias}

$\mathtt{PyBird}$ uses a different set of bias parameters, stochastic term coefficients, and counterterms that we denote as $\Big \{b_1, \tilde{b}_2, \tilde{b}_3, \tilde{b}_5, \tilde{b}_8, c_1^{\mathrm{St}}, c_2^{\mathrm{St}}, c_3^{\mathrm{St}}, c_6^\mathrm{St}, c_1^{(222)},  c_1^h, c_1^\pi, c_1^{\pi v}, c_3^{\pi v}  \Big \}$. Here $\tilde{b}_2, \tilde{b}_3, \tilde{b}_5, \tilde{b}_8$ are the nonlinear EFTofLSS bias parameters \citep{Senatore2015, AnguloEtal2015, FujitaEtal2020, DAmicoEtal2024, DAmicoEtal2024B}. Additionally, $c_1^{\mathrm{St}}, c_2^{\mathrm{St}}, c_3^{\mathrm{St}}, c_6^\mathrm{St}, \text{ and } c_1^{(222)}$ are the stochastic terms coefficients. The first three parameters, $c_1^{\mathrm{St}}, c_2^{\mathrm{St}},  \text{ and } c_3^{\mathrm{St}}$, contribute to the shot noise of the power spectrum through terms proportional to a constant, $k^2/k_{\rm M}^2$, and $\mu^2 k^2/k_{\rm M}^2$, respectively, for some scale $k_{\rm M}$. These same parameters also appear in the bispectrum shot-noise contribution proportional to $1/\bar{n}$, through the specific functions detailed in Appendix D.3 of \cite{DAmicoEtal2024B}, while $c_6^\mathrm{St}$ is proportional to $\alpha_1$ in \eq{eq:B_stochastic}. We also include $c_1^{(222)}$, a parameter that controls the deviation from Poisson shot noise $1/\bar{n}^2$ in \eq{eq:B_stochastic}. Finally, the dimensionless counterterms $c_1^h, c_1^\pi, c_1^{\pi v},  \text{ and } c_3^{\pi v}$ are proportional to $k^2/k_{\rm M}^2$, $f \mu^2 k^2/k_{\rm R}^2$, $f^2 \mu^4 k^2/k_{\rm R}^2$, and $f^2 \mu^2 k^2/k_{\rm R}^2$, respectively, for some scale $k_{\rm R}$. In the analysis, we use $k_{\rm M} = 0.7 \kMpc$ and $k_{\rm R} = 0.25 \kMpc$.

In the $\mathtt{PyBird}$ analysis, we adopt the flat priors on the cosmological parameters and linear galaxy bias from Table~\ref{tab:priors} and we impose flat priors on the nonlinear bias parameters, $\tilde{b}_2 \sim \mathcal{U}[-10, 10]$ and $\tilde{b}_5 \sim \mathcal{U}[-10, 10]$. The remaining parameters, which enter the theoretical model linearly, are marginalised over analytically, assuming the Gaussian priors given in Table~\ref{tab:PyBirdBias}. The widths of the priors for the stochastic parameters $c_1^h$ and $c_6^\mathrm{St}$ were chosen to approximate the flat priors used for other codes, as described in Table \ref{tab:priors}. Here the central value for $c_6^\mathrm{st}$ is proportional to the linear bias as $c_6^\mathrm{st} = b_1 (1+\alpha_1)$. For the other parameters, we adopt broad, as uninformative as possible, priors. In addition, since the hexadecapole of the power spectrum is excluded from the analysis, $c_1^{\pi v}$ is set to zero. The FoG counterterm $c_1^\mathrm{FoG}$ in the bispectrum is also set to zero.

\begin{table}[t!]
\caption{Gaussian priors $\mathcal{N}[\bar{\theta}, \sigma]$, centered on $\bar{\theta}$ with a standard deviation $\sigma$, on the model parameters analytically marginalised in the $\texttt{PyBird}$ runs.}
  \renewcommand{\arraystretch}{1.3}
  \centering
  \begin{tabular}{lcc}
    \hline \hline
     & Parameter & Prior\\
     \hline
    \multirow{1}{*}{Galaxy bias} & $\tilde{b}_3$,  $\tilde{b}_5$, $\tilde{b}_8$ & \small{$\mathcal{N}[0, 10]$}\\
    \cline{2-3}
    \hline
    \multirow{1}{*}{Counterterms} & $c_1^h,\, c_1^\pi,\, c_3^{\pi v}$ & \small{$\mathcal{N}[0, 10]$}\\    
    \hline
    \multirow{4}{*}{Shot noise} & $c_1^\mathrm{St}$ & \small{$\mathcal{N}[1, 2]$} \\
    & $c_2^\mathrm{St},\, (2/3) f c_3^\mathrm{St}$ & \small{$\mathcal{N}[0, 10]$}\\
    & $2 \, c_6^\mathrm{St}$ & $\mathcal{N}[1.386, 2]$\\
    & $c_1^{(222)}$ & $\mathcal{N}[1, 2]$\\
    \hline
  \end{tabular}
  \label{tab:PyBirdBias}
\end{table}

\end{appendix}

\end{document}

%% file: authors.tex
\newcommand{\orcid}[1]{} 
\author{Euclid Collaboration: K.~Pardede\orcid{0000-0002-7728-8220}\thanks{\email{kevinfrc@jpl.nasa.gov}}\inst{\ref{aff1},\ref{aff2},\ref{aff3},\ref{aff4},\ref{aff5}}
\and A.~Eggemeier\orcid{0000-0002-1841-8910}\inst{\ref{aff6}}
\and D.~Alkhanishvili\orcid{0000-0001-6450-6914}\inst{\ref{aff6}}
\and E.~Sefusatti\orcid{0000-0003-0473-1567}\inst{\ref{aff7},\ref{aff4},\ref{aff8}}
\and A.~Moradinezhad~Dizgah\orcid{0000-0001-8841-9989}\inst{\ref{aff9}}
\and L.~Christoph\orcid{0009-0005-5324-0731}\inst{\ref{aff9},\ref{aff10}}
\and A.~Chudaykin\inst{\ref{aff11}}
\and M.~K{\"a}rcher\orcid{0000-0001-5868-647X}\inst{\ref{aff12}}
\and D.~Linde\orcid{0000-0001-7192-1067}\inst{\ref{aff1}}
\and M.~Marinucci\orcid{0000-0003-1159-3756}\inst{\ref{aff13},\ref{aff14}}
\and C.~Porciani\orcid{0000-0002-7797-2508}\inst{\ref{aff6}}
\and A.~Veropalumbo\orcid{0000-0003-2387-1194}\inst{\ref{aff15},\ref{aff16},\ref{aff17}}
\and M.~Crocce\orcid{0000-0002-9745-6228}\inst{\ref{aff18},\ref{aff19}}
\and M.~S.~Cagliari\orcid{0000-0002-2912-9233}\inst{\ref{aff9}}
\and B.~Camacho~Quevedo\orcid{0000-0002-8789-4232}\inst{\ref{aff4},\ref{aff2},\ref{aff7}}
\and L.~Castiblanco\orcid{0000-0002-2324-7335}\inst{\ref{aff20},\ref{aff21}}
\and E.~Castorina\inst{\ref{aff12},\ref{aff22}}
\and G.~D'Amico\orcid{0000-0002-8183-1214}\inst{\ref{aff23},\ref{aff1}}
\and V.~Desjacques\orcid{0000-0003-2062-8172}\inst{\ref{aff24}}
\and A.~Farina\orcid{0009-0000-3420-929X}\inst{\ref{aff17},\ref{aff15},\ref{aff16}}
\and G.~Gambardella\orcid{0009-0001-1281-1746}\inst{\ref{aff18},\ref{aff19}}
\and M.~Guidi\orcid{0000-0001-9408-1101}\inst{\ref{aff25},\ref{aff26}}
\and F.~Janssen\orcid{0009-0000-1616-5799}\inst{\ref{aff10},\ref{aff27}}
\and J.~Lesgourgues\orcid{0000-0001-7627-353X}\inst{\ref{aff10}}
\and C.~Moretti\orcid{0000-0003-3314-8936}\inst{\ref{aff7},\ref{aff4},\ref{aff8}}
\and A.~Pezzotta\orcid{0000-0003-0726-2268}\inst{\ref{aff15}}
\and A.~Pugno\inst{\ref{aff6}}
\and J.~Salvalaggio\orcid{0000-0002-1431-5607}\inst{\ref{aff7},\ref{aff4},\ref{aff28},\ref{aff8}}
\and B.~Altieri\orcid{0000-0003-3936-0284}\inst{\ref{aff29}}
\and S.~Andreon\orcid{0000-0002-2041-8784}\inst{\ref{aff15}}
\and N.~Auricchio\orcid{0000-0003-4444-8651}\inst{\ref{aff26}}
\and M.~Baldi\orcid{0000-0003-4145-1943}\inst{\ref{aff25},\ref{aff26},\ref{aff30}}
\and S.~Bardelli\orcid{0000-0002-8900-0298}\inst{\ref{aff26}}
\and P.~Battaglia\orcid{0000-0002-7337-5909}\inst{\ref{aff26}}
\and A.~Biviano\orcid{0000-0002-0857-0732}\inst{\ref{aff7},\ref{aff4}}
\and M.~Brescia\orcid{0000-0001-9506-5680}\inst{\ref{aff31},\ref{aff32}}
\and S.~Camera\orcid{0000-0003-3399-3574}\inst{\ref{aff33},\ref{aff34},\ref{aff35}}
\and G.~Ca\~nas-Herrera\orcid{0000-0003-2796-2149}\inst{\ref{aff36},\ref{aff37}}
\and V.~Capobianco\orcid{0000-0002-3309-7692}\inst{\ref{aff35}}
\and C.~Carbone\orcid{0000-0003-0125-3563}\inst{\ref{aff38}}
\and V.~F.~Cardone\inst{\ref{aff39},\ref{aff40}}
\and J.~Carretero\orcid{0000-0002-3130-0204}\inst{\ref{aff41},\ref{aff42}}
\and S.~Casas\orcid{0000-0002-4751-5138}\inst{\ref{aff10},\ref{aff43}}
\and M.~Castellano\orcid{0000-0001-9875-8263}\inst{\ref{aff39}}
\and G.~Castignani\orcid{0000-0001-6831-0687}\inst{\ref{aff26}}
\and S.~Cavuoti\orcid{0000-0002-3787-4196}\inst{\ref{aff32},\ref{aff44}}
\and K.~C.~Chambers\orcid{0000-0001-6965-7789}\inst{\ref{aff45}}
\and A.~Cimatti\inst{\ref{aff46}}
\and C.~Colodro-Conde\inst{\ref{aff47}}
\and G.~Congedo\orcid{0000-0003-2508-0046}\inst{\ref{aff36}}
\and C.~J.~Conselice\orcid{0000-0003-1949-7638}\inst{\ref{aff48}}
\and L.~Conversi\orcid{0000-0002-6710-8476}\inst{\ref{aff49},\ref{aff29}}
\and Y.~Copin\orcid{0000-0002-5317-7518}\inst{\ref{aff50}}
\and F.~Courbin\orcid{0000-0003-0758-6510}\inst{\ref{aff51},\ref{aff52},\ref{aff53}}
\and H.~M.~Courtois\orcid{0000-0003-0509-1776}\inst{\ref{aff54}}
\and A.~Da~Silva\orcid{0000-0002-6385-1609}\inst{\ref{aff55},\ref{aff56}}
\and H.~Degaudenzi\orcid{0000-0002-5887-6799}\inst{\ref{aff57}}
\and S.~de~la~Torre\inst{\ref{aff58}}
\and G.~De~Lucia\orcid{0000-0002-6220-9104}\inst{\ref{aff7}}
\and H.~Dole\orcid{0000-0002-9767-3839}\inst{\ref{aff59}}
\and F.~Dubath\orcid{0000-0002-6533-2810}\inst{\ref{aff57}}
\and X.~Dupac\inst{\ref{aff29}}
\and S.~Escoffier\orcid{0000-0002-2847-7498}\inst{\ref{aff60}}
\and M.~Farina\orcid{0000-0002-3089-7846}\inst{\ref{aff61}}
\and R.~Farinelli\inst{\ref{aff26}}
\and F.~Faustini\orcid{0000-0001-6274-5145}\inst{\ref{aff39},\ref{aff62}}
\and S.~Ferriol\inst{\ref{aff50}}
\and F.~Finelli\orcid{0000-0002-6694-3269}\inst{\ref{aff26},\ref{aff63}}
\and P.~Fosalba\orcid{0000-0002-1510-5214}\inst{\ref{aff19},\ref{aff18}}
\and S.~Fotopoulou\orcid{0000-0002-9686-254X}\inst{\ref{aff64}}
\and N.~Fourmanoit\orcid{0009-0005-6816-6925}\inst{\ref{aff60}}
\and M.~Frailis\orcid{0000-0002-7400-2135}\inst{\ref{aff7}}
\and E.~Franceschi\orcid{0000-0002-0585-6591}\inst{\ref{aff26}}
\and M.~Fumana\orcid{0000-0001-6787-5950}\inst{\ref{aff38}}
\and S.~Galeotta\orcid{0000-0002-3748-5115}\inst{\ref{aff7}}
\and K.~George\orcid{0000-0002-1734-8455}\inst{\ref{aff65}}
\and B.~Gillis\orcid{0000-0002-4478-1270}\inst{\ref{aff36}}
\and C.~Giocoli\orcid{0000-0002-9590-7961}\inst{\ref{aff26},\ref{aff30}}
\and J.~Gracia-Carpio\inst{\ref{aff66}}
\and A.~Grazian\orcid{0000-0002-5688-0663}\inst{\ref{aff67}}
\and F.~Grupp\inst{\ref{aff66},\ref{aff68}}
\and S.~V.~H.~Haugan\orcid{0000-0001-9648-7260}\inst{\ref{aff69}}
\and W.~Holmes\inst{\ref{aff5}}
\and F.~Hormuth\inst{\ref{aff70}}
\and A.~Hornstrup\orcid{0000-0002-3363-0936}\inst{\ref{aff71},\ref{aff72}}
\and K.~Jahnke\orcid{0000-0003-3804-2137}\inst{\ref{aff73}}
\and M.~Jhabvala\inst{\ref{aff74}}
\and B.~Joachimi\orcid{0000-0001-7494-1303}\inst{\ref{aff75}}
\and S.~Kermiche\orcid{0000-0002-0302-5735}\inst{\ref{aff60}}
\and A.~Kiessling\orcid{0000-0002-2590-1273}\inst{\ref{aff5}}
\and B.~Kubik\orcid{0009-0006-5823-4880}\inst{\ref{aff50}}
\and M.~Kunz\orcid{0000-0002-3052-7394}\inst{\ref{aff11}}
\and H.~Kurki-Suonio\orcid{0000-0002-4618-3063}\inst{\ref{aff76},\ref{aff77}}
\and A.~M.~C.~Le~Brun\orcid{0000-0002-0936-4594}\inst{\ref{aff78}}
\and S.~Ligori\orcid{0000-0003-4172-4606}\inst{\ref{aff35}}
\and P.~B.~Lilje\orcid{0000-0003-4324-7794}\inst{\ref{aff69}}
\and V.~Lindholm\orcid{0000-0003-2317-5471}\inst{\ref{aff76},\ref{aff77}}
\and I.~Lloro\orcid{0000-0001-5966-1434}\inst{\ref{aff79}}
\and G.~Mainetti\orcid{0000-0003-2384-2377}\inst{\ref{aff80}}
\and D.~Maino\inst{\ref{aff12},\ref{aff38},\ref{aff22}}
\and E.~Maiorano\orcid{0000-0003-2593-4355}\inst{\ref{aff26}}
\and O.~Mansutti\orcid{0000-0001-5758-4658}\inst{\ref{aff7}}
\and S.~Marcin\inst{\ref{aff81}}
\and O.~Marggraf\orcid{0000-0001-7242-3852}\inst{\ref{aff6}}
\and K.~Markovic\orcid{0000-0001-6764-073X}\inst{\ref{aff5}}
\and M.~Martinelli\orcid{0000-0002-6943-7732}\inst{\ref{aff39},\ref{aff40}}
\and N.~Martinet\orcid{0000-0003-2786-7790}\inst{\ref{aff58}}
\and F.~Marulli\orcid{0000-0002-8850-0303}\inst{\ref{aff82},\ref{aff26},\ref{aff30}}
\and R.~J.~Massey\orcid{0000-0002-6085-3780}\inst{\ref{aff83}}
\and E.~Medinaceli\orcid{0000-0002-4040-7783}\inst{\ref{aff26}}
\and S.~Mei\orcid{0000-0002-2849-559X}\inst{\ref{aff84},\ref{aff85}}
\and M.~Melchior\inst{\ref{aff86}}
\and Y.~Mellier\thanks{Deceased}\inst{\ref{aff87},\ref{aff88}}
\and M.~Meneghetti\orcid{0000-0003-1225-7084}\inst{\ref{aff26},\ref{aff30}}
\and E.~Merlin\orcid{0000-0001-6870-8900}\inst{\ref{aff39}}
\and G.~Meylan\inst{\ref{aff89}}
\and A.~Mora\orcid{0000-0002-1922-8529}\inst{\ref{aff90}}
\and M.~Moresco\orcid{0000-0002-7616-7136}\inst{\ref{aff82},\ref{aff26}}
\and L.~Moscardini\orcid{0000-0002-3473-6716}\inst{\ref{aff82},\ref{aff26},\ref{aff30}}
\and C.~Neissner\orcid{0000-0001-8524-4968}\inst{\ref{aff91},\ref{aff42}}
\and S.-M.~Niemi\orcid{0009-0005-0247-0086}\inst{\ref{aff92}}
\and J.~W.~Nightingale\orcid{0000-0002-8987-7401}\inst{\ref{aff20}}
\and C.~Padilla\orcid{0000-0001-7951-0166}\inst{\ref{aff91}}
\and S.~Paltani\orcid{0000-0002-8108-9179}\inst{\ref{aff57}}
\and F.~Pasian\orcid{0000-0002-4869-3227}\inst{\ref{aff7}}
\and K.~Pedersen\inst{\ref{aff93}}
\and W.~J.~Percival\orcid{0000-0002-0644-5727}\inst{\ref{aff94},\ref{aff95},\ref{aff96}}
\and V.~Pettorino\orcid{0000-0002-4203-9320}\inst{\ref{aff92}}
\and S.~Pires\orcid{0000-0002-0249-2104}\inst{\ref{aff97}}
\and G.~Polenta\orcid{0000-0003-4067-9196}\inst{\ref{aff62}}
\and M.~Poncet\inst{\ref{aff98}}
\and L.~A.~Popa\inst{\ref{aff99}}
\and F.~Raison\orcid{0000-0002-7819-6918}\inst{\ref{aff66}}
\and A.~Renzi\orcid{0000-0001-9856-1970}\inst{\ref{aff13},\ref{aff14},\ref{aff26}}
\and J.~Rhodes\orcid{0000-0002-4485-8549}\inst{\ref{aff5}}
\and G.~Riccio\inst{\ref{aff32}}
\and E.~Romelli\orcid{0000-0003-3069-9222}\inst{\ref{aff7}}
\and M.~Roncarelli\orcid{0000-0001-9587-7822}\inst{\ref{aff26}}
\and R.~Saglia\orcid{0000-0003-0378-7032}\inst{\ref{aff68},\ref{aff66}}
\and Z.~Sakr\orcid{0000-0002-4823-3757}\inst{\ref{aff100},\ref{aff101},\ref{aff102}}
\and A.~G.~S\'anchez\orcid{0000-0003-1198-831X}\inst{\ref{aff66}}
\and D.~Sapone\orcid{0000-0001-7089-4503}\inst{\ref{aff103}}
\and B.~Sartoris\orcid{0000-0003-1337-5269}\inst{\ref{aff68},\ref{aff7}}
\and P.~Schneider\orcid{0000-0001-8561-2679}\inst{\ref{aff6}}
\and A.~Secroun\orcid{0000-0003-0505-3710}\inst{\ref{aff60}}
\and G.~Seidel\orcid{0000-0003-2907-353X}\inst{\ref{aff73}}
\and S.~Serrano\orcid{0000-0002-0211-2861}\inst{\ref{aff19},\ref{aff104},\ref{aff18}}
\and E.~Sihvola\orcid{0000-0003-1804-7715}\inst{\ref{aff105}}
\and P.~Simon\inst{\ref{aff6}}
\and C.~Sirignano\orcid{0000-0002-0995-7146}\inst{\ref{aff13},\ref{aff14}}
\and G.~Sirri\orcid{0000-0003-2626-2853}\inst{\ref{aff30}}
\and A.~Spurio~Mancini\orcid{0000-0001-5698-0990}\inst{\ref{aff106}}
\and L.~Stanco\orcid{0000-0002-9706-5104}\inst{\ref{aff14}}
\and J.~Steinwagner\orcid{0000-0001-7443-1047}\inst{\ref{aff66}}
\and P.~Tallada-Cresp\'{i}\orcid{0000-0002-1336-8328}\inst{\ref{aff41},\ref{aff42}}
\and I.~Tereno\orcid{0000-0002-4537-6218}\inst{\ref{aff55},\ref{aff107}}
\and N.~Tessore\orcid{0000-0002-9696-7931}\inst{\ref{aff108}}
\and S.~Toft\orcid{0000-0003-3631-7176}\inst{\ref{aff109},\ref{aff110}}
\and R.~Toledo-Moreo\orcid{0000-0002-2997-4859}\inst{\ref{aff111}}
\and F.~Torradeflot\orcid{0000-0003-1160-1517}\inst{\ref{aff42},\ref{aff41}}
\and I.~Tutusaus\orcid{0000-0002-3199-0399}\inst{\ref{aff18},\ref{aff19},\ref{aff101}}
\and J.~Valiviita\orcid{0000-0001-6225-3693}\inst{\ref{aff76},\ref{aff77}}
\and T.~Vassallo\orcid{0000-0001-6512-6358}\inst{\ref{aff7}}
\and G.~Verdoes~Kleijn\orcid{0000-0001-5803-2580}\inst{\ref{aff112}}
\and Y.~Wang\orcid{0000-0002-4749-2984}\inst{\ref{aff113}}
\and J.~Weller\orcid{0000-0002-8282-2010}\inst{\ref{aff68},\ref{aff66}}
\and G.~Zamorani\orcid{0000-0002-2318-301X}\inst{\ref{aff26}}
\and F.~M.~Zerbi\inst{\ref{aff15}}
\and E.~Zucca\orcid{0000-0002-5845-8132}\inst{\ref{aff26}}
\and V.~Allevato\orcid{0000-0001-7232-5152}\inst{\ref{aff32}}
\and M.~Ballardini\orcid{0000-0003-4481-3559}\inst{\ref{aff114},\ref{aff115},\ref{aff26}}
\and A.~Boucaud\orcid{0000-0001-7387-2633}\inst{\ref{aff84}}
\and E.~Bozzo\orcid{0000-0002-8201-1525}\inst{\ref{aff57}}
\and C.~Burigana\orcid{0000-0002-3005-5796}\inst{\ref{aff116},\ref{aff63}}
\and R.~Cabanac\orcid{0000-0001-6679-2600}\inst{\ref{aff101}}
\and M.~Calabrese\orcid{0000-0002-2637-2422}\inst{\ref{aff117},\ref{aff38}}
\and A.~Cappi\inst{\ref{aff118},\ref{aff26}}
\and T.~Castro\orcid{0000-0002-6292-3228}\inst{\ref{aff7},\ref{aff8},\ref{aff4},\ref{aff119}}
\and J.~A.~Escartin~Vigo\inst{\ref{aff66}}
\and L.~Gabarra\orcid{0000-0002-8486-8856}\inst{\ref{aff120}}
\and J.~Garc\'ia-Bellido\orcid{0000-0002-9370-8360}\inst{\ref{aff121}}
\and V.~Gautard\inst{\ref{aff122}}
\and S.~Hemmati\orcid{0000-0003-2226-5395}\inst{\ref{aff113}}
\and J.~Macias-Perez\orcid{0000-0002-5385-2763}\inst{\ref{aff123}}
\and R.~Maoli\orcid{0000-0002-6065-3025}\inst{\ref{aff124},\ref{aff39}}
\and J.~Mart\'{i}n-Fleitas\orcid{0000-0002-8594-569X}\inst{\ref{aff125}}
\and N.~Mauri\orcid{0000-0001-8196-1548}\inst{\ref{aff46},\ref{aff30}}
\and R.~B.~Metcalf\orcid{0000-0003-3167-2574}\inst{\ref{aff82},\ref{aff26}}
\and P.~Monaco\orcid{0000-0003-2083-7564}\inst{\ref{aff28},\ref{aff7},\ref{aff8},\ref{aff4}}
\and M.~P\"ontinen\orcid{0000-0001-5442-2530}\inst{\ref{aff76}}
\and I.~Risso\orcid{0000-0003-2525-7761}\inst{\ref{aff15},\ref{aff16}}
\and V.~Scottez\orcid{0009-0008-3864-940X}\inst{\ref{aff87},\ref{aff126}}
\and M.~Sereno\orcid{0000-0003-0302-0325}\inst{\ref{aff26},\ref{aff30}}
\and M.~Tenti\orcid{0000-0002-4254-5901}\inst{\ref{aff30}}
\and M.~Tucci\inst{\ref{aff57}}
\and M.~Viel\orcid{0000-0002-2642-5707}\inst{\ref{aff4},\ref{aff7},\ref{aff2},\ref{aff8},\ref{aff119}}
\and M.~Wiesmann\orcid{0009-0000-8199-5860}\inst{\ref{aff69}}
\and Y.~Akrami\orcid{0000-0002-2407-7956}\inst{\ref{aff121},\ref{aff127}}
\and I.~T.~Andika\orcid{0000-0001-6102-9526}\inst{\ref{aff65}}
\and G.~Angora\orcid{0000-0002-0316-6562}\inst{\ref{aff32},\ref{aff114}}
\and S.~Anselmi\orcid{0000-0002-3579-9583}\inst{\ref{aff14},\ref{aff13},\ref{aff128}}
\and M.~Archidiacono\orcid{0000-0003-4952-9012}\inst{\ref{aff12},\ref{aff22}}
\and F.~Atrio-Barandela\orcid{0000-0002-2130-2513}\inst{\ref{aff129}}
\and E.~Aubourg\orcid{0000-0002-5592-023X}\inst{\ref{aff84},\ref{aff130}}
\and L.~Bazzanini\orcid{0000-0003-0727-0137}\inst{\ref{aff114},\ref{aff26}}
\and J.~Bel\inst{\ref{aff131}}
\and D.~Bertacca\orcid{0000-0002-2490-7139}\inst{\ref{aff13},\ref{aff67},\ref{aff14}}
\and M.~Bethermin\orcid{0000-0002-3915-2015}\inst{\ref{aff132}}
\and F.~Beutler\orcid{0000-0003-0467-5438}\inst{\ref{aff36}}
\and A.~Blanchard\orcid{0000-0001-8555-9003}\inst{\ref{aff101}}
\and L.~Blot\orcid{0000-0002-9622-7167}\inst{\ref{aff133},\ref{aff78}}
\and M.~Bonici\orcid{0000-0002-8430-126X}\inst{\ref{aff94},\ref{aff38}}
\and S.~Borgani\orcid{0000-0001-6151-6439}\inst{\ref{aff28},\ref{aff4},\ref{aff7},\ref{aff8},\ref{aff119}}
\and M.~L.~Brown\orcid{0000-0002-0370-8077}\inst{\ref{aff48}}
\and S.~Bruton\orcid{0000-0002-6503-5218}\inst{\ref{aff27}}
\and A.~Calabro\orcid{0000-0003-2536-1614}\inst{\ref{aff39}}
\and F.~Caro\inst{\ref{aff39}}
\and C.~S.~Carvalho\inst{\ref{aff107}}
\and F.~Cogato\orcid{0000-0003-4632-6113}\inst{\ref{aff82},\ref{aff26}}
\and S.~Conseil\orcid{0000-0002-3657-4191}\inst{\ref{aff50}}
\and A.~R.~Cooray\orcid{0000-0002-3892-0190}\inst{\ref{aff134}}
\and S.~Davini\orcid{0000-0003-3269-1718}\inst{\ref{aff16}}
\and G.~Desprez\orcid{0000-0001-8325-1742}\inst{\ref{aff112}}
\and A.~D\'iaz-S\'anchez\orcid{0000-0003-0748-4768}\inst{\ref{aff135}}
\and S.~Di~Domizio\orcid{0000-0003-2863-5895}\inst{\ref{aff17},\ref{aff16}}
\and J.~M.~Diego\orcid{0000-0001-9065-3926}\inst{\ref{aff136}}
\and V.~Duret\orcid{0009-0009-0383-4960}\inst{\ref{aff60}}
\and M.~Y.~Elkhashab\orcid{0000-0001-9306-2603}\inst{\ref{aff7},\ref{aff8},\ref{aff28},\ref{aff4}}
\and A.~Enia\orcid{0000-0002-0200-2857}\inst{\ref{aff26}}
\and Y.~Fang\orcid{0000-0002-0334-6950}\inst{\ref{aff68}}
\and A.~Finoguenov\orcid{0000-0002-4606-5403}\inst{\ref{aff76}}
\and A.~Fontana\orcid{0000-0003-3820-2823}\inst{\ref{aff39}}
\and F.~Fontanot\orcid{0000-0003-4744-0188}\inst{\ref{aff7},\ref{aff4}}
\and A.~Franco\orcid{0000-0002-4761-366X}\inst{\ref{aff137},\ref{aff138},\ref{aff139}}
\and K.~Ganga\orcid{0000-0001-8159-8208}\inst{\ref{aff84}}
\and T.~Gasparetto\orcid{0000-0002-7913-4866}\inst{\ref{aff39}}
\and E.~Gaztanaga\orcid{0000-0001-9632-0815}\inst{\ref{aff18},\ref{aff19},\ref{aff140}}
\and F.~Giacomini\orcid{0000-0002-3129-2814}\inst{\ref{aff30}}
\and F.~Gianotti\orcid{0000-0003-4666-119X}\inst{\ref{aff26}}
\and G.~Gozaliasl\orcid{0000-0002-0236-919X}\inst{\ref{aff141},\ref{aff76}}
\and A.~Gruppuso\orcid{0000-0001-9272-5292}\inst{\ref{aff26},\ref{aff30}}
\and C.~M.~Gutierrez\orcid{0000-0001-7854-783X}\inst{\ref{aff47},\ref{aff142}}
\and A.~Hall\orcid{0000-0002-3139-8651}\inst{\ref{aff36}}
\and C.~Hern\'andez-Monteagudo\orcid{0000-0001-5471-9166}\inst{\ref{aff142},\ref{aff47}}
\and H.~Hildebrandt\orcid{0000-0002-9814-3338}\inst{\ref{aff143}}
\and J.~Hjorth\orcid{0000-0002-4571-2306}\inst{\ref{aff93}}
\and J.~J.~E.~Kajava\orcid{0000-0002-3010-8333}\inst{\ref{aff144},\ref{aff145},\ref{aff146}}
\and Y.~Kang\orcid{0009-0000-8588-7250}\inst{\ref{aff57}}
\and V.~Kansal\orcid{0000-0002-4008-6078}\inst{\ref{aff147},\ref{aff148}}
\and D.~Karagiannis\orcid{0000-0002-4927-0816}\inst{\ref{aff114},\ref{aff149}}
\and K.~Kiiveri\inst{\ref{aff105}}
\and J.~Kim\orcid{0000-0003-2776-2761}\inst{\ref{aff120}}
\and C.~C.~Kirkpatrick\inst{\ref{aff105}}
\and S.~Kruk\orcid{0000-0001-8010-8879}\inst{\ref{aff29}}
\and M.~Lattanzi\orcid{0000-0003-1059-2532}\inst{\ref{aff115}}
\and L.~Legrand\orcid{0000-0003-0610-5252}\inst{\ref{aff150},\ref{aff151}}
\and M.~Lembo\orcid{0000-0002-5271-5070}\inst{\ref{aff88}}
\and F.~Lepori\orcid{0009-0000-5061-7138}\inst{\ref{aff152}}
\and G.~Leroy\orcid{0009-0004-2523-4425}\inst{\ref{aff153},\ref{aff83}}
\and G.~F.~Lesci\orcid{0000-0002-4607-2830}\inst{\ref{aff82},\ref{aff26}}
\and T.~I.~Liaudat\orcid{0000-0002-9104-314X}\inst{\ref{aff130}}
\and S.~J.~Liu\orcid{0000-0001-7680-2139}\inst{\ref{aff61}}
\and M.~Magliocchetti\orcid{0000-0001-9158-4838}\inst{\ref{aff61}}
\and C.~J.~A.~P.~Martins\orcid{0000-0002-4886-9261}\inst{\ref{aff154},\ref{aff155}}
\and L.~Maurin\orcid{0000-0002-8406-0857}\inst{\ref{aff59}}
\and M.~Miluzio\inst{\ref{aff29},\ref{aff156}}
\and G.~Morgante\inst{\ref{aff26}}
\and S.~Nadathur\orcid{0000-0001-9070-3102}\inst{\ref{aff140}}
\and K.~Naidoo\orcid{0000-0002-9182-1802}\inst{\ref{aff140},\ref{aff73}}
\and P.~Natoli\orcid{0000-0003-0126-9100}\inst{\ref{aff114},\ref{aff115}}
\and A.~Navarro-Alsina\orcid{0000-0002-3173-2592}\inst{\ref{aff6}}
\and S.~Nesseris\orcid{0000-0002-0567-0324}\inst{\ref{aff121}}
\and L.~Pagano\orcid{0000-0003-1820-5998}\inst{\ref{aff114},\ref{aff115}}
\and D.~Paoletti\orcid{0000-0003-4761-6147}\inst{\ref{aff26},\ref{aff63}}
\and F.~Passalacqua\orcid{0000-0002-8606-4093}\inst{\ref{aff13},\ref{aff14}}
\and K.~Paterson\orcid{0000-0001-8340-3486}\inst{\ref{aff73}}
\and L.~Patrizii\inst{\ref{aff30}}
\and R.~Paviot\orcid{0009-0002-8108-3460}\inst{\ref{aff97}}
\and A.~Pisani\orcid{0000-0002-6146-4437}\inst{\ref{aff60}}
\and D.~Potter\orcid{0000-0002-0757-5195}\inst{\ref{aff152}}
\and G.~W.~Pratt\inst{\ref{aff97}}
\and S.~Quai\orcid{0000-0002-0449-8163}\inst{\ref{aff82},\ref{aff26}}
\and M.~Radovich\orcid{0000-0002-3585-866X}\inst{\ref{aff67}}
\and K.~Rojas\orcid{0000-0003-1391-6854}\inst{\ref{aff81}}
\and W.~Roster\orcid{0000-0002-9149-6528}\inst{\ref{aff66}}
\and S.~Sacquegna\orcid{0000-0002-8433-6630}\inst{\ref{aff157}}
\and M.~Sahl\'en\orcid{0000-0003-0973-4804}\inst{\ref{aff158}}
\and D.~B.~Sanders\orcid{0000-0002-1233-9998}\inst{\ref{aff45}}
\and E.~Sarpa\orcid{0000-0002-1256-655X}\inst{\ref{aff2},\ref{aff119},\ref{aff7}}
\and A.~Schneider\orcid{0000-0001-7055-8104}\inst{\ref{aff152}}
\and D.~Sciotti\orcid{0009-0008-4519-2620}\inst{\ref{aff39},\ref{aff40}}
\and E.~Sellentin\inst{\ref{aff159},\ref{aff37}}
\and L.~C.~Smith\orcid{0000-0002-3259-2771}\inst{\ref{aff160}}
\and J.~G.~Sorce\orcid{0000-0002-2307-2432}\inst{\ref{aff161},\ref{aff59}}
\and K.~Tanidis\orcid{0000-0001-9843-5130}\inst{\ref{aff120}}
\and C.~Tao\orcid{0000-0001-7961-8177}\inst{\ref{aff60}}
\and F.~Tarsitano\orcid{0000-0002-5919-0238}\inst{\ref{aff162},\ref{aff57}}
\and G.~Testera\inst{\ref{aff16}}
\and R.~Teyssier\orcid{0000-0001-7689-0933}\inst{\ref{aff163}}
\and S.~Tosi\orcid{0000-0002-7275-9193}\inst{\ref{aff17},\ref{aff16},\ref{aff15}}
\and A.~Troja\orcid{0000-0003-0239-4595}\inst{\ref{aff13},\ref{aff14}}
\and D.~Vergani\orcid{0000-0003-0898-2216}\inst{\ref{aff26}}
\and F.~Vernizzi\orcid{0000-0003-3426-2802}\inst{\ref{aff164}}
\and G.~Verza\orcid{0000-0002-1886-8348}\inst{\ref{aff3},\ref{aff165}}
\and P.~Vielzeuf\orcid{0000-0003-2035-9339}\inst{\ref{aff60}}
\and S.~Vinciguerra\orcid{0009-0005-4018-3184}\inst{\ref{aff58}}
\and N.~A.~Walton\orcid{0000-0003-3983-8778}\inst{\ref{aff160}}
\and A.~H.~Wright\orcid{0000-0001-7363-7932}\inst{\ref{aff143}}}
										   
\institute{INFN Gruppo Collegato di Parma, Viale delle Scienze 7/A 43124 Parma, Italy\label{aff1}
\and
SISSA, International School for Advanced Studies, Via Bonomea 265, 34136 Trieste TS, Italy\label{aff2}
\and
International Centre for Theoretical Physics (ICTP), Strada Costiera 11, 34151 Trieste, Italy\label{aff3}
\and
IFPU, Institute for Fundamental Physics of the Universe, via Beirut 2, 34151 Trieste, Italy\label{aff4}
\and
Jet Propulsion Laboratory, California Institute of Technology, 4800 Oak Grove Drive, Pasadena, CA, 91109, USA\label{aff5}
\and
Universit\"at Bonn, Argelander-Institut f\"ur Astronomie, Auf dem H\"ugel 71, 53121 Bonn, Germany\label{aff6}
\and
INAF-Osservatorio Astronomico di Trieste, Via G. B. Tiepolo 11, 34143 Trieste, Italy\label{aff7}
\and
INFN, Sezione di Trieste, Via Valerio 2, 34127 Trieste TS, Italy\label{aff8}
\and
Laboratoire d'Annecy-le-Vieux de Physique Theorique, CNRS \& Universite Savoie Mont Blanc, 9 Chemin de Bellevue, BP 110, Annecy-le-Vieux, 74941 ANNECY Cedex, France\label{aff9}
\and
Institute for Theoretical Particle Physics and Cosmology (TTK), RWTH Aachen University, 52056 Aachen, Germany\label{aff10}
\and
Universit\'e de Gen\`eve, D\'epartement de Physique Th\'eorique and Centre for Astroparticle Physics, 24 quai Ernest-Ansermet, CH-1211 Gen\`eve 4, Switzerland\label{aff11}
\and
Dipartimento di Fisica "Aldo Pontremoli", Universit\`a degli Studi di Milano, Via Celoria 16, 20133 Milano, Italy\label{aff12}
\and
Dipartimento di Fisica e Astronomia "G. Galilei", Universit\`a di Padova, Via Marzolo 8, 35131 Padova, Italy\label{aff13}
\and
INFN-Padova, Via Marzolo 8, 35131 Padova, Italy\label{aff14}
\and
INAF-Osservatorio Astronomico di Brera, Via Brera 28, 20122 Milano, Italy\label{aff15}
\and
INFN-Sezione di Genova, Via Dodecaneso 33, 16146, Genova, Italy\label{aff16}
\and
Dipartimento di Fisica, Universit\`a di Genova, Via Dodecaneso 33, 16146, Genova, Italy\label{aff17}
\and
Institute of Space Sciences (ICE, CSIC), Campus UAB, Carrer de Can Magrans, s/n, 08193 Barcelona, Spain\label{aff18}
\and
Institut d'Estudis Espacials de Catalunya (IEEC),  Edifici RDIT, Campus UPC, 08860 Castelldefels, Barcelona, Spain\label{aff19}
\and
School of Mathematics, Statistics and Physics, Newcastle University, Herschel Building, Newcastle-upon-Tyne, NE1 7RU, UK\label{aff20}
\and
Fakult\"at f\"ur Physik, Universit\"at Bielefeld, Postfach 100131, 33501 Bielefeld, Germany\label{aff21}
\and
INFN-Sezione di Milano, Via Celoria 16, 20133 Milano, Italy\label{aff22}
\and
Dipartimento di Scienze Matematiche, Fisiche e Informatiche, Universit\`a di Parma, Viale delle Scienze 7/A 43124 Parma, Italy\label{aff23}
\and
Technion Israel Institute of Technology, Israel\label{aff24}
\and
Dipartimento di Fisica e Astronomia, Universit\`a di Bologna, Via Gobetti 93/2, 40129 Bologna, Italy\label{aff25}
\and
INAF-Osservatorio di Astrofisica e Scienza dello Spazio di Bologna, Via Piero Gobetti 93/3, 40129 Bologna, Italy\label{aff26}
\and
California Institute of Technology, 1200 E California Blvd, Pasadena, CA 91125, USA\label{aff27}
\and
Dipartimento di Fisica - Sezione di Astronomia, Universit\`a di Trieste, Via Tiepolo 11, 34131 Trieste, Italy\label{aff28}
\and
ESAC/ESA, Camino Bajo del Castillo, s/n., Urb. Villafranca del Castillo, 28692 Villanueva de la Ca\~nada, Madrid, Spain\label{aff29}
\and
INFN-Sezione di Bologna, Viale Berti Pichat 6/2, 40127 Bologna, Italy\label{aff30}
\and
Department of Physics "E. Pancini", University Federico II, Via Cinthia 6, 80126, Napoli, Italy\label{aff31}
\and
INAF-Osservatorio Astronomico di Capodimonte, Via Moiariello 16, 80131 Napoli, Italy\label{aff32}
\and
Dipartimento di Fisica, Universit\`a degli Studi di Torino, Via P. Giuria 1, 10125 Torino, Italy\label{aff33}
\and
INFN-Sezione di Torino, Via P. Giuria 1, 10125 Torino, Italy\label{aff34}
\and
INAF-Osservatorio Astrofisico di Torino, Via Osservatorio 20, 10025 Pino Torinese (TO), Italy\label{aff35}
\and
Institute for Astronomy, University of Edinburgh, Royal Observatory, Blackford Hill, Edinburgh EH9 3HJ, UK\label{aff36}
\and
Leiden Observatory, Leiden University, Einsteinweg 55, 2333 CC Leiden, The Netherlands\label{aff37}
\and
INAF-IASF Milano, Via Alfonso Corti 12, 20133 Milano, Italy\label{aff38}
\and
INAF-Osservatorio Astronomico di Roma, Via Frascati 33, 00078 Monteporzio Catone, Italy\label{aff39}
\and
INFN-Sezione di Roma, Piazzale Aldo Moro, 2 - c/o Dipartimento di Fisica, Edificio G. Marconi, 00185 Roma, Italy\label{aff40}
\and
Centro de Investigaciones Energ\'eticas, Medioambientales y Tecnol\'ogicas (CIEMAT), Avenida Complutense 40, 28040 Madrid, Spain\label{aff41}
\and
Port d'Informaci\'{o} Cient\'{i}fica, Campus UAB, C. Albareda s/n, 08193 Bellaterra (Barcelona), Spain\label{aff42}
\and
Deutsches Zentrum f\"ur Luft- und Raumfahrt e. V. (DLR), Linder H\"ohe, 51147 K\"oln, Germany\label{aff43}
\and
INFN section of Naples, Via Cinthia 6, 80126, Napoli, Italy\label{aff44}
\and
Institute for Astronomy, University of Hawaii, 2680 Woodlawn Drive, Honolulu, HI 96822, USA\label{aff45}
\and
Dipartimento di Fisica e Astronomia "Augusto Righi" - Alma Mater Studiorum Universit\`a di Bologna, Viale Berti Pichat 6/2, 40127 Bologna, Italy\label{aff46}
\and
Instituto de Astrof\'{\i}sica de Canarias, E-38205 La Laguna, Tenerife, Spain\label{aff47}
\and
Jodrell Bank Centre for Astrophysics, Department of Physics and Astronomy, University of Manchester, Oxford Road, Manchester M13 9PL, UK\label{aff48}
\and
European Space Agency/ESRIN, Largo Galileo Galilei 1, 00044 Frascati, Roma, Italy\label{aff49}
\and
Universit\'e Claude Bernard Lyon 1, CNRS/IN2P3, IP2I Lyon, UMR 5822, Villeurbanne, F-69100, France\label{aff50}
\and
Institut de Ci\`{e}ncies del Cosmos (ICCUB), Universitat de Barcelona (IEEC-UB), Mart\'{i} i Franqu\`{e}s 1, 08028 Barcelona, Spain\label{aff51}
\and
Instituci\'o Catalana de Recerca i Estudis Avan\c{c}ats (ICREA), Passeig de Llu\'{\i}s Companys 23, 08010 Barcelona, Spain\label{aff52}
\and
Institut de Ciencies de l'Espai (IEEC-CSIC), Campus UAB, Carrer de Can Magrans, s/n Cerdanyola del Vall\'es, 08193 Barcelona, Spain\label{aff53}
\and
UCB Lyon 1, CNRS/IN2P3, IUF, IP2I Lyon, 4 rue Enrico Fermi, 69622 Villeurbanne, France\label{aff54}
\and
Departamento de F\'isica, Faculdade de Ci\^encias, Universidade de Lisboa, Edif\'icio C8, Campo Grande, PT1749-016 Lisboa, Portugal\label{aff55}
\and
Instituto de Astrof\'isica e Ci\^encias do Espa\c{c}o, Faculdade de Ci\^encias, Universidade de Lisboa, Campo Grande, 1749-016 Lisboa, Portugal\label{aff56}
\and
Department of Astronomy, University of Geneva, ch. d'Ecogia 16, 1290 Versoix, Switzerland\label{aff57}
\and
Aix-Marseille Universit\'e, CNRS, CNES, LAM, Marseille, France\label{aff58}
\and
Universit\'e Paris-Saclay, CNRS, Institut d'astrophysique spatiale, 91405, Orsay, France\label{aff59}
\and
Aix-Marseille Universit\'e, CNRS/IN2P3, CPPM, Marseille, France\label{aff60}
\and
INAF-Istituto di Astrofisica e Planetologia Spaziali, via del Fosso del Cavaliere, 100, 00100 Roma, Italy\label{aff61}
\and
Space Science Data Center, Italian Space Agency, via del Politecnico snc, 00133 Roma, Italy\label{aff62}
\and
INFN-Bologna, Via Irnerio 46, 40126 Bologna, Italy\label{aff63}
\and
School of Physics, HH Wills Physics Laboratory, University of Bristol, Tyndall Avenue, Bristol, BS8 1TL, UK\label{aff64}
\and
University Observatory, LMU Faculty of Physics, Scheinerstr.~1, 81679 Munich, Germany\label{aff65}
\and
Max Planck Institute for Extraterrestrial Physics, Giessenbachstr. 1, 85748 Garching, Germany\label{aff66}
\and
INAF-Osservatorio Astronomico di Padova, Via dell'Osservatorio 5, 35122 Padova, Italy\label{aff67}
\and
Universit\"ats-Sternwarte M\"unchen, Fakult\"at f\"ur Physik, Ludwig-Maximilians-Universit\"at M\"unchen, Scheinerstr.~1, 81679 M\"unchen, Germany\label{aff68}
\and
Institute of Theoretical Astrophysics, University of Oslo, P.O. Box 1029 Blindern, 0315 Oslo, Norway\label{aff69}
\and
Felix Hormuth Engineering, Goethestr. 17, 69181 Leimen, Germany\label{aff70}
\and
Technical University of Denmark, Elektrovej 327, 2800 Kgs. Lyngby, Denmark\label{aff71}
\and
Cosmic Dawn Center (DAWN), Denmark\label{aff72}
\and
Max-Planck-Institut f\"ur Astronomie, K\"onigstuhl 17, 69117 Heidelberg, Germany\label{aff73}
\and
NASA Goddard Space Flight Center, Greenbelt, MD 20771, USA\label{aff74}
\and
Department of Physics and Astronomy, University College London, Gower Street, London WC1E 6BT, UK\label{aff75}
\and
Department of Physics, P.O. Box 64, University of Helsinki, 00014 Helsinki, Finland\label{aff76}
\and
Helsinki Institute of Physics, Gustaf H{\"a}llstr{\"o}min katu 2, University of Helsinki, 00014 Helsinki, Finland\label{aff77}
\and
Laboratoire d'etude de l'Univers et des phenomenes eXtremes, Observatoire de Paris, Universit\'e PSL, Sorbonne Universit\'e, CNRS, 92190 Meudon, France\label{aff78}
\and
SKAO, Jodrell Bank, Lower Withington, Macclesfield SK11 9FT, UK\label{aff79}
\and
Centre de Calcul de l'IN2P3/CNRS, 21 avenue Pierre de Coubertin 69627 Villeurbanne Cedex, France\label{aff80}
\and
University of Applied Sciences and Arts of Northwestern Switzerland, School of Computer Science, 5210 Windisch, Switzerland\label{aff81}
\and
Dipartimento di Fisica e Astronomia "Augusto Righi" - Alma Mater Studiorum Universit\`a di Bologna, via Piero Gobetti 93/2, 40129 Bologna, Italy\label{aff82}
\and
Department of Physics, Institute for Computational Cosmology, Durham University, South Road, Durham, DH1 3LE, UK\label{aff83}
\and
Universit\'e Paris Cit\'e, CNRS, Astroparticule et Cosmologie, 75013 Paris, France\label{aff84}
\and
CNRS-UCB International Research Laboratory, Centre Pierre Bin\'etruy, IRL2007, CPB-IN2P3, Berkeley, USA\label{aff85}
\and
University of Applied Sciences and Arts of Northwestern Switzerland, School of Engineering, 5210 Windisch, Switzerland\label{aff86}
\and
Institut d'Astrophysique de Paris, 98bis Boulevard Arago, 75014, Paris, France\label{aff87}
\and
Institut d'Astrophysique de Paris, UMR 7095, CNRS, and Sorbonne Universit\'e, 98 bis boulevard Arago, 75014 Paris, France\label{aff88}
\and
Institute of Physics, Laboratory of Astrophysics, Ecole Polytechnique F\'ed\'erale de Lausanne (EPFL), Observatoire de Sauverny, 1290 Versoix, Switzerland\label{aff89}
\and
Telespazio UK S.L. for European Space Agency (ESA), Camino bajo del Castillo, s/n, Urbanizacion Villafranca del Castillo, Villanueva de la Ca\~nada, 28692 Madrid, Spain\label{aff90}
\and
Institut de F\'{i}sica d'Altes Energies (IFAE), The Barcelona Institute of Science and Technology, Campus UAB, 08193 Bellaterra (Barcelona), Spain\label{aff91}
\and
European Space Agency/ESTEC, Keplerlaan 1, 2201 AZ Noordwijk, The Netherlands\label{aff92}
\and
DARK, Niels Bohr Institute, University of Copenhagen, Jagtvej 155, 2200 Copenhagen, Denmark\label{aff93}
\and
Waterloo Centre for Astrophysics, University of Waterloo, Waterloo, Ontario N2L 3G1, Canada\label{aff94}
\and
Department of Physics and Astronomy, University of Waterloo, Waterloo, Ontario N2L 3G1, Canada\label{aff95}
\and
Perimeter Institute for Theoretical Physics, Waterloo, Ontario N2L 2Y5, Canada\label{aff96}
\and
Universit\'e Paris-Saclay, Universit\'e Paris Cit\'e, CEA, CNRS, AIM, 91191, Gif-sur-Yvette, France\label{aff97}
\and
Centre National d'Etudes Spatiales -- Centre spatial de Toulouse, 18 avenue Edouard Belin, 31401 Toulouse Cedex 9, France\label{aff98}
\and
Institute of Space Science, Str. Atomistilor, nr. 409 M\u{a}gurele, Ilfov, 077125, Romania\label{aff99}
\and
Institut f\"ur Theoretische Physik, University of Heidelberg, Philosophenweg 16, 69120 Heidelberg, Germany\label{aff100}
\and
Institut de Recherche en Astrophysique et Plan\'etologie (IRAP), Universit\'e de Toulouse, CNRS, UPS, CNES, 14 Av. Edouard Belin, 31400 Toulouse, France\label{aff101}
\and
Universit\'e St Joseph; Faculty of Sciences, Beirut, Lebanon\label{aff102}
\and
Departamento de F\'isica, FCFM, Universidad de Chile, Blanco Encalada 2008, Santiago, Chile\label{aff103}
\and
Satlantis, University Science Park, Sede Bld 48940, Leioa-Bilbao, Spain\label{aff104}
\and
Department of Physics and Helsinki Institute of Physics, Gustaf H\"allstr\"omin katu 2, University of Helsinki, 00014 Helsinki, Finland\label{aff105}
\and
Department of Physics, Royal Holloway, University of London, Surrey TW20 0EX, UK\label{aff106}
\and
Instituto de Astrof\'isica e Ci\^encias do Espa\c{c}o, Faculdade de Ci\^encias, Universidade de Lisboa, Tapada da Ajuda, 1349-018 Lisboa, Portugal\label{aff107}
\and
Mullard Space Science Laboratory, University College London, Holmbury St Mary, Dorking, Surrey RH5 6NT, UK\label{aff108}
\and
Cosmic Dawn Center (DAWN)\label{aff109}
\and
Niels Bohr Institute, University of Copenhagen, Jagtvej 128, 2200 Copenhagen, Denmark\label{aff110}
\and
Universidad Polit\'ecnica de Cartagena, Departamento de Electr\'onica y Tecnolog\'ia de Computadoras,  Plaza del Hospital 1, 30202 Cartagena, Spain\label{aff111}
\and
Kapteyn Astronomical Institute, University of Groningen, PO Box 800, 9700 AV Groningen, The Netherlands\label{aff112}
\and
Caltech/IPAC, 1200 E. California Blvd., Pasadena, CA 91125, USA\label{aff113}
\and
Dipartimento di Fisica e Scienze della Terra, Universit\`a degli Studi di Ferrara, Via Giuseppe Saragat 1, 44122 Ferrara, Italy\label{aff114}
\and
Istituto Nazionale di Fisica Nucleare, Sezione di Ferrara, Via Giuseppe Saragat 1, 44122 Ferrara, Italy\label{aff115}
\and
INAF, Istituto di Radioastronomia, Via Piero Gobetti 101, 40129 Bologna, Italy\label{aff116}
\and
Astronomical Observatory of the Autonomous Region of the Aosta Valley (OAVdA), Loc. Lignan 39, I-11020, Nus (Aosta Valley), Italy\label{aff117}
\and
Universit\'e C\^{o}te d'Azur, Observatoire de la C\^{o}te d'Azur, CNRS, Laboratoire Lagrange, Bd de l'Observatoire, CS 34229, 06304 Nice cedex 4, France\label{aff118}
\and
ICSC - Centro Nazionale di Ricerca in High Performance Computing, Big Data e Quantum Computing, Via Magnanelli 2, Bologna, Italy\label{aff119}
\and
Department of Physics, Oxford University, Keble Road, Oxford OX1 3RH, UK\label{aff120}
\and
Instituto de F\'isica Te\'orica UAM-CSIC, Campus de Cantoblanco, 28049 Madrid, Spain\label{aff121}
\and
CEA Saclay, DFR/IRFU, Service d'Astrophysique, Bat. 709, 91191 Gif-sur-Yvette, France\label{aff122}
\and
Univ. Grenoble Alpes, CNRS, Grenoble INP, LPSC-IN2P3, 53, Avenue des Martyrs, 38000, Grenoble, France\label{aff123}
\and
Dipartimento di Fisica, Sapienza Universit\`a di Roma, Piazzale Aldo Moro 2, 00185 Roma, Italy\label{aff124}
\and
Aurora Technology for European Space Agency (ESA), Camino bajo del Castillo, s/n, Urbanizacion Villafranca del Castillo, Villanueva de la Ca\~nada, 28692 Madrid, Spain\label{aff125}
\and
ICL, Junia, Universit\'e Catholique de Lille, LITL, 59000 Lille, France\label{aff126}
\and
CERCA/ISO, Department of Physics, Case Western Reserve University, 10900 Euclid Avenue, Cleveland, OH 44106, USA\label{aff127}
\and
Laboratoire Univers et Th\'eorie, Observatoire de Paris, Universit\'e PSL, Universit\'e Paris Cit\'e, CNRS, 92190 Meudon, France\label{aff128}
\and
Departamento de F{\'\i}sica Fundamental. Universidad de Salamanca. Plaza de la Merced s/n. 37008 Salamanca, Spain\label{aff129}
\and
IRFU, CEA, Universit\'e Paris-Saclay 91191 Gif-sur-Yvette Cedex, France\label{aff130}
\and
Aix-Marseille Universit\'e, Universit\'e de Toulon, CNRS, CPT, Marseille, France\label{aff131}
\and
Universit\'e de Strasbourg, CNRS, Observatoire astronomique de Strasbourg, UMR 7550, 67000 Strasbourg, France\label{aff132}
\and
Center for Data-Driven Discovery, Kavli IPMU (WPI), UTIAS, The University of Tokyo, Kashiwa, Chiba 277-8583, Japan\label{aff133}
\and
Department of Physics \& Astronomy, University of California Irvine, Irvine CA 92697, USA\label{aff134}
\and
Departamento F\'isica Aplicada, Universidad Polit\'ecnica de Cartagena, Campus Muralla del Mar, 30202 Cartagena, Murcia, Spain\label{aff135}
\and
Instituto de F\'isica de Cantabria, Edificio Juan Jord\'a, Avenida de los Castros, 39005 Santander, Spain\label{aff136}
\and
INFN, Sezione di Lecce, Via per Arnesano, CP-193, 73100, Lecce, Italy\label{aff137}
\and
Department of Mathematics and Physics E. De Giorgi, University of Salento, Via per Arnesano, CP-I93, 73100, Lecce, Italy\label{aff138}
\and
INAF-Sezione di Lecce, c/o Dipartimento Matematica e Fisica, Via per Arnesano, 73100, Lecce, Italy\label{aff139}
\and
Institute of Cosmology and Gravitation, University of Portsmouth, Portsmouth PO1 3FX, UK\label{aff140}
\and
Department of Computer Science, Aalto University, PO Box 15400, Espoo, FI-00 076, Finland\label{aff141}
\and
Universidad de La Laguna, Dpto. Astrof\'\i sica, E-38206 La Laguna, Tenerife, Spain\label{aff142}
\and
Ruhr University Bochum, Faculty of Physics and Astronomy, Astronomical Institute (AIRUB), German Centre for Cosmological Lensing (GCCL), 44780 Bochum, Germany\label{aff143}
\and
Department of Physics and Astronomy, Vesilinnantie 5, University of Turku, 20014 Turku, Finland\label{aff144}
\and
Finnish Centre for Astronomy with ESO (FINCA), Quantum, Vesilinnantie 5, University of Turku, 20014 Turku, Finland\label{aff145}
\and
Serco for European Space Agency (ESA), Camino bajo del Castillo, s/n, Urbanizacion Villafranca del Castillo, Villanueva de la Ca\~nada, 28692 Madrid, Spain\label{aff146}
\and
ARC Centre of Excellence for Dark Matter Particle Physics, Melbourne, Australia\label{aff147}
\and
Centre for Astrophysics \& Supercomputing, Swinburne University of Technology,  Hawthorn, Victoria 3122, Australia\label{aff148}
\and
Department of Physics and Astronomy, University of the Western Cape, Bellville, Cape Town, 7535, South Africa\label{aff149}
\and
DAMTP, Centre for Mathematical Sciences, Wilberforce Road, Cambridge CB3 0WA, UK\label{aff150}
\and
Kavli Institute for Cosmology Cambridge, Madingley Road, Cambridge, CB3 0HA, UK\label{aff151}
\and
Department of Astrophysics, University of Zurich, Winterthurerstrasse 190, 8057 Zurich, Switzerland\label{aff152}
\and
Department of Physics, Centre for Extragalactic Astronomy, Durham University, South Road, Durham, DH1 3LE, UK\label{aff153}
\and
Centro de Astrof\'{\i}sica da Universidade do Porto, Rua das Estrelas, 4150-762 Porto, Portugal\label{aff154}
\and
Instituto de Astrof\'isica e Ci\^encias do Espa\c{c}o, Universidade do Porto, CAUP, Rua das Estrelas, PT4150-762 Porto, Portugal\label{aff155}
\and
HE Space for European Space Agency (ESA), Camino bajo del Castillo, s/n, Urbanizacion Villafranca del Castillo, Villanueva de la Ca\~nada, 28692 Madrid, Spain\label{aff156}
\and
INAF - Osservatorio Astronomico d'Abruzzo, Via Maggini, 64100, Teramo, Italy\label{aff157}
\and
Theoretical astrophysics, Department of Physics and Astronomy, Uppsala University, Box 516, 751 37 Uppsala, Sweden\label{aff158}
\and
Mathematical Institute, University of Leiden, Einsteinweg 55, 2333 CA Leiden, The Netherlands\label{aff159}
\and
Institute of Astronomy, University of Cambridge, Madingley Road, Cambridge CB3 0HA, UK\label{aff160}
\and
Univ. Lille, CNRS, Centrale Lille, UMR 9189 CRIStAL, 59000 Lille, France\label{aff161}
\and
Institute for Particle Physics and Astrophysics, Dept. of Physics, ETH Zurich, Wolfgang-Pauli-Strasse 27, 8093 Zurich, Switzerland\label{aff162}
\and
Department of Astrophysical Sciences, Peyton Hall, Princeton University, Princeton, NJ 08544, USA\label{aff163}
\and
Institut de Physique Th\'eorique, CEA, CNRS, Universit\'e Paris-Saclay 91191 Gif-sur-Yvette Cedex, France\label{aff164}
\and
Center for Computational Astrophysics, Flatiron Institute, 162 5th Avenue, 10010, New York, NY, USA\label{aff165}}    

%% file: main.bbl
\begin{thebibliography}{125}
\expandafter\ifx\csname natexlab\endcsname\relax\def\natexlab#1{#1}\fi

\bibitem[{{Abidi} \& {Baldauf}(2018)}]{AbidiBaldauf2018}
{Abidi}, M.~M. \& {Baldauf}, T. 2018, \jcap, 07, 029

\bibitem[{{Agarwal} {et~al.}(2021){Agarwal}, {Desjacques}, {Jeong}, \&
  {Schmidt}}]{AgarwalEtal2021}
{Agarwal}, N., {Desjacques}, V., {Jeong}, D., \& {Schmidt}, F. 2021, \jcap, 03,
  021

\bibitem[{{Alcock} \& {Paczy{\'n}ski}(1979)}]{AlcockPaczynski1979}
{Alcock}, C. \& {Paczy{\'n}ski}, B. 1979, \nat, 281, 358

\bibitem[{{Alkhanishvili} {et~al.}(2022){Alkhanishvili}, {Porciani},
  {Sefusatti}, {Biagetti}, {Lazanu}, {Oddo}, \&
  {Yankelevich}}]{AlkhanishviliEtal2022}
{Alkhanishvili}, D., {Porciani}, C., {Sefusatti}, E., {et~al.} 2022, \mnras,
  512, 4961

\bibitem[{{Anastasiou} {et~al.}(2024){Anastasiou}, {Bragan{\c{c}}a},
  {Senatore}, \& {Zheng}}]{AnastasiouEtal2024}
{Anastasiou}, C., {Bragan{\c{c}}a}, D. P.~L., {Senatore}, L., \& {Zheng}, H.
  2024, \jhep, 01, 002

\bibitem[{{Angulo} {et~al.}(2015{\natexlab{a}}){Angulo}, {Fasiello},
  {Senatore}, \& {Vlah}}]{AnguloEtal2015}
{Angulo}, R., {Fasiello}, M., {Senatore}, L., \& {Vlah}, Z. 2015{\natexlab{a}},
  \jcap, 09, 029

\bibitem[{{Angulo} {et~al.}(2015{\natexlab{b}}){Angulo}, {Foreman},
  {Schmittfull}, \& {Senatore}}]{AnguloEtal2015B}
{Angulo}, R.~E., {Foreman}, S., {Schmittfull}, M., \& {Senatore}, L.
  2015{\natexlab{b}}, \jcap, 10, 039

\bibitem[{Audren {et~al.}(2013)Audren, Lesgourgues, Benabed, \&
  Prunet}]{AudrenEtal2012}
Audren, B., Lesgourgues, J., Benabed, K., \& Prunet, S. 2013, JCAP, 02, 001

\bibitem[{{Bakx} {et~al.}(2025){Bakx}, {Ivanov}, {Philcox}, \&
  {Vlah}}]{BaksEtal2025A}
{Bakx}, T., {Ivanov}, M.~M., {Philcox}, O. H.~E., \& {Vlah}, Z. 2025,
  arXiv:2507.22110

\bibitem[{{Baldauf} {et~al.}(2015{\natexlab{a}}){Baldauf}, {Mercolli},
  {Mirbabayi}, \& {Pajer}}]{BaldaufEtal2015A}
{Baldauf}, T., {Mercolli}, L., {Mirbabayi}, M., \& {Pajer}, E.
  2015{\natexlab{a}}, \jcap, 05, 007

\bibitem[{{Baldauf} {et~al.}(2015{\natexlab{b}}){Baldauf}, {Mirbabayi},
  {Simonovi{\'c}}, \& {Zaldarriaga}}]{BaldaufEtal2015B}
{Baldauf}, T., {Mirbabayi}, M., {Simonovi{\'c}}, M., \& {Zaldarriaga}, M.
  2015{\natexlab{b}}, \prd, 92, 043514

\bibitem[{{Baldauf} {et~al.}(2012){Baldauf}, {Seljak}, {Desjacques}, \&
  {McDonald}}]{BaldaufEtal2012}
{Baldauf}, T., {Seljak}, U., {Desjacques}, V., \& {McDonald}, P. 2012, \prd,
  86, 083540

\bibitem[{{Barreira}(2019)}]{Barreira2019}
{Barreira}, A. 2019, \jcap, 03, 008

\bibitem[{{Barreira}(2020)}]{Barreira2020}
{Barreira}, A. 2020, \jcap, 12, 031

\bibitem[{{Barreira} {et~al.}(2021){Barreira}, {Lazeyras}, \&
  {Schmidt}}]{BarreiraLazeyrasSchmidt2021}
{Barreira}, A., {Lazeyras}, T., \& {Schmidt}, F. 2021, \jcap, 08, 029

\bibitem[{{Baumann} {et~al.}(2012){Baumann}, {Nicolis}, {Senatore}, \&
  {Zaldarriaga}}]{BaumannEtal2012}
{Baumann}, D., {Nicolis}, A., {Senatore}, L., \& {Zaldarriaga}, M. 2012, \jcap,
  07, 051

\bibitem[{{Baumgart} \& {Fry}(1991)}]{BaumgartFry1991}
{Baumgart}, D.~J. \& {Fry}, J.~N. 1991, \apj, 375, 25

\bibitem[{Bernardeau {et~al.}(2002)Bernardeau, Colombi, Gazta{\~n}aga, \&
  Scoccimarro}]{BernardeauEtal2002}
Bernardeau, F., Colombi, S., Gazta{\~n}aga, E., \& Scoccimarro, R. 2002,
  \physrep, 367, 1

\bibitem[{{Biagetti} {et~al.}(2022){Biagetti}, {Castiblanco}, {Nore{\~n}a}, \&
  {Sefusatti}}]{BiagettiEtal2022}
{Biagetti}, M., {Castiblanco}, L., {Nore{\~n}a}, J., \& {Sefusatti}, E. 2022,
  \jcap, 09, 009

\bibitem[{{Blas} {et~al.}(2016){Blas}, {Garny}, {Ivanov}, \&
  {Sibiryakov}}]{BlasEtal2016}
{Blas}, D., {Garny}, M., {Ivanov}, M.~M., \& {Sibiryakov}, S. 2016, \jcap, 07,
  052

\bibitem[{Brinckmann \& Lesgourgues(2019)}]{BrinckmannLesgourgues2018}
Brinckmann, T. \& Lesgourgues, J. 2019, Phys. Dark Univ., 24, 100260

\bibitem[{{Buchner} {et~al.}(2014){Buchner}, {Georgakakis}, {Nandra}, {Hsu},
  {Rangel}, {Brightman}, {Merloni}, {Salvato}, {Donley}, \&
  {Kocevski}}]{BuchnerEtal2014}
{Buchner}, J., {Georgakakis}, A., {Nandra}, K., {et~al.} 2014, \aap, 564, A125

\bibitem[{{Byun} {et~al.}(2017){Byun}, {Eggemeier}, {Regan}, {Seery}, \&
  {Smith}}]{ByunEtal2017}
{Byun}, J., {Eggemeier}, A., {Regan}, D., {Seery}, D., \& {Smith}, R.~E. 2017,
  \mnras, 471, 1581

\bibitem[{{Cabass} {et~al.}(2022{\natexlab{a}}){Cabass}, {Ivanov}, {Philcox},
  {Simonovi{\'c}}, \& {Zaldarriaga}}]{CabassEtal2022B}
{Cabass}, G., {Ivanov}, M.~M., {Philcox}, O. H.~E., {Simonovi{\'c}}, M., \&
  {Zaldarriaga}, M. 2022{\natexlab{a}}, \prd, 106, 043506

\bibitem[{{Cabass} {et~al.}(2022{\natexlab{b}}){Cabass}, {Ivanov}, {Philcox},
  {Simonovi{\'c}}, \& {Zaldarriaga}}]{CabassEtal2022}
{Cabass}, G., {Ivanov}, M.~M., {Philcox}, O. H.~E., {Simonovi{\'c}}, M., \&
  {Zaldarriaga}, M. 2022{\natexlab{b}}, \prl, 129, 021301

\bibitem[{{Cagliari} {et~al.}(2025){Cagliari}, {Barberi-Squarotti}, {Pardede},
  {Castorina}, \& {D'Amico}}]{CagliariEtal2025}
{Cagliari}, M.~S., {Barberi-Squarotti}, M., {Pardede}, K., {Castorina}, E., \&
  {D'Amico}, G. 2025, \jcap, 07, 043

\bibitem[{{Carrasco} {et~al.}(2012){Carrasco}, {Hertzberg}, \&
  {Senatore}}]{CarrascoHertzbergSenatore2012}
{Carrasco}, J.~J.~M., {Hertzberg}, M.~P., \& {Senatore}, L. 2012, \jhep, 9, 82

\bibitem[{Chan {et~al.}(2012)Chan, Scoccimarro, \&
  Sheth}]{ChanScoccimarroSheth2012}
Chan, K.~C., Scoccimarro, R., \& Sheth, R.~K. 2012, \prd, 85, 083509

\bibitem[{{Chen} {et~al.}(2021){Chen}, {Vlah}, {Castorina}, \&
  {White}}]{ChenEtal2021}
{Chen}, S.-F., {Vlah}, Z., {Castorina}, E., \& {White}, M. 2021, \jcap, 03, 100

\bibitem[{{Chudaykin} {et~al.}(2020){Chudaykin}, {Ivanov}, {Philcox}, \&
  {Simonovi{\'c}}}]{ChudaykinEtal2020}
{Chudaykin}, A., {Ivanov}, M.~M., {Philcox}, O. H.~E., \& {Simonovi{\'c}}, M.
  2020, \prd, 102, 063533

\bibitem[{Cooray \& Sheth(2002)}]{CooraySheth2002}
Cooray, A. \& Sheth, R.~K. 2002, \physrep, 372, 1

\bibitem[{Crocce \& Scoccimarro(2008)}]{CrocceScoccimarro2008}
Crocce, M. \& Scoccimarro, R. 2008, \prd, 77, 023533

\bibitem[{{D'Amico} {et~al.}(2024{\natexlab{a}}){D'Amico}, {Donath},
  {Lewandowski}, {Senatore}, \& {Zhang}}]{DAmicoEtal2024}
{D'Amico}, G., {Donath}, Y., {Lewandowski}, M., {Senatore}, L., \& {Zhang}, P.
  2024{\natexlab{a}}, \jcap, 05, 059

\bibitem[{{D'Amico} {et~al.}(2024{\natexlab{b}}){D'Amico}, {Donath},
  {Lewandowski}, {Senatore}, \& {Zhang}}]{DAmicoEtal2024B}
{D'Amico}, G., {Donath}, Y., {Lewandowski}, M., {Senatore}, L., \& {Zhang}, P.
  2024{\natexlab{b}}, \jcap, 07, 041

\bibitem[{{D'Amico} {et~al.}(2020){D'Amico}, {Gleyzes}, {Kokron}, {Markovic},
  {Senatore}, {Zhang}, {Beutler}, \& {Gil-Mar{\'\i}n}}]{DAmicoEtal2020}
{D'Amico}, G., {Gleyzes}, J., {Kokron}, N., {et~al.} 2020, \jcap, 05, 005

\bibitem[{{D'Amico} {et~al.}(2025){D'Amico}, {Lewandowski}, {Senatore}, \&
  {Zhang}}]{DAmicoEtal2025}
{D'Amico}, G., {Lewandowski}, M., {Senatore}, L., \& {Zhang}, P. 2025, \prd,
  111, 063514

\bibitem[{{D'Amico} {et~al.}(2021){D'Amico}, {Senatore}, \&
  {Zhang}}]{DAmicoSenatoreZhang2021}
{D'Amico}, G., {Senatore}, L., \& {Zhang}, P. 2021, \jcap, 01, 006

\bibitem[{{DESI Collaboration} {et~al.}(2016){DESI Collaboration}, {Aghamousa},
  {Aguilar}, {Ahlen}, {Alam}, {Allen}, {Allende Prieto}, {Annis}, {Bailey},
  {Balland}, \& et~al.}]{AghamousaEtal2016}
{DESI Collaboration}, {Aghamousa}, A., {Aguilar}, J., {et~al.} 2016, arXiv:
  1110.3193 \eprint[ArXiv]{1611.00036}

\bibitem[{{Desjacques} {et~al.}(2018){Desjacques}, {Jeong}, \&
  {Schmidt}}]{DesjacquesJeongSchmidt2018}
{Desjacques}, V., {Jeong}, D., \& {Schmidt}, F. 2018, \physrep, 733, 1

\bibitem[{{Eggemeier} {et~al.}(2023){Eggemeier}, {Camacho-Quevedo}, {Pezzotta},
  {Crocce}, {Scoccimarro}, \& {S{\'a}nchez}}]{EggemeierEtal2023}
{Eggemeier}, A., {Camacho-Quevedo}, B., {Pezzotta}, A., {et~al.} 2023, \mnras,
  519, 2962

\bibitem[{{Eggemeier} {et~al.}(2025){Eggemeier}, {Lee}, {Scoccimarro},
  {Camacho-Quevedo}, {Pezzotta}, {Crocce}, \&
  {S{\'a}nchez}}]{EggemeierEtal2025}
{Eggemeier}, A., {Lee}, N., {Scoccimarro}, R., {et~al.} 2025, \prd, 112, 063532

\bibitem[{{Eggemeier} {et~al.}(2020){Eggemeier}, {Scoccimarro}, {Crocce},
  {Pezzotta}, \& {S{\'a}nchez}}]{EggemeierEtal2020}
{Eggemeier}, A., {Scoccimarro}, R., {Crocce}, M., {Pezzotta}, A., \&
  {S{\'a}nchez}, A.~G. 2020, \prd, 102, 103530

\bibitem[{{Eggemeier} {et~al.}(2019){Eggemeier}, {Scoccimarro}, \&
  {Smith}}]{EggemeierScoccimarroSmith2019}
{Eggemeier}, A., {Scoccimarro}, R., \& {Smith}, R.~E. 2019, \prd, 99, 123514

\bibitem[{{Eggemeier} {et~al.}(2021){Eggemeier}, {Scoccimarro}, {Smith},
  {Crocce}, {Pezzotta}, \& {S{\'a}nchez}}]{EggemeierEtal2021}
{Eggemeier}, A., {Scoccimarro}, R., {Smith}, R.~E., {et~al.} 2021, \prd, 103,
  123550

\bibitem[{{Eisenstein} {et~al.}(2007){Eisenstein}, {Seo}, \&
  {White}}]{EisensteinSeoWhite2007}
{Eisenstein}, D.~J., {Seo}, H.-J., \& {White}, M. 2007, \apj, 664, 660

\bibitem[{Euclid Collaboration:~{Blanchard} {et~al.}(2020)Euclid
  Collaboration:~{Blanchard}, {Camera}, {Carbone}, {Cardone}, {Casas},
  {Clesse}, {Ili{\'c}}, {Kilbinger}, {Kitching}, {Kunz}, {Lacasa}, {Linder},
  {Majerotto}, {Markovi{\v{c}}}, {Martinelli}, {Pettorino}, {Pourtsidou},
  {Sakr}, {S{\'a}nchez}, {Sapone}, {Tutusaus}, {Yahia-Cherif}, {Yankelevich},
  {Andreon}, {Aussel}, {Balaguera-Antol{\'\i}nez}, {Baldi}, {Bardelli},
  {Bender}, {Biviano}, {Bonino}, {Boucaud}, {Bozzo}, {Branchini}, {Brau-Nogue},
  {Brescia}, {Brinchmann}, {Burigana}, {Cabanac}, {Capobianco}, {Cappi},
  {Carretero}, {Carvalho}, {Casas}, {Castander}, {Castellano}, {Cavuoti},
  {Cimatti}, {Cledassou}, {Colodro-Conde}, {Congedo}, {Conselice}, {Conversi},
  {Copin}, {Corcione}, {Coupon}, {Courtois}, {Cropper}, {Da Silva}, {de la
  Torre}, {Di Ferdinando}, {Dubath}, {Ducret}, {Duncan}, {Dupac}, {Dusini},
  {Fabbian}, {Fabricius}, {Farrens}, {Fosalba}, {Fotopoulou}, {Fourmanoit},
  {Frailis}, {Franceschi}, {Franzetti}, {Fumana}, {Galeotta}, {Gillard},
  {Gillis}, {Giocoli}, {G{\'o}mez-Alvarez}, {Graci{\'a}-Carpio}, {Grupp},
  {Guzzo}, {Hoekstra}, {Hormuth}, {Israel}, {Jahnke}, {Keihanen}, {Kermiche},
  {Kirkpatrick}, {Kohley}, {Kubik}, {Kurki-Suonio}, {Ligori}, {Lilje}, {Lloro},
  {Maino}, {Maiorano}, {Marggraf}, {Martinet}, {Marulli}, {Massey},
  {Medinaceli}, {Mei}, {Mellier}, {Metcalf}, {Metge}, {Meylan}, {Moresco},
  {Moscardini}, {Munari}, {Nichol}, {Niemi}, {Nucita}, {Padilla}, {Paltani},
  {Pasian}, {Percival}, {Pires}, {Polenta}, {Poncet}, {Pozzetti}, {Racca},
  {Raison}, {Renzi}, {Rhodes}, {Romelli}, {Roncarelli}, {Rossetti}, {Saglia},
  {Schneider}, {Scottez}, {Secroun}, {Sirri}, {Stanco}, {Starck}, {Sureau},
  {Tallada-Cresp{\'\i}}, {Tavagnacco}, {Taylor}, {Tenti}, {Tereno},
  {Toledo-Moreo}, {Torradeflot}, {Valenziano}, {Vassallo}, {Verdoes Kleijn},
  {Viel}, {Wang}, {Zacchei}, {Zoubian}, \& {Zucca}}]{BlanchardEtal2020}
Euclid Collaboration:~{Blanchard}, A., {Camera}, S., {Carbone}, C., {et~al.}
  2020, \aap, 642, A191

\bibitem[{Euclid Collaboration:~{Guidi} {et~al.}(2025)Euclid
  Collaboration:~{Guidi}, {Veropalumbo}, {Pugno}, {Moresco}, {Sefusatti},
  {Porciani}, {Branchini}, {Breton}, {Camacho Quevedo}, {Crocce}, {de la
  Torre}, {Desjacques}, {Eggemeier}, {Farina}, {K{\"a}rcher}, {Linde},
  {Marinucci}, {Moradinezhad Dizgah}, {Moretti}, {Pardede}, {Pezzotta},
  {Sarpa}, {Amara}, {Andreon}, {Auricchio}, {Baccigalupi}, {Bagot}, {Baldi},
  {Bardelli}, {Battaglia}, {Biviano}, {Brescia}, {Camera}, {Ca{\~n}as-Herrera},
  {Capobianco}, {Carbone}, {Cardone}, {Carretero}, {Castellano}, {Castignani},
  {Cavuoti}, {Chambers}, {Cimatti}, {Colodro-Conde}, {Congedo}, {Conversi},
  {Copin}, {Courbin}, {Courtois}, {Da Silva}, {Degaudenzi}, {De Lucia}, {Dole},
  {Douspis}, {Dubath}, {Dupac}, {Dusini}, {Escoffier}, {Farina}, {Farinelli},
  {Faustini}, {Ferriol}, {Finelli}, {Fosalba}, {Fotopoulou}, {Frailis},
  {Franceschi}, {Fumana}, {Galeotta}, {Gillis}, {Giocoli}, {Gracia-Carpio},
  {Grazian}, {Grupp}, {Guzzo}, {Haugan}, {Holmes}, {Hormuth}, {Hornstrup},
  {Jahnke}, {Jhabvala}, {Joachimi}, {Keih{\"a}nen}, {Kermiche}, {Kiessling},
  {Kubik}, {K{\"u}mmel}, {Kunz}, {Kurki-Suonio}, {Le Brun}, {Ligori}, {Lilje},
  {Lindholm}, {Lloro}, {Mainetti}, {Maino}, {Maiorano}, {Mansutti}, {Marcin},
  {Marggraf}, {Markovic}, {Martinelli}, {Martinet}, {Marulli}, {Massey},
  {Medinaceli}, {Mei}, {Melchior}, {Mellier}, {Meneghetti}, {Merlin}, {Meylan},
  {Mora}, {Morin}, {Moscardini}, {Munari}, {Neissner}, {Niemi}, {Padilla},
  {Paltani}, {Pasian}, {Pedersen}, {Percival}, {Pettorino}, {Pires}, {Polenta},
  {Poncet}, {Popa}, {Raison}, {Rebolo}, {Renzi}, {Rhodes}, {Riccio}, {Romelli},
  {Roncarelli}, {Saglia}, {Sakr}, {S{\'a}nchez}, {Sapone}, {Sartoris},
  {Schewtschenko}, {Schneider}, {Schrabback}, {Scodeggio}, {Secroun}, {Seidel},
  {Seiffert}, {Serrano}, {Simon}, {Sirignano}, {Sirri}, {Spurio Mancini},
  {Stanco}, {Steinwagner}, {Tallada-Cresp{\'\i}}, {Tavagnacco}, {Taylor},
  {Tereno}, {Tessore}, {Toft}, {Toledo-Moreo}, {Torradeflot}, {Tsyganov},
  {Tutusaus}, {Valenziano}, {Valiviita}, {Vassallo}, {Verdoes Kleijn}, {Wang},
  {Weller}, {Zamorani}, {Zerbi}, {Zucca}, {Allevato}, {Ballardini},
  {Bolzonella}, {Bozzo}, {Burigana}, {Cabanac}, {Calabrese}, {Cappi}, {Di
  Ferdinando}, {Escartin Vigo}, {Gabarra}, {Mart{\'\i}n-Fleitas}, {Matthew},
  {Maturi}, {Mauri}, {Metcalf}, {Nucita}, {P{\"o}ntinen}, {Risso}, {Scottez},
  {Sereno}, {Tenti}, {Viel}, {Wiesmann}, {Akrami}, \& {Andika}}]{GuidiEtal2025}
Euclid Collaboration:~{Guidi}, M., {Veropalumbo}, A., {Pugno}, A., {et~al.}
  2025, arXiv:2506.22257

\bibitem[{{Euclid Collaboration}:~{Mellier} {et~al.}(2025){Euclid
  Collaboration}:~{Mellier}, {Abdurro'uf}, {Acevedo Barroso}, {Ach{\'u}carro},
  {Adamek}, {Adam}, {Addison}, {Aghanim}, {Aguena}, {Ajani}, {Akrami},
  {Al-Bahlawan}, {Alavi}, {Albuquerque}, {Alestas}, {Alguero}, {Allaoui},
  {Allen}, {Allevato}, {Alonso-Tetilla}, {Altieri}, {Alvarez-Candal}, {Alvi},
  {Amara}, {Amendola}, {Amiaux}, {Andika}, {Andreon}, {Andrews}, {Angora},
  {Angulo}, {Annibali}, {Anselmi}, {Anselmi}, {Arcari}, {Archidiacono},
  {Aric{\`o}}, {Arnaud}, {Arnouts}, {Asgari}, {Asorey}, {Atayde}, {Atek},
  {Atrio-Barandela}, {Aubert}, {Aubourg}, {Auphan}, {Auricchio}, {Aussel},
  {Aussel}, {Avelino}, {Avgoustidis}, {Avila}, {Awan}, {Azzollini},
  {Baccigalupi}, {Bachelet}, {Bacon}, {Baes}, {Bagley}, {Bahr-Kalus},
  {Balaguera-Antolinez}, {Balbinot}, {Balcells}, {Baldi}, {Baldry}, {Balestra},
  {Ballardini}, {Ballester}, {Balogh}, {Ba{\~n}ados}, {Barbier}, {Bardelli},
  {Baron}, {Barreiro}, {Barrena}, {Barriere}, {Barros}, {Barthelemy},
  {Bartolo}, {Basset}, {Battaglia}, {Battisti}, {Baugh}, {Baumont},
  {Bazzanini}, {Beaulieu}, {Beckmann}, {Belikov}, {Bel}, {Bellagamba}, {Bella},
  {Bellini}, {Benabed}, {Bender}, {Benevento}, {Bennett}, {Benson},
  {Bergamini}, {Bermejo-Climent}, {Bernardeau}, {Bertacca}, {Berthe},
  {Berthier}, {Bethermin}, {Beutler}, {Bevillon}, {Bhargava}, {Bhatawdekar},
  {Bianchi}, {Bisigello}, {Biviano}, {Blake}, {Blanchard}, {Blazek}, {Blot},
  {Bosco}, {Bodendorf}, {Boenke}, {B{\"o}hringer}, {Boldrini}, {Bolzonella},
  {Bonchi}, {Bonici}, {Bonino}, {Bonino}, {Bonvin}, {Bon}, {Booth}, {Borgani},
  {Borlaff}, {Borsato}, {Bose}, {Botticella}, {Boucaud}, {Bouche}, {Boucher},
  {Boutigny}, {Bouvard}, {Bouwens}, {Bouy}, {Bowler}, {Bozza}, {Bozzo},
  {Branchini}, {Brando}, {Brau-Nogue}, {Brekke}, {Bremer}, {Brescia}, {Breton},
  {Brinchmann}, {Brinckmann}, {Brockley-Blatt}, {Brodwin}, {Brouard}, {Brown},
  {Bruton}, {Bucko}, {Buddelmeijer}, {Buenadicha}, {Buitrago}, {Burger},
  {Burigana}, {Busillo}, {Busonero}, {Cabanac}, {Cabayol-Garcia}, {Cagliari},
  {Caillat}, {Caillat}, {Calabrese}, {Calabro}, {Calderone}, {Calura}, {Camacho
  Quevedo}, {Camera}, {Campos}, {Ca{\~n}as-Herrera}, {Candini}, {Cantiello},
  {Capobianco}, {Cappellaro}, {Cappelluti}, {Cappi}, {Caputi}, {Cara},
  {Carbone}, {Cardone}, {Carella}, {Carlberg}, {Carle}, {Carminati}, {Caro},
  {Carrasco}, {Carretero}, {Carrilho}, {Carron Duque}, \&
  {Carry}}]{MellierEtal2025}
{Euclid Collaboration}:~{Mellier}, Y., {Abdurro'uf}, {Acevedo Barroso}, J.~A.,
  {et~al.} 2025, \aap, 697, A1

\bibitem[{Euclid Collaboration:~{Pezzotta} {et~al.}(2024)Euclid
  Collaboration:~{Pezzotta}, {Moretti}, {Zennaro}, {Moradinezhad Dizgah},
  {Crocce}, {Sefusatti}, {Ferrero}, {Pardede}, {Eggemeier}, {Barreira},
  {Angulo}, {Marinucci}, {Camacho Quevedo}, {de la Torre}, {Alkhanishvili},
  {Biagetti}, {Breton}, {Castorina}, {D'Amico}, {Desjacques}, {Guidi},
  {K{\"a}rcher}, {Oddo}, {Pellejero Ibanez}, {Porciani}, {Pugno},
  {Salvalaggio}, {Sarpa}, {Veropalumbo}, {Vlah}, {Amara}, {Andreon},
  {Auricchio}, {Baldi}, {Bardelli}, {Bender}, {Bodendorf}, {Bonino},
  {Branchini}, {Brescia}, {Brinchmann}, {Camera}, {Capobianco}, {Carbone},
  {Cardone}, {Carretero}, {Casas}, {Castander}, {Castellano}, {Cavuoti},
  {Cimatti}, {Congedo}, {Conselice}, {Conversi}, {Copin}, {Corcione},
  {Courbin}, {Courtois}, {Da Silva}, {Degaudenzi}, {Di Giorgio}, {Dinis},
  {Dupac}, {Dusini}, {Ealet}, {Farina}, {Farrens}, {Fosalba}, {Frailis},
  {Franceschi}, {Galeotta}, {Gillis}, {Giocoli}, {Granett}, {Grazian}, {Grupp},
  {Guzzo}, {Haugan}, {Hormuth}, {Hornstrup}, {Jahnke}, {Joachimi},
  {Keih{\"a}nen}, {Kermiche}, {Kiessling}, {Kilbinger}, {Kitching}, {Kubik},
  {Kunz}, {Kurki-Suonio}, {Ligori}, {Lilje}, {Lindholm}, {Lloro}, {Maiorano},
  {Mansutti}, {Marggraf}, {Markovic}, {Martinet}, {Marulli}, {Massey},
  {Medinaceli}, {Mellier}, {Meneghetti}, {Merlin}, {Meylan}, {Moresco},
  {Moscardini}, {Munari}, {Niemi}, {Padilla}, {Paltani}, {Pasian}, {Pedersen},
  {Percival}, {Pettorino}, {Pires}, {Polenta}, {Pollack}, {Poncet}, {Popa},
  {Pozzetti}, {Raison}, {Renzi}, {Rhodes}, {Riccio}, {Romelli}, {Roncarelli},
  {Rossetti}, {Saglia}, {Sapone}, {Sartoris}, {Schneider}, {Schrabback},
  {Secroun}, {Seidel}, {Seiffert}, {Serrano}, {Sirignano}, {Sirri}, {Stanco},
  {Surace}, {Tallada-Cresp{\'\i}}, {Taylor}, {Tereno}, {Toledo-Moreo},
  {Torradeflot}, {Tutusaus}, {Valentijn}, {Valenziano}, {Vassallo}, {Wang},
  {Weller}, {Zamorani}, {Zoubian}, {Zucca}, {Biviano}, {Bozzo}, {Burigana},
  {Colodro-Conde}, {Di Ferdinando}, {Mainetti}, {Martinelli}, {Mauri}, {Sakr},
  {Scottez}, {Tenti}, {Viel}, {Wiesmann}, {Akrami}, {Allevato}, {Anselmi},
  {Baccigalupi}, {Ballardini}, {Bernardeau}, {Blanchard}, {Borgani}, {Bruton},
  {Cabanac}, {Cappi}, {Carvalho}, {Castignani}, {Castro}, {Ca{\~n}as-Herrera},
  {Chambers}, {Contarini}, {Cooray}, {Coupon}, {Davini}, {De Lucia}, {Desprez},
  {Di Domizio}, {Dole}, {D{\'\i}az-S{\'a}nchez}, {Escartin Vigo}, {Escoffier},
  {Ferreira}, {Finelli}, {Gabarra}, {Ganga}, {Garc{\'\i}a-Bellido},
  {Giacomini}, {Gozaliasl}, {Hall}, {Ili{\'c}}, {Joudaki}, {Kajava}, {Kansal},
  {Kirkpatrick}, {Legrand}, {Loureiro}, {Macias-Perez}, {Magliocchetti},
  {Mannucci}, {Maoli}, {Martins}, {Matthew}, {Maurin}, {Metcalf}, {Migliaccio},
  {Monaco}, {Morgante}, {Nadathur}, {Walton}, {Patrizii}, {Popa}, {Potter},
  {Pourtsidou}, {P{\"o}ntinen}, {Risso}, {Rocci}, {Sahl{\'e}n}, {S{\'a}nchez},
  {Schneider}, {Sereno}, {Simon}, {Spurio Mancini}, {Steinwagner}, {Testera},
  {Teyssier}, {Toft}, {Tosi}, {Troja}, {Tucci}, {Valiviita}, {Vergani},
  {Verza}, \& {Vielzeuf}}]{PezzottaEtal2024}
Euclid Collaboration:~{Pezzotta}, A., {Moretti}, C., {Zennaro}, M., {et~al.}
  2024, \aap, 687, A216

\bibitem[{{Feldman} {et~al.}(2001){Feldman}, Frieman, Fry, \&
  Scoccimarro}]{FeldmanEtal2001}
{Feldman}, H.~A., Frieman, J.~A., Fry, J.~N., \& Scoccimarro, R. 2001, Physical
  Review Letters, 86, 1434

\bibitem[{{Feroz} \& {Hobson}(2008)}]{FerozHobson2008}
{Feroz}, F. \& {Hobson}, M.~P. 2008, \mnras, 384, 449

\bibitem[{{Feroz} {et~al.}(2009){Feroz}, {Hobson}, \&
  {Bridges}}]{FerozHobsonBridges2009}
{Feroz}, F., {Hobson}, M.~P., \& {Bridges}, M. 2009, \mnras, 398, 1601

\bibitem[{{Feroz} {et~al.}(2019){Feroz}, {Hobson}, {Cameron}, \&
  {Pettitt}}]{FerozEtal2019}
{Feroz}, F., {Hobson}, M.~P., {Cameron}, E., \& {Pettitt}, A.~N. 2019, The Open
  Journal of Astrophysics, 2, 10

\bibitem[{Fry(1984)}]{Fry1984}
Fry, J.~N. 1984, \apj, 279, 499

\bibitem[{Fry(1996)}]{Fry1996}
Fry, J.~N. 1996, \apjl, 461, L65

\bibitem[{Fry \& Gazta{\~n}aga(1993)}]{FryGaztanaga1993}
Fry, J.~N. \& Gazta{\~n}aga, E. 1993, \apj, 413, 447

\bibitem[{{Fujita} {et~al.}(2020){Fujita}, {Mauerhofer}, {Senatore}, {Vlah}, \&
  {Angulo}}]{FujitaEtal2020}
{Fujita}, T., {Mauerhofer}, V., {Senatore}, L., {Vlah}, Z., \& {Angulo}, R.
  2020, \jcap, 01, 009

\bibitem[{{Gagrani} \& {Samushia}(2017)}]{GagraniSamushia2017}
{Gagrani}, P. \& {Samushia}, L. 2017, \mnras, 467, 928

\bibitem[{{Gil-Mar{\'{\i}}n} {et~al.}(2017){Gil-Mar{\'{\i}}n}, {Percival},
  {Verde}, {Brownstein}, {Chuang}, {Kitaura}, {Rodr{\'{\i}}guez-Torres}, \&
  {Olmstead}}]{GilMarinEtal2017}
{Gil-Mar{\'{\i}}n}, H., {Percival}, W.~J., {Verde}, L., {et~al.} 2017, \mnras,
  465, 1757

\bibitem[{Grieb {et~al.}(2016)Grieb, S{\'a}nchez, Salazar-Albornoz, \&
  Dalla~Vecchia}]{GriebEtal2016}
Grieb, J.~N., S{\'a}nchez, A.~G., Salazar-Albornoz, S., \& Dalla~Vecchia, C.
  2016, \mnras, 457, 1577

\bibitem[{{Gualdi} {et~al.}(2021){Gualdi}, {Gil-Mar{\'\i}n}, \&
  {Verde}}]{GualdiGilMarinVerde2021}
{Gualdi}, D., {Gil-Mar{\'\i}n}, H., \& {Verde}, L. 2021, \jcap, 07, 008

\bibitem[{{Gualdi} {et~al.}(2018){Gualdi}, {Manera}, {Joachimi}, \&
  {Lahav}}]{GualdiEtal2018}
{Gualdi}, D., {Manera}, M., {Joachimi}, B., \& {Lahav}, O. 2018, \mnras, 476,
  4045

\bibitem[{{Gualdi} \& {Verde}(2020)}]{GualdiVerde2020}
{Gualdi}, D. \& {Verde}, L. 2020, \jcap, 06, 041

\bibitem[{{Hahn} {et~al.}(2024){Hahn}, {Eickenberg}, {Ho}, {Hou}, {Lemos},
  {Massara}, {Modi}, {Dizgah}, {Parker}, {Blancard}, \& {SimBIG
  Collaboration}}]{HahnEtal2024}
{Hahn}, C., {Eickenberg}, M., {Ho}, S., {et~al.} 2024, \prd, 109, 083534

\bibitem[{{Hahn} \& {Villaescusa-Navarro}(2021)}]{HahnVillaescusaNavarro2021}
{Hahn}, C. \& {Villaescusa-Navarro}, F. 2021, \jcap, 04, 029

\bibitem[{{Hahn} {et~al.}(2020){Hahn}, {Villaescusa-Navarro}, {Castorina}, \&
  {Scoccimarro}}]{HahnEtal2020}
{Hahn}, C., {Villaescusa-Navarro}, F., {Castorina}, E., \& {Scoccimarro}, R.
  2020, \jcap, 03, 040

\bibitem[{{Hamilton}(2000)}]{Hamilton2000}
{Hamilton}, A.~J.~S. 2000, \mnras, 312, 257

\bibitem[{{Hashimoto} {et~al.}(2017){Hashimoto}, {Rasera}, \&
  {Taruya}}]{HashimotoRaseraTaruya2017}
{Hashimoto}, I., {Rasera}, Y., \& {Taruya}, A. 2017, \prd, 96, 043526

\bibitem[{{Hivon} {et~al.}(1995){Hivon}, {Bouchet}, {Colombi}, \&
  {Juszkiewicz}}]{HivonEtal1995}
{Hivon}, E., {Bouchet}, F.~R., {Colombi}, S., \& {Juszkiewicz}, R. 1995, \aap,
  298, 643

\bibitem[{{Ivanov} {et~al.}(2023){Ivanov}, {Philcox}, {Cabass}, {Nishimichi},
  {Simonovi{\'c}}, \& {Zaldarriaga}}]{IvanovEtal2023}
{Ivanov}, M.~M., {Philcox}, O. H.~E., {Cabass}, G., {et~al.} 2023, \prd, 107,
  083515

\bibitem[{{Ivanov} {et~al.}(2022){Ivanov}, {Philcox}, {Nishimichi},
  {Simonovi{\'c}}, {Takada}, \& {Zaldarriaga}}]{IvanovEtal2022B}
{Ivanov}, M.~M., {Philcox}, O. H.~E., {Nishimichi}, T., {et~al.} 2022, \prd,
  105, 063512

\bibitem[{{Ivanov} \& {Sibiryakov}(2018)}]{IvanovSibiryakov2018}
{Ivanov}, M.~M. \& {Sibiryakov}, S. 2018, \jcap, 07, 053

\bibitem[{{Ivezic} {et~al.}(2009){Ivezic}, {Tyson}, {Axelrod}, {Burke},
  {Claver}, {Cook}, {Kahn}, {Lupton}, {Monet}, {Pinto}, {Strauss}, {Stubbs},
  {Jones}, {Saha}, {Scranton}, {Smith}, \& {LSST
  Collaboration}}]{IvezicEtal2009}
{Ivezic}, Z., {Tyson}, J.~A., {Axelrod}, T., {et~al.} 2009, in American
  Astronomical Society Meeting Abstracts, Vol. 213, American Astronomical
  Society Meeting Abstracts, 460.03

\bibitem[{{Laureijs} {et~al.}(2011){Laureijs}, {Amiaux}, {Arduini},
  {Augu{\`e}res}, {Brinchmann}, {Cole}, {Cropper}, {Dabin}, {Duvet}, {Ealet},
  \& et~al.}]{LaureijsEtal2011}
{Laureijs}, R., {Amiaux}, J., {Arduini}, S., {et~al.} 2011, arXiv: 1110.3193
  \eprint[arXiv]{1110.3193}

\bibitem[{{Lazeyras} {et~al.}(2016){Lazeyras}, Wagner, Baldauf, \&
  Schmidt}]{LazeyrasEtal2016}
{Lazeyras}, T., Wagner, C., Baldauf, T., \& Schmidt, F. 2016, \jcap, 02, 018

\bibitem[{{Lewandowski} \& {Senatore}(2020)}]{LewandowskiSenatore2020}
{Lewandowski}, M. \& {Senatore}, L. 2020, \jcap, 03, 018

\bibitem[{{Linde} {et~al.}(2024){Linde}, {Dizgah}, {Radermacher}, {Casas}, \&
  {Lesgourgues}}]{LindeEtal2024}
{Linde}, D., {Dizgah}, A.~M., {Radermacher}, C., {Casas}, S., \& {Lesgourgues},
  J. 2024, \jcap, 07, 068

\bibitem[{Matarrese {et~al.}(1997)Matarrese, Verde, \&
  Heavens}]{MatarreseVerdeHeavens1997}
Matarrese, S., Verde, L., \& Heavens, A.~F. 1997, \mnras, 290, 651

\bibitem[{{McDonald} \& {Roy}(2009)}]{McDonaldRoy2009}
{McDonald}, P. \& {Roy}, A. 2009, \jcap, 08, 020

\bibitem[{{McEwen} {et~al.}(2016){McEwen}, {Fang}, {Hirata}, \&
  {Blazek}}]{McEwenEtal2016}
{McEwen}, J.~E., {Fang}, X., {Hirata}, C.~M., \& {Blazek}, J.~A. 2016, \jcap,
  09, 015

\bibitem[{{Moradinezhad Dizgah} {et~al.}(2021){Moradinezhad Dizgah},
  {Biagetti}, {Sefusatti}, {Desjacques}, \&
  {Nore{\~n}a}}]{MoradinezhadDizgahEtal2021}
{Moradinezhad Dizgah}, A., {Biagetti}, M., {Sefusatti}, E., {Desjacques}, V.,
  \& {Nore{\~n}a}, J. 2021, \jcap, 05, 015

\bibitem[{{Moretti} {et~al.}(2023){Moretti}, {Tsedrik}, {Carrilho}, \&
  {Pourtsidou}}]{MorettiEtal2023}
{Moretti}, C., {Tsedrik}, M., {Carrilho}, P., \& {Pourtsidou}, A. 2023, \jcap,
  12, 025

\bibitem[{{Novell-Masot} {et~al.}(2025){Novell-Masot}, {Gil-Mar{\'\i}n},
  {Verde}, {Aguilar}, {Ahlen}, {Bailey}, {BenZvi}, {Bianchi}, {Brooks},
  {Buckley-Geer}, {Carnero Rosell}, {Chaussidon}, {Claybaugh}, {Cole}, {Cuceu},
  {Dawson}, {de la Macorra}, {Demina}, {Dey}, {Dey}, {Doel}, {Ferraro},
  {Font-Ribera}, {Forero-Romero}, {Gazta{\~n}aga}, {Gontcho A Gontcho},
  {Gonzalez-Morales}, {Gutierrez}, {Herrera-Alcantar}, {Honscheid}, {Howlett},
  {Juneau}, {Kehoe}, {Kirkby}, {Kisner}, {Kremin}, {Lamman}, {Landriau}, {Le
  Guillou}, {Levi}, {Magneville}, {Manera}, {Meisner}, {Miquel}, {Moustakas},
  {Mu{\~n}oz-Guti{\'e}rrez}, {Myers}, {Nadathur}, {Niz}, {Noriega}, {Percival},
  {Poppett}, {Prada}, {P{\'e}rez-R{\`a}fols}, {Ross}, {Rossi}, {Samushia},
  {Sanchez}, {Schlegel}, {Schubnell}, {Seo}, {Silber}, {Sprayberry},
  {Tarl{\'e}}, {Vargas-Maga{\~n}a}, {Weaver}, {Zarrouk}, {Zhou}, \&
  {Zou}}]{NovellMasotEtal2025}
{Novell-Masot}, S., {Gil-Mar{\'\i}n}, H., {Verde}, L., {et~al.} 2025, \jcap,
  06, 005

\bibitem[{{Oddo} {et~al.}(2021){Oddo}, {Rizzo}, {Sefusatti}, {Porciani}, \&
  {Monaco}}]{OddoEtal2021}
{Oddo}, A., {Rizzo}, F., {Sefusatti}, E., {Porciani}, C., \& {Monaco}, P. 2021,
  \jcap, 11, 038

\bibitem[{{Oddo} {et~al.}(2020){Oddo}, {Sefusatti}, {Porciani}, {Monaco}, \&
  {S{\'a}nchez}}]{OddoEtal2020}
{Oddo}, A., {Sefusatti}, E., {Porciani}, C., {Monaco}, P., \& {S{\'a}nchez},
  A.~G. 2020, \jcap, 03, 056

\bibitem[{{Perko} {et~al.}(2016){Perko}, {Senatore}, {Jennings}, \&
  {Wechsler}}]{PerkoEtal2016A}
{Perko}, A., {Senatore}, L., {Jennings}, E., \& {Wechsler}, R.~H. 2016, arXiv:
  1610.09321 \eprint[arXiv]{1610.09321}

\bibitem[{{Pezzotta} {et~al.}(2021){Pezzotta}, {Crocce}, {Eggemeier},
  {S{\'a}nchez}, \& {Scoccimarro}}]{PezzottaEtal2021}
{Pezzotta}, A., {Crocce}, M., {Eggemeier}, A., {S{\'a}nchez}, A.~G., \&
  {Scoccimarro}, R. 2021, \prd, 104, 043531

\bibitem[{{Pezzotta} {et~al.}(2025){Pezzotta}, {Eggemeier}, {Gambardella},
  {Finkbeiner}, {S{\'a}nchez}, {Camacho Quevedo}, {Crocce}, {Lee},
  {Parimbelli}, \& {Scoccimarro}}]{PezzottaEtal2025}
{Pezzotta}, A., {Eggemeier}, A., {Gambardella}, G., {et~al.} 2025, \prd, 112,
  023520

\bibitem[{{Philcox} \& {Ivanov}(2022)}]{PhilcoxIvanov2022}
{Philcox}, O. H.~E. \& {Ivanov}, M.~M. 2022, \prd, 105, 043517

\bibitem[{{Philcox} {et~al.}(2022){Philcox}, {Ivanov}, {Cabass},
  {Simonovi{\'c}}, {Zaldarriaga}, \& {Nishimichi}}]{PhilcoxEtal2022}
{Philcox}, O. H.~E., {Ivanov}, M.~M., {Cabass}, G., {et~al.} 2022, \prd, 106,
  043530

\bibitem[{{Philcox} {et~al.}(2021){Philcox}, {Ivanov}, {Zaldarriaga},
  {Simonovi{\'c}}, \& {Schmittfull}}]{PhilcoxEtal2021}
{Philcox}, O. H.~E., {Ivanov}, M.~M., {Zaldarriaga}, M., {Simonovi{\'c}}, M.,
  \& {Schmittfull}, M. 2021, \prd, 103, 043508

\bibitem[{{Potter} {et~al.}(2017){Potter}, {Stadel}, \&
  {Teyssier}}]{PotterStadelTeyssier2017}
{Potter}, D., {Stadel}, J., \& {Teyssier}, R. 2017, Computational Astrophysics
  and Cosmology, 4, 2

\bibitem[{{Pozzetti} {et~al.}(2016){Pozzetti}, {Hirata}, {Geach}, {Cimatti},
  {Baugh}, {Cucciati}, {Merson}, {Norberg}, \& {Shi}}]{PozzettiEtal2016}
{Pozzetti}, L., {Hirata}, C.~M., {Geach}, J.~E., {et~al.} 2016, \aap, 590, A3

\bibitem[{{Rizzo} {et~al.}(2023){Rizzo}, {Moretti}, {Pardede}, {Eggemeier},
  {Oddo}, {Sefusatti}, {Porciani}, \& {Monaco}}]{RizzoEtal2023}
{Rizzo}, F., {Moretti}, C., {Pardede}, K., {et~al.} 2023, \jcap, 01, 031

\bibitem[{{Salvalaggio} {et~al.}(2024){Salvalaggio}, {Castiblanco},
  {Nore{\~n}a}, {Sefusatti}, \& {Monaco}}]{SalvalaggioEtal2024}
{Salvalaggio}, J., {Castiblanco}, L., {Nore{\~n}a}, J., {Sefusatti}, E., \&
  {Monaco}, P. 2024, \jcap, 08, 046

\bibitem[{Schmittfull {et~al.}(2015)Schmittfull, Baldauf, \&
  Seljak}]{SchmittfullBaldaufSeljak2015}
Schmittfull, M., Baldauf, T., \& Seljak, U. 2015, \prd, 91, 043530

\bibitem[{{Schmittfull} \& {Moradinezhad
  Dizgah}(2021)}]{SchmittfullMoradinezhad2021}
{Schmittfull}, M. \& {Moradinezhad Dizgah}, A. 2021, \jcap, 03, 020

\bibitem[{Scoccimarro(2000)}]{Scoccimarro2000A}
Scoccimarro, R. 2000, \apj, 542, 1

\bibitem[{Scoccimarro {et~al.}(1998)Scoccimarro, Colombi, Fry, Frieman, Hivon,
  \& Melott}]{ScoccimarroEtal1998}
Scoccimarro, R., Colombi, S., Fry, J.~N., {et~al.} 1998, \apj, 496, 586

\bibitem[{Scoccimarro {et~al.}(1999{\natexlab{a}})Scoccimarro, Couchman, \&
  Frieman}]{ScoccimarroCouchmanFrieman1999}
Scoccimarro, R., Couchman, H. M.~P., \& Frieman, J.~A. 1999{\natexlab{a}},
  \apj, 517, 531

\bibitem[{Scoccimarro {et~al.}(2001)Scoccimarro, {Feldman}, Fry, \&
  Frieman}]{ScoccimarroEtal2001B}
Scoccimarro, R., {Feldman}, H.~A., Fry, J.~N., \& Frieman, J.~A. 2001, \apj,
  546, 652

\bibitem[{{Scoccimarro} \& {Frieman}(1996)}]{ScoccimarroFrieman1996}
{Scoccimarro}, R. \& {Frieman}, J. 1996, \apjs, 105, 37

\bibitem[{Scoccimarro {et~al.}(2004)Scoccimarro, Sefusatti, \&
  Zaldarriaga}]{ScoccimarroSefusattiZaldarriaga2004}
Scoccimarro, R., Sefusatti, E., \& Zaldarriaga, M. 2004, \prd, 69, 103513

\bibitem[{Scoccimarro {et~al.}(1999{\natexlab{b}})Scoccimarro, Zaldarriaga, \&
  Hui}]{ScoccimarroZaldarriagaHui1999}
Scoccimarro, R., Zaldarriaga, M., \& Hui, L. 1999{\natexlab{b}}, \apj, 527, 1

\bibitem[{Sefusatti(2009)}]{Sefusatti2009}
Sefusatti, E. 2009, \prd, 80, 123002

\bibitem[{Sefusatti {et~al.}(2006)Sefusatti, Crocce, Pueblas, \&
  Scoccimarro}]{SefusattiEtal2006}
Sefusatti, E., Crocce, M., Pueblas, S., \& Scoccimarro, R. 2006, \prd, 74,
  023522

\bibitem[{{Sefusatti} {et~al.}(2016){Sefusatti}, {Crocce}, {Scoccimarro}, \&
  {Couchman}}]{SefusattiEtal2016}
{Sefusatti}, E., {Crocce}, M., {Scoccimarro}, R., \& {Couchman}, H.~M.~P. 2016,
  \mnras, 460, 3624

\bibitem[{Sefusatti \& Komatsu(2007)}]{SefusattiKomatsu2007}
Sefusatti, E. \& Komatsu, E. 2007, \prd, 76, 083004

\bibitem[{Sefusatti \& Scoccimarro(2005)}]{SefusattiScoccimarro2005}
Sefusatti, E. \& Scoccimarro, R. 2005, \prd, 71, 063001

\bibitem[{{Senatore}(2015)}]{Senatore2015}
{Senatore}, L. 2015, \jcap, 11, 007

\bibitem[{Senatore \& Zaldarriaga(2015)}]{SenatoreZaldarriaga2015}
Senatore, L. \& Zaldarriaga, M. 2015, \jcap, 02, 013

\bibitem[{{Sheth} {et~al.}(2013){Sheth}, {Chan}, \&
  {Scoccimarro}}]{ShethChanScoccimarro2013}
{Sheth}, R.~K., {Chan}, K.~C., \& {Scoccimarro}, R. 2013, \prd, 87, 083002

\bibitem[{{Simonovi{\'c}} {et~al.}(2018){Simonovi{\'c}}, {Baldauf},
  {Zaldarriaga}, {Carrasco}, \& {Kollmeier}}]{SimonovicEtal2018}
{Simonovi{\'c}}, M., {Baldauf}, T., {Zaldarriaga}, M., {Carrasco}, J.~J., \&
  {Kollmeier}, J.~A. 2018, \jcap, 04, 030

\bibitem[{{Sugiyama} {et~al.}(2019){Sugiyama}, {Saito}, {Beutler}, \&
  {Seo}}]{SugiyamaEtal2019}
{Sugiyama}, N.~S., {Saito}, S., {Beutler}, F., \& {Seo}, H.-J. 2019, \mnras,
  484, 364

\bibitem[{{Tinker} {et~al.}(2010){Tinker}, {Robertson}, {Kravtsov}, {Klypin},
  {Warren}, {Yepes}, \& {Gottl{\"o}ber}}]{TinkerEtal2010}
{Tinker}, J.~L., {Robertson}, B.~E., {Kravtsov}, A.~V., {et~al.} 2010, \apj,
  724, 878

\bibitem[{Verde {et~al.}(2002)Verde, {Heavens}, {Percival}, Matarrese, {Baugh},
  {Bland-Hawthorn}, {Bridges}, {Cannon}, {Cole}, {Colless}, {Collins}, {Couch},
  {Dalton}, {De Propris}, {Driver}, {Efstathiou}, {Ellis}, {Frenk},
  {Glazebrook}, {Jackson}, {Lahav}, {Lewis}, {Lumsden}, {Maddox}, {Madgwick},
  {Norberg}, {Peacock}, {Peterson}, {Sutherland}, \& {Taylor}}]{VerdeEtal2002}
Verde, L., {Heavens}, A.~F., {Percival}, W.~J., {et~al.} 2002, \mnras, 335, 432

\bibitem[{Verde {et~al.}(2000)Verde, Wang, Heavens, \&
  Kamionkowski}]{VerdeEtal2000}
Verde, L., Wang, L., Heavens, A.~F., \& Kamionkowski, M. 2000, \mnras, 313, 141

\bibitem[{{Vlah} {et~al.}(2016){Vlah}, {Seljak}, {Yat Chu}, \&
  {Feng}}]{VlahEtal2016}
{Vlah}, Z., {Seljak}, U., {Yat Chu}, M., \& {Feng}, Y. 2016, \jcap, 03, 057

\bibitem[{{Voivodic} \& {Barreira}(2021)}]{VoivodicBarreira2021}
{Voivodic}, R. \& {Barreira}, A. 2021, \jcap, 05, 069

\bibitem[{{Wadekar} \& {Scoccimarro}(2020)}]{WadekarScoccimarro2020}
{Wadekar}, D. \& {Scoccimarro}, R. 2020, \prd, 102, 123517

\bibitem[{{Wang}(2008)}]{Wang2008}
{Wang}, Y. 2008, \prd, 77, 123525

\bibitem[{{Wang} {et~al.}(2022){Wang}, {Zhai}, {Alavi}, {Massara}, {Pisani},
  {Benson}, {Hirata}, {Samushia}, {Weinberg}, {Colbert}, {Dor{\'e}}, {Eifler},
  {Heinrich}, {Ho}, {Krause}, {Padmanabhan}, {Spergel}, \&
  {Teplitz}}]{WangEtal2022}
{Wang}, Y., {Zhai}, Z., {Alavi}, A., {et~al.} 2022, \apj, 928, 1

\bibitem[{{Yankelevich} \& {Porciani}(2019)}]{YankelevichPorciani2019}
{Yankelevich}, V. \& {Porciani}, C. 2019, \mnras, 483, 2078

\bibitem[{{Zhang} {et~al.}(2022){Zhang}, {D'Amico}, {Senatore}, {Zhao}, \&
  {Cai}}]{ZhangEtal2022}
{Zhang}, P., {D'Amico}, G., {Senatore}, L., {Zhao}, C., \& {Cai}, Y. 2022,
  \jcap, 02, 036

\bibitem[{{Zuntz} {et~al.}(2015){Zuntz}, {Paterno}, {Jennings}, {Rudd},
  {Manzotti}, {Dodelson}, {Bridle}, {Sehrish}, \&
  {Kowalkowski}}]{ZuntzEtal2015}
{Zuntz}, J., {Paterno}, M., {Jennings}, E., {et~al.} 2015, Astronomy and
  Computing, 12, 45

\end{thebibliography}
